\documentclass[a4paper,fleqn]{cas-sc}

\usepackage[numbers]{natbib}
\usepackage{chemformula}
\usepackage[version=4]{mhchem}
\usepackage{threeparttable}
\usepackage{graphicx} 
\usepackage{epstopdf}
\usepackage{algorithm}
\usepackage{algorithmic}
\usepackage{subcaption}
\usepackage{stfloats}
\usepackage{framed} 
\usepackage{multicol} 
\usepackage{nomencl} 
\makenomenclature
\usepackage{etoolbox}
\renewcommand\nomgroup[1]{%
\item[\bfseries
\ifstrequal{#1}{A}{Nomenclature}{%
\ifstrequal{#1}{D}{Abbreviations}{%
\ifstrequal{#1}{C}{Subscripts}{%
\ifstrequal{#1}{B}{Superscripts}{}}}}%
]\vspace{10pt}}
\renewcommand*\nompreamble{\begin{multicols}{2}}
\renewcommand*\nompostamble{\end{multicols}}

\def\tsc#1{\csdef{#1}{\textsc{\lowercase{#1}}\xspace}}
\tsc{WGM}
\tsc{QE}
\tsc{EP}
\tsc{PMS}
\tsc{BEC}
\tsc{DE}

\begin{document}
\let\WriteBookmarks\relax
\def\floatpagepagefraction{1}
\def\textpagefraction{.001}
\shorttitle{Applied Energy}
\shortauthors{Yuxuan Gu et~al.}

\title [mode = title]{A simplified electro-chemical lithium-ion battery model applicable for in situ monitoring and online control}



\author[1]{Yuxuan Gu}[style=chinese]
\fnmark[1]


\address[1]{Department of Electrical Engineering, Tsinghua University, Beijing, 100084, China.}

\author[2]{Jianxiao Wang}[style=chinese]
\address[2]{School of Electrical and Electronic Engineering, North China Electric Power University, Beijing, 102206, China}
\fnmark[2]

\author[1]{Yuanbo Chen}[style=chinese]
\fnmark[3]

\author[3]{Zhongwei Deng}[style=chinese]
\address[3]{College of Mechanical and Vehicle Engineering, Chongqing University, Chongqing, 400044, China}
\fnmark[4]

\author[1]{Hongye Guo}[style=chinese]
\fnmark[5]

\author[1]{Kedi Zheng}[style=chinese]
\fnmark[6]

\author[1]{Qixin Chen}[style=chinese,
                       orcid=0000-0002-3733-8641]
\cormark[1]
\fnmark[7]






\cortext[cor1]{Corresponding author}
\fntext[fn1]{This work was supported by the China National Key R\&D Program (2021YFB2401201).}


\begin{abstract}
The penetration of lithium-ion batteries (LIBs) in transport, energy and communication systems is increasing rapidly. A meticulous LIB model applicable for precise in situ monitoring and convenient online control is sought to bridge the gap between research and applications.
\textcolor{blue}{On the basis of the classic pseudo-two-dimensional (P2D) model, a simplified electro-chemical model for LIBs that is adaptive to variant working environments and materials is proposed.}
\textcolor{blue}{Specifically, a bottom-up approach is adopted to decompose the complex
P2D model into decoupled sub-models, including the time-variant parameter model, solution-phase migration model, solid-phase diffusion model, reaction distribution model and output model.}
\textcolor{blue}{The simplification schemes of different sub-models are developed independently and finally reassembled.}
\textcolor{blue}{For ease of online simulation and control in real-world implementations, a discrete-time state-space realization of the proposed model is derived.}
\textcolor{blue}{A full-cycle simulation framework, including the initialization process, stabilization method and closed-loop correction scheme, is designed as well.}
\textcolor{blue}{Numerical experiments for the commonly used NCM
and LFP cells in different operating scenarios demonstrate that the proposed model can accurately predict battery output along with the spatial distribution of internal states with limited computation resources, which provides opportunities for degradation analysis and meticulous management of LIBs in practice.}
\end{abstract}




\maketitle

\section{Introduction}
Lithium-ion batteries (LIBs) have become the dominant energy source in various applications, such as electric vehicles and grid-level energy storage. 
The advantages of LIBs include high power and energy density, long lifespan, high efficiency, low self-discharge, and non-memory effect. With the widespread usage of LIBs, security and economic concerns are rapidly rising as well. 
Generally, LIBs are monitored and controlled by a BMS
to achieve safe, efficient and reliable operation. 
The BMS estimates the SOC, SOP
and SOH of LIBs based on the battery model and output measurements and then generates optimal control actions.
For an advanced BMS, the underlying battery model should have the following features. First, it should be able to provide information of internal states such as potentials, \ce{Li^+} concentrations, reaction rates, etc., for meticulous management. Second, it can be easily converted to state-space representations for online control. Last, it should be simple, with low requirements on the processor and memory since an LIB pack can usually contain hundreds of cells.


Existing LIB models can be categorized into three groups: data-driven, empirical, and electro-chemical. Data-driven models (black-box models) are usually fitted on experimental data by statistical methods to predict battery dynamics~\cite{li_big_2019}. Empirical models usually refer to the ECM, which uses a series of resistors and capacitors to mimic a battery~\cite{hu_comparative_2012, hu_model-based_2014, ding_improved_2019}. Electro-chemical models use a set of PDAEs to depict chemical and physical processes at the microscale inside the battery cell~\cite{Doyle_1993}. Compared with the former two groups, electro-chemical models can give mechanistic interpretations of the battery and are adaptive to a wide range of working scenarios. However, electro-chemical models are usually complex and difficult to transform into state-space models for control, which impedes their widespread usage.

A representative of electro-chemical models is the so-called P2D model, which was proposed by Doyle et al. \cite{Doyle_1993} and later became the original source of subsequent models. Based on the porous electrode theory and concentrated solution theory, the P2D model depicts the diffusion/migration of ions in the electrode/electrolyte and their intercalation at the solid-solution interface. Since the seminal work of~\cite{Doyle_1993}, many works have refined the P2D model, such as incorporating the double layer capacitance~\cite{legrand_including_2014,chu_control-oriented_2019,zhang_electrochemical_2020}, constant-phase-element dynamics~\cite{chu_control-oriented_2019}, ageing factors~\cite{zhang_electrochemical_2020}, and varying parameters~\cite{farkhondeh_mathematical_2011, gao_implementation_2021}. However, these refinements further increase the complexity. To enable the practical usage of electro-chemical models, a plethora of works have focused on model reduction techniques, which can be categorized into three approaches: numerical, analytical, and hybrid.

Numerical approaches focus on developing highly efficient computation methods for PDAEs in the P2D model. Mathematically, PDAEs are spatially discretized into ordinary differential and algebraic equations and then solved iteratively. For the discretization process, finite-difference~\cite{xiong_electrochemical_2018}, control-volume formulation~\cite{farkhondeh_mathematical_2011}, Crank--Nicolson~\cite{Doyle2010DesignAS}, forward time-central space approximation~\cite{ringbeck_uncertainty-aware_2020} and asymptotic reduction~\cite{hennessy_asymptotic_2020} method have been proposed. Ref.~\cite{Cai_2009} used proper orthogonal decomposition to solve the whole model, which was also used by~\cite{ZHAO2018440} to calculate solid-phase potentials. Ref.~\cite{zou_framework_2016} developed a solution scheme based on singular perturbation and averaging theory. However, numerically reduced models still have high orders (30-100 orders). In addition, the deficiency of a control-oriented view precludes online implementation of these models.

\begin{table*}[!t]   
    \begin{framed}
    \printnomenclature
    \nomenclature[A]{$m$}{battery weight (kg)}
    \nomenclature[A]{$L$}{electrode or separator thickness (m)}
    \nomenclature[A]{$A$}{electrode or separator projected area (m$^2$)}
    \nomenclature[A]{$A_{\rm surf}$}{battery surface area (m$^2$)}
    \nomenclature[A]{$D$}{diffusion coefficient (m$^2$/s)}
    \nomenclature[A]{$\sigma$}{electronic conductivity (S/m)}
    \nomenclature[A]{$\kappa$}{ionic conductivity (S/m)}
    \nomenclature[A]{$t_+^0$}{transfer number (dimensionless)}
    \nomenclature[A]{$R_c$}{battery contact resistance ($\Omega$)}
    \nomenclature[A]{$R_f$}{SEI film resistance ($\Omega\cdot$m$^2$)}
    \nomenclature[A]{$R_s$}{radius of the active particle (m)}
    \nomenclature[A]{$a_s$}{specific surface area per volume (1/m)}
    \nomenclature[A]{$M$}{molar mass (kg/mol)}
    \nomenclature[A]{$\rho$}{density (kg/m$^3$)}
    \nomenclature[A]{$\varepsilon$}{volume fraction (dimensionless)}
    \nomenclature[A]{$c$}{\ce{Li^+} concentration (mol/m$^3$)} 
    \nomenclature[A]{$C_p$}{battery heat capacity (J/kg/K)}
    \nomenclature[A]{$C_Q$}{battery charge capacity (mAh)}
    \nomenclature[A]{$h_c$}{battery heat transfer coefficient (W/m$^2$/K)}
    \nomenclature[A]{$k_r$}{reaction rate coefficient (A$\cdot$m$^{2.5}$/mol$^{1.5}$)}
    \nomenclature[A]{$k_d$}{time constant coefficient of diffusion (dimensionless)}
    \nomenclature[A]{$F$}{Faraday constant (96485 C/mol)}
    \nomenclature[A]{$R$}{gas constant (8.314 J/mol/K)}
    \nomenclature[A]{$p$}{Bruggeman coefficient (dimensionless)}
    \nomenclature[A]{$j_n$}{reaction rate, also named the pore-wall flux (mol/m$^2$/s)}
    \nomenclature[A]{$i_0$}{exchange current density (A/m$^2$)}
    \nomenclature[A]{$x$}{coordinate along the thickness direction (m)}
    \nomenclature[A]{$r$}{coordinate along the radius direction (m)}
    \nomenclature[A]{$t$}{time (s)}
    \nomenclature[A]{$\Phi$}{electrical potential (V)}
    \nomenclature[A]{$U_{\mathrm{OCP}}$}{equilibrium potential (V)}
    \nomenclature[A]{$U_{\mathrm{OCV}}$}{open circuit voltage (V)}
    \nomenclature[A]{$\eta$}{over-potential of reaction (V)}
    \nomenclature[A]{$T$}{battery temperature (K)}
    \nomenclature[A]{$E_A$}{activation energy (J/mol)}
    \nomenclature[A]{$\kappa_D$}{diffusional conductivity (J/C)}
    \nomenclature[A]{$\alpha_a$}{anodic transfer coefficient (dimensionless)}
    \nomenclature[A]{$\alpha_c$}{cathodic transfer coefficient (dimensionless)}
    \nomenclature[A]{$\theta$}{\ce{Li^+} stoichiometry (dimensionless)}
    \nomenclature[A]{$Q$}{\ce{Li^+} quantity (mol)}
    \nomenclature[A]{$i$}{current density (A/m$^2$)}
    \nomenclature[A]{$I$}{applied current on the battery (A)}
    \nomenclature[A]{$\tau$}{time constant (s)}
    \nomenclature[A]{$V_t$}{terminal voltage (V)}
    \nomenclature[A]{$T_{\mathrm{amb}}$}{ambient temperature (K)}
    \nomenclature[A]{$H$}{heat (J)}

    \nomenclature[B]{$+$}{positive electrode}
    \nomenclature[B]{$-$}{negative electrode}
    \nomenclature[B]{sep}{separator}
    \nomenclature[B]{eff}{effective}
    \nomenclature[B]{ref}{value at the reference temperature}
    
    \nomenclature[C]{$s$}{solid-phase (active particles)}
    \nomenclature[C]{$ss$}{solid-phase surface}
    \nomenclature[C]{$e$}{solution-phase (electrolyte)}
    \nomenclature[C]{$s-e$}{solution-solid interface}
    \nomenclature[C]{$l$}{time slot index}
    \nomenclature[C]{max}{maximum}
    \nomenclature[C]{min}{minimum}
    \nomenclature[C]{0}{initial state}
    
    \nomenclature[D]{LIB}{lithium-ion battery}
    \nomenclature[D]{BMS}{battery management system}
    \nomenclature[D]{SOC}{state of charge}
    \nomenclature[D]{SOP}{state of power}
    \nomenclature[D]{SOH}{state of health}
    \nomenclature[D]{ECM}{equivalent circuit model}
    \nomenclature[D]{PDAE}{partial differential and algebraic equation}
    \nomenclature[D]{P2D}{pseudo-two-dimensional}
    \nomenclature[D]{ESP}{extended single particle}
    \end{framed}
\end{table*}

Analytical approaches aim to find approximate expressions for concerned states in the battery, which are obtained by either intuitive assumptions or rigorous derivation. To approximate the solution-phase \ce{Li^+} concentration, constants~ \cite{lyu_situ_2019,hu_control_2020}, parabolic or cubic polynomials~\cite{khaleghi_rahimian_extension_2013,han_simplification_2015_1, wu_evaluation_2021, wang_lithium-ion_2020, li_reduced-order_2021,li_development_2019} and residue grouping~\cite{ZHAO2018440,bi_adaptive_2020,fan_systematic_2020} are commonly used. Sinusoidal and exponential functions were also tried in~\cite{DAO2012329,li_aging_2020}. To approximate the solid-phase surface \ce{Li^+} concentration, \textcolor{blue}{existing research can be categorized into three approaches.} The first approach simplifies the transfer function of the solid-phase surface \ce{Li^+} concentration in the frequency domain and then obtains its reduced state-space realization. The representative approach is the Pad\'{e} approximation, which was first proposed by Forman et~al.~\cite{Forman_2011} and then widely used in subsequent research~\cite{gao_implementation_2021,ZHAO2018440,wu_evaluation_2021,bi_adaptive_2020,DAO2012329,9207759}. It uses a rational polynomial to approximate the original transfer function by Taylor expansion. Since the Pad\'{e} approximation is accurate only at low frequencies, ref.~\cite{smith_control_2007,li_physics-based_2017} determined the coefficients of rational polynomials by fitting the frequency response over a wider frequency band. Ref.~\cite{han_simplification_2015_1,feng_co-estimation_2020} determined the coefficients by fitting the state trajectories in the time domain. Upon state-space realization, these models are generally equivalent to the combination of several first-order inertial processes. \textcolor{blue}{Note also that ref.~\cite{li_physics-based_2017,wang_experimental_2020} used fractional-order representations to replace the first-order processes to achieve high accuracy in recent years.} The second approach uses the realization algorithm (xRA) to directly obtain the state-space representation~\cite{chu_control-oriented_2019,  LEE2012430}. The xRA can generate a discrete-time state-space realization with a unit-pulse response similar to that of the original transfer function. The third approach directly approximates the time domain and assumes a polynomial distribution of \ce{Li^+} concentrations along the r-axis~\cite{wang_lithium-ion_2020,li_reduced-order_2021,li_development_2019,li_aging_2020,klein_electrochemical_2013,hu_linear_2012}. To approximate the reaction rate, constants, stepwise lines~\cite{li_physics-based_2017}, parabolic polynomials~\cite{deng_polynomial_2018,han_simplification_2015_2,luo_new_2013} and cubic polynomials~\cite{li_control-oriented_2021} have been proposed. \textcolor{blue}{To conclude, analytical approaches have generally focused on approximating \ce{Li^+} concentrations but have rarely discussed the reaction rate in detail.}

Hybrid approaches treat some part of the model with numerical methods and the remaining part with analytical methods, e.g., applying the Pad\'{e} approximation for solid-phase and finite-difference for the solution phase~\cite{li_electrochemical_2020,Subramanian_2009}.

Upon reviewing the aforementioned research and attempting several highly cited models, some problems emerge to be solved.
First, for solution-phase \ce{Li^+} concentrations, previous works mainly focused on the spatial distribution. To develop a control-oriented model, time trajectory modelling and discrete-time state-space realization require consumption.
Second, for solid-phase \ce{Li^+} concentrations, we find that the simplified state-space representations obtained by the Pad\'{e} approximation, response or time-series optimization, or xRA are likely to suffer from oscillations. Developing an accurate and stable method is necessary for practical usage.
Third, for the reaction rate distribution, polynomial approximations used by existing models are based on intuitive assumptions and are not adaptive to various scenarios. As the key to in situ monitoring, degradation prediction and lumped-state estimation (SOC, SOP, SOH)~\cite{allam_online_2020}, a rigorous mathematical formula is sought.
Fourth, for the whole cell, a model considering coupled electrical, chemical, physical, and thermal dynamics along with time-variant parameters and a full cycle simulation framework containing the initialization process, stabilization method, and closed-loop correction scheme are desired for real-world applications.

To bridge the gaps mentioned above, a high-fidelity simplified electro-chemical model along with a simulation framework are proposed in this work. Parameters sensitive to temperatures or concentrations are extracted and then modelled by the Arrhenius law and empirical formulas. By taking the ensemble average strategy, the solution-phase migration is simplified with two coupled first-order inertial processes derived from mass conservation and Fick's law. For solid-phase diffusion, we also take the ensemble average strategy and find the suitable time constants of first-order inertial processes to approximate the surface \ce{Li^+} concentrations that achieve a balance between accuracy, stabilization and simplicity. For the reaction rate distribution, the conceptual content is fully considered, and a rigorous mathematical expression without intuitive assumptions is derived by simultaneously solving chemical equations and electrical equations. For the output, the terminal voltage is derived based on the obtained reaction rate distribution formula, and the cell temperature is derived based on a lumped thermal model. The above sub-models are assembled to obtain the final simplified model and its discrete-time state-space realization. For full-cycle simulation of the battery, a initialization process, stabilization method and closed-loop correction scheme composed of basic operators or simple optimization that require low computational resources are designed. For validation, the proposed model is compared with the full-order P2D model, a classic ESP model and a well-cited advanced ESP model~\cite{han_simplification_2015_1,han_simplification_2015_2} under different working scenarios for commonly used NCM and LFPO cells.

The contributions of this work are fourfold.
\begin{itemize}
\item \textcolor{blue}{A bottom-up approach is designed to construct the simplified electro-chemical lithium-ion battery model by decomposing the sophisticated P2D model into decoupled sub-models, including the time-variant parameter model, solution-phase migration model, solid-phase diffusion model, reaction distribution model and output model, which makes the model not only adaptive to variant working environments and materials but also reserves potential for future upgrades.}
\item \textcolor{blue}{Decoupled sub-models are simplified independently according their specific characteristics. The ensemble average strategy is used to derive the simplified solution-phase migration model and solid-phase diffusion model. The rigorous mathematical expression of the reaction rate distribution is derived by simultaneously solving chemical equations and electrical equations. The terminal voltage is derived based on the obtained reaction rate distribution formula, and the cell temperature is derived based on a lumped thermal model.}
\item \textcolor{blue}{Decoupled sub-models are assembled to obtain the final battery model. For ease of online simulation and control in real-world implementations, a discrete-time state-space realization of the model is derived. A full-cycle simulation framework including the initialization process, stabilization method and closed-loop correction scheme that requires low computational resources is designed.}
\item \textcolor{blue}{For validation, comprehensive numerical experiments are conducted. Specifically, the proposed model is tested under different scenarios for commonly used NCM and LFPO cells, including galvanostatic current and dynamic current protocols, low-temperature and high-temperature environments, and low and high C-rates. Comparison against two highly cited ESP models reveals the superiority of this work, which provides opportunities for degradation analysis and meticulous management of batteries in practice.}
\end{itemize}

The rest of this paper is organized as follows: Section~\ref{sec:model} introduces the bottom-up approach to construct the simplified lithium-ion battery model. Section~\ref{sec:method} describes the discrete-time state-space realization of the model and designs a full-cycle simulation framework, including a initialization process, stabilization method and closed-loop correction scheme. Section~\ref{sec:case} presents the results of numerical experiments. Section~\ref{sec:con} draws conclusions.

\section{\textcolor{blue}{Bottom-up modelling approach}}\label{sec:model}
\textcolor{blue}{
The simplified model is used to provide internal states of the battery for in situ monitoring, so we start from the classic full-order P2D model instead of the ECM. By taking a bottom-up approach, the sophisticated P2D model is decomposed into decoupled sub-models first and reassembled to obtain the final simplified model. Generally, the P2D model is appropriate for the battery with the following settings}:
\begin{itemize}
\item The electrodes have a porous structure where the solid phase is mainly composed of active particles and the solution phase is filled by the electrolyte. 
The separator is a perforated microplastic that insulates electrons but allows ions to pass.
\item During the charge, lithium ions deintercalate from active particles in the negative electrode, migrate through the electrolyte and pass through the separator, finally intercalating into active particles in the positive electrode.
Meanwhile, electrons are transported through the current collector from the negative current collector to the positive in the external circuit. During discharge, ions and electrons are transported in reverse directions. 
\item \textcolor{blue}{The diffusion of \ce{Li^+} in the active particle and the migration of \ce{Li^+} in the electrolyte obey Fick's second law, i.e., the cell should be made of intercalation electrode materials such as \ce{LiNi_xMn_yCo_{1-x-y}O_2}, \ce{LiFePO_4}, \ce{LiCoO_2}, \ce{LiMn_2O_4}, and Graphite (\ce{LiC_6}), and the electrolyte should satisfy concentrated solution theory such as the commonly used PC-EC-DMC solvent~\cite{smith_control_2007}.}
\end{itemize}
\textcolor{blue}{According to existing industrial practice, many commercial LIBs meet the above requirements, thus guaranteeing the applicability of the P2D model and its derivatives.}

Typically, \textcolor{blue}{whether the battery cell is cylindrical or prismatic, its micro-structure is sandwich-like,}, i.e., terminals of the cell are current collectors connecting the external circuit, between which lie three domains in order: negative electrode, separator and positive electrode, as shown in Fig. 1. The basic formulas of the P2D model are given below, where \textcolor{blue}{ Eqs. (\ref{equ:solution-migration}) and (\ref{equ:solid-diffusion}) depict the diffusion of \ce{Li^+} in the solution phase and solid phase, respectively. Eqs. (\ref{equ:Butler-Volmer}-\ref{equ:solution-phi}) establish the spatial distribution of the chemical reaction rates and potentials across the thickness direction.}
\begin{framed}\noindent
\begin{minipage}{0.5\textwidth}
    \label{fig:cell}
    \centering
	\includegraphics[width=1\textwidth]{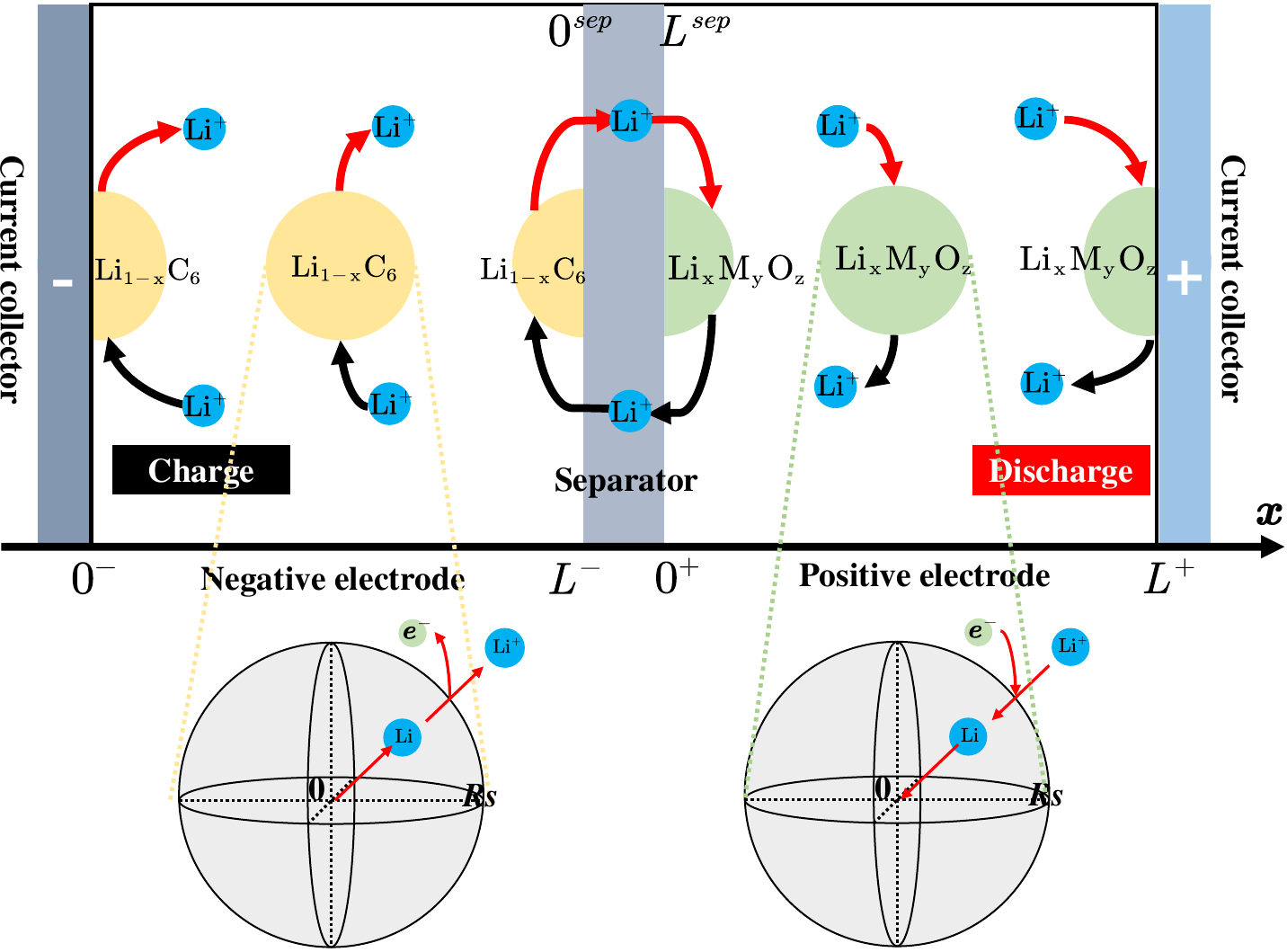}
\captionof{figure}{LIB cell structure and working mechanism.}
\end{minipage}%
\begin{minipage}{0.5\textwidth}
\begin{eqnarray}
    \label{equ:solution-migration}
    & \varepsilon_{e}^{\pm} \frac{\partial }{\partial t} c_e = \frac{\partial}{\partial x}\left(D_e^{\mathrm{eff}} \frac{\partial }{\partial x} c_e \right) + a_s^{\pm}\left(1-t_+^0\right)j_n, \\
    \nonumber
    & \quad\\
    \label{equ:solid-diffusion}
    & \frac{\partial }{\partial t} c_s = \frac{D_s}{r^2} \frac{\partial}{\partial r}\left(r^2 \frac{\partial }{\partial r} c_s \right), \\
    \nonumber
    & \quad\\
    \label{equ:Butler-Volmer}
    & j_n = \frac{i_0}{F} \left(\exp\left(\frac{ \alpha_a F\eta}{RT}\right)-\exp\left(\frac{-\alpha_c F \eta}{RT}\right) \right), \\
    \nonumber
    & \quad\\
    \label{equ:solid-phi}
    & \frac{\partial}{\partial x} \left( \sigma^{\mathrm{eff}} \frac{\partial}{\partial x} \phi_s \right) = a_s F j_n, \\
    \nonumber
    & \quad\\
    \label{equ:solution-phi}
    & \frac{\partial}{\partial x} \left( \kappa^{\mathrm{eff}} \frac{\partial}{\partial x} \phi_e \right) + \frac{\partial}{\partial x} \left( \kappa_D^{\mathrm{eff}} \frac{\partial}{\partial x} \ln{c_e} \right) = -a_s F j_n.
\end{eqnarray}
\end{minipage}
\end{framed}

\textcolor{blue}{In a bottom-up approach, we separately establish six simplified analytical sub-models, depicting the temperature- or concentration-incorporating
parameters, solution-phase migration, solid-phase diffusion, reaction rate distribution, potential distribution and thermal conservation. Finally, they are reassembled to form the final LIB model, which is applicable for in situ monitoring and online control simultaneously.}

\subsection{Parameters}\label{subsec:params}
\textcolor{blue}{Generally, parameters involved in an electro-chemical battery model can be categorized into two groups: parameters related to the manufacturing and parameters related to the physical or chemical properties of the battery material, as shown in Fig.~\ref{fig:params}. To improve the model fidelity, we further divide property parameters into constant and time-variant parameters. The splitting criterion is whether the parameter is affected by the \ce{Li^+} concentration or the temperature. In this part, the modelling of time-variant parameters is introduced in detail.}
\begin{figure}[!htb]
    \centering
\begin{minipage}{.3\textwidth}\label{fig:params}
        \centering
        \includegraphics[width=\textwidth]{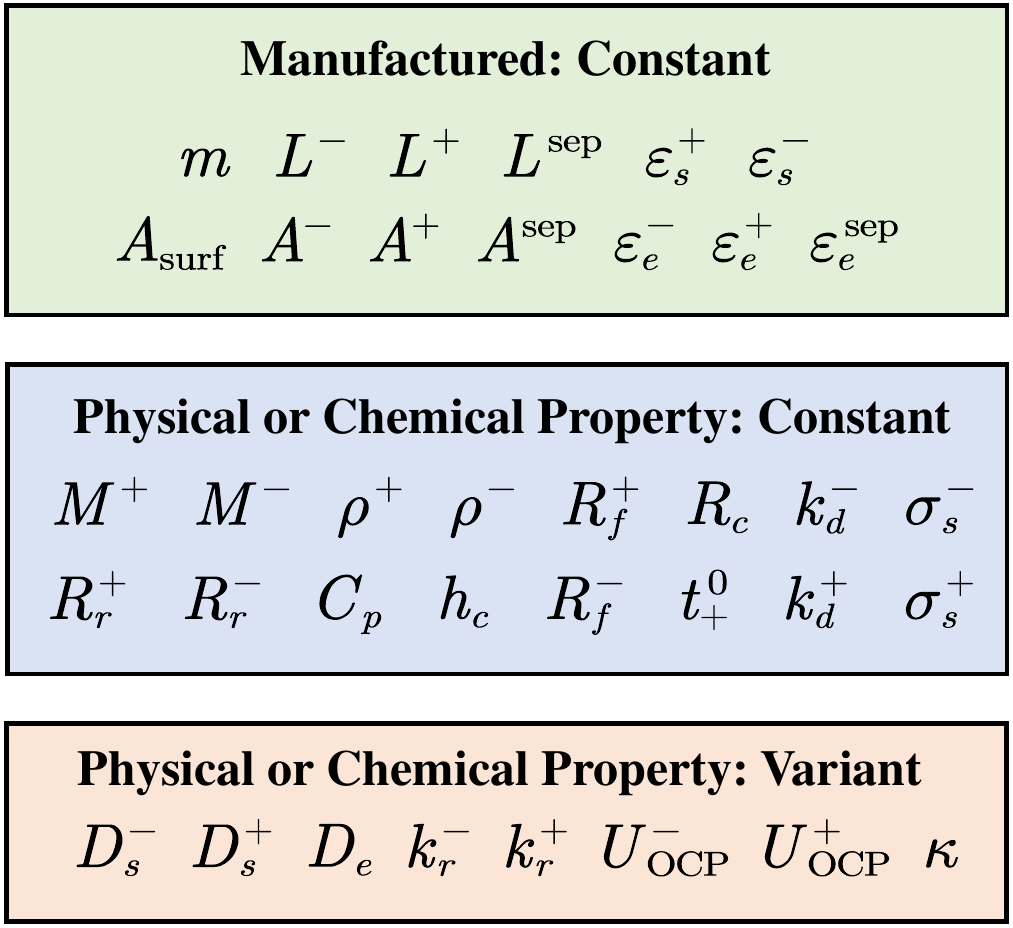}
\subcaption{Parameter categories. }\label{fig:params}
\end{minipage}%
\begin{minipage}{0.32\textwidth}\label{fig:Uocp}
        \centering
        \includegraphics[width=\textwidth]{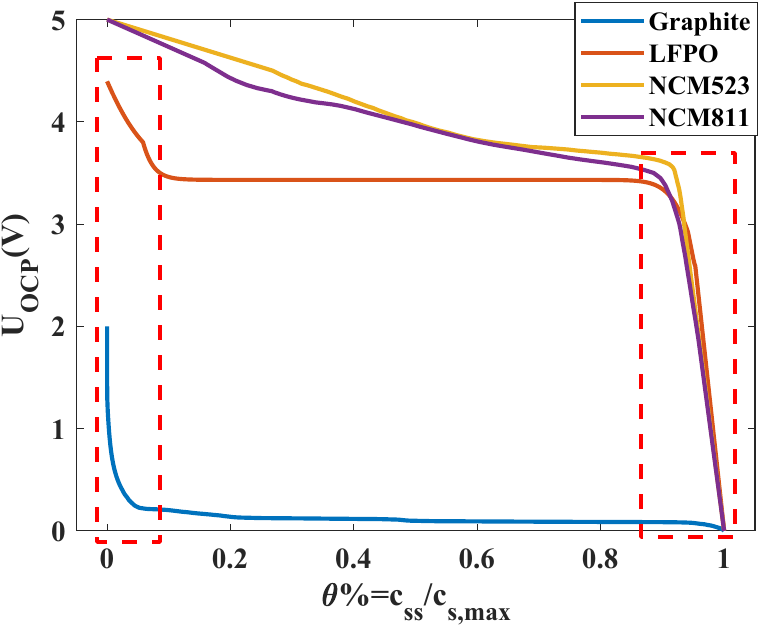}
\subcaption{Equilibrium potentials of commonly used electrode materials at 298 K}\label{fig:Uocp}
\end{minipage}
\caption{Parameters involved in constructing the proposed electro-chemical model.}
\end{figure}

\textcolor{blue}{First, we introduce the modelling of variant parameters in the solution phase. Existing commercial LIBs commonly use a similar mixture solvent as the electrolyte, e.g., PC-EC-DMC. Thus, many studies have investigated such electrolyte solvents deeply and have provided empirical formulas to describe their diffusion and conductivity properties~\cite{khaleghi_rahimian_extension_2013,wang_lithium-ion_2020,Valo_en_2005,torchio_lionsimba_2016,xu_fast_2019}, which are adopted in this work as well:}
\begin{equation}
    \label{equ:de_kappa}
    \lg D_e = -8.43-\frac{54}{T-229-0.005c_e}-\frac{2.2}{10^4}c_e.
\end{equation}
\begin{equation}
\label{equ:kappa}
\kappa = \frac{c_e}{10^4}\left( \left( \frac{0.494}{10^6}c_e^{2}+\frac{0.668}{10^3}c_{e}-10.5 \right)
    +\left(\frac{-8.86}{10^{10}}c_{e}^{2}-\frac{1.78}{10^5}c_{e}+0.074 \right)T +\left( \frac{2.8}{10^8}c_{e}-\frac{6.96}{10^5}\right) T^{2}\right)^2.
\end{equation}

\textcolor{blue}{Second, we introduce the modelling of time-variant parameters in the solid phase. However, different from the electrolyte, the specific values of the solid-phase parameters are not the same for different materials used by the specific battery. To maintain the generality of the proposed model, general expressions that capture the basic characteristics of commonly used intercalation materials are designed.} According to previous research, the solid-phase diffusion coefficient $D_s$ is related to the \ce{Li^+} concentration in the active particle and the cell temperature~\cite{RENGANATHAN2011442}. For simplicity, we decouple the impact of two factors. By denoting the bulk-averaged concentration of \ce{Li^+} in an active particle by $\bar{c}_s$, a linear approximation formula is adopted to describe the relation between $D_s$ and $\bar{c}_s$ first, as shown in the first formula of Eq. (\ref{equ:ds}):
\begin{equation}
    \centering 
    \label{equ:ds}
    D_s =  k_{D_s} \frac{\bar{c}_s}{c_{s,\mathrm{max}}} + b_{D_s}, k_{D_s}=exp \left(-\frac{E_{A,k_{D_s}}}{R}\left(\frac{1}{T}-\frac{1}{T^{\mathrm{ref}}}\right)  \right)k_{D_s}^{\mathrm{ref}}, b_{D_s}=exp \left(-\frac{E_{A,b_{D_s}}}{R}\left(\frac{1}{T}-\frac{1}{T^{\mathrm{ref}}}\right)  \right)b_{D_s}^{\mathrm{ref}}. 
\end{equation}
Next, the thermodynamic variation of $D_s$ is introduced by applying the Arrhenius law to the linearity coefficients, as shown in the last two formulas of Eq. (\ref{equ:ds}), where $E_{A,k_{D_s}}$ and $E_{A,b_{D_s}}$ are the activation energy and $k_{D_s}^{\mathrm{ref}}$ and $b_{D_s}^{\mathrm{ref}}$ are values at the reference temperature, i.e., 298 K.

\textcolor{blue}{Last, we introduce the modelling of variant parameters depicting the intercalation reaction occurring at the interface of the solid-phase and solution-phase.} The chemical kinetics parameter, the reaction rate coefficient $k_r$, determines how fast the reaction takes place and obeys the Arrhenius law when the temperature varies. Denoting the activation energy by $E_{A,k_r}$ and the value at the reference temperature by $k_{r}^{\mathrm{ref}}$, the formula of $k_r$ is given by:
\begin{equation}
    \label{equ:kr}
    k_r = \exp \left( -\frac{E_{A,k_r}}{R}\left( \frac{1}{T}-\frac{1}{T^{\mathrm{ref}}} \right)  \right)k_{r}^{\mathrm{ref}}.
\end{equation}
\textcolor{blue}{The chemical thermodynamics parameter, the equilibrium potential $U_{\mathrm{OCP}}$, determines whether the reaction can take place. Different from $k_r$, $U_{\mathrm{OCP}}$ is mainly determined by the \ce{Li^+} stoichiometry at the surface of active particles. However, the relationship between these two parameters is non-linear and complex. Thus, we construct the look-up table for commonly used active materials in this work, as shown in Fig.~\ref{fig:Uocp}. Once the surface \ce{Li^+} stoichiometry $\theta_{ss}$ is determined, the corresponding $U_{\mathrm{OCP}}$ can be obtained by interpolation in the curves. The original data are extracted from the experiment~\cite{AutoLion}. Note that $U_{\mathrm{OCP}}$ is also slightly affected by the temperature. However, previous studies found that the order of magnitude of the change in $U_{\mathrm{OCP}}$ with temperature, $\frac{{\rm d} U_{\mathrm{OCP}}}{{\rm d} T}$, is approximately $10^{-4}$ V/K~\cite{torchio_lionsimba_2016}. This impact is neglected in this work for simplicity.
}

The remaining parameters are assumed to be constant values. However, in the long term, some parameters can vary with degradation. However, this work mainly focuses on real-time simulations, and degradation identification and analysis will be investigated in future research. \textcolor{blue}{Note that the modelling of the parameters introduced above is appropriate for the negative electrode, positive electrode and separator domains, so the superscripts $+$,$-$ and $\mathrm{sep}$ are omitted for notation simplicity.}

\subsection{Solution-phase migration model}\label{sec:solution-phase migration model}
\textcolor{blue}{Based on the law of material conservation}, the migration of \ce{Li^+} in the electrolyte along the thickness is depicted by Eq. (\ref{equ:solution-migration}). In this section, we introduce the simplified migration model derived from Eq. (\ref{equ:solution-migration}).

First, integrate the LHS and RHS of Eq. (\ref{equ:solution-migration}) along the x-axis over the electrode domain yields:
\begin{equation}
    \label{equ:solution-integration}
    \frac{1}{A^{\pm}} \frac{\partial \int_{0^{\pm}}^{L^{\pm}} A^{\pm}\varepsilon_e^{\pm} c_e(x,t) dx}{\partial t} =  D_e^{\mathrm{eff},\pm}(t) \frac{\partial c_e(x,t)}{\partial x} \bigg|_{0^{\pm}}^{L^{\pm}}  + \frac{(1-t_+^0)}{F}\int_{0^{\pm}}^{L^{\pm}} a_s^{\pm} F j_n(x,t)dx.
\end{equation}

\textcolor{blue}{As commonly adopted in existing research~\cite{han_simplification_2015_1,wang_lithium-ion_2020}, parabolic polynomials are used in this work to approximate the spatial distribution of $c_e$}, i.e., $c_e(x,t) = a_e^-(t)x^2+b_e^-(t)$ for $x\in [0^-, L^-]$ in the negative electrode and $c_e(x,t) =a_e^+(t)(x-L^+)^2 + b_e^+(t)$ for $x\in [0^+, L^+]$ in the positive electrode. Since the separator domain is very thin compared with the electrode domain, we apply linear approximation to represent $c_e(x,t)$ in this domain to avoid high complexity, i.e., $c_e(x,t) =a_e^{\mathrm{sep}}(t) x + b_e^{\mathrm{sep}}(t)$ for $x\in [0^{\mathrm{sep}}, L^{\mathrm{sep}}]$.

Now, we note that the numerator of the first term on the LHS of Eq. (\ref{equ:solution-integration}) is equal to the total quantity of \ce{Li^+} in the solution phase among the positive and negative electrode domains, denoted by $Q_e^{\pm}(t)$, respectively. Substituting the expressions of $c_e$ in the negative electrode and positive electrode into $Q_e^{\pm}(t)$ yields:
\begin{equation}
    \label{equ:Qe}
    \int_{0^{\pm}}^{L^{\pm}} A^{\pm}\varepsilon_e^{\pm} c_e(x,t) dx = Q_e^{\pm}(t) = A^{\pm}\varepsilon_e^{\pm} \left( \frac{1}{3}a_e^{\pm}(t)(L^{\pm})^3+b_e^{\pm}(t)L^{\pm} \right)
\end{equation}

We now turn to the RHS of Eq. (\ref{equ:solution-integration}). \textcolor{blue}{The first term equals $2A^{\pm}L^{\pm}D_e^{\mathrm{eff},\pm}(t)a_e^{\pm}(t)$ by substituting the expressions of $c_e$ into the original formula. According to Faraday's law, the second term equals the difference between current densities in the solid phase at two sides of the electrode, i.e., $\int_{0^{\pm}}^{L^{\pm}} a_s^{\pm} F j_n(x,t)dx = i_s(L^{\pm},t)-i_s(0^{\pm},t)$. Since the current densities in the solid-phase and solution-phase obey KCL, the boundary conditions of $i_s$ in the negative and positive electrodes are expressed by $i_s(0^-,t)=i_s(L^+,t)=\frac{I(t)}{A^{\pm}}$ and $i_s(0^+,t)=i_s(L^-,t)=0$. Thus, we have $\int_{0^{\pm}}^{L^{\pm}} a_s^{\pm} F j_n(x,t)dx = \mp \frac{I(t)}{A^{\pm}}$.} Substituting the above terms into Eq. (\ref{equ:solution-integration}) yields:
\begin{equation}
    \label{equ:Qe_diff}
    \frac{d Q_e^{\pm}(t)}{\rm{d}t} = 2A^{\pm}L^{\pm}D_e^{\mathrm{eff},\pm}(t)a_e^{\pm}(t) \mp \frac{(1-t_+^0)}{F}I(t).
\end{equation}

\textcolor{blue}{
Based on the material conservation law, the \ce{Li^+} concentration and flux are continuous at the boundaries between the negative electrode, the separator and the positive electrode, i.e., $c_e(L^-,t)=c_e(0^{\mathrm{sep}},t)$, $c_e(L^{\mathrm{sep}},t)=c_e(0^+,t)$, $D_e^{\mathrm{eff},-}(t) \frac{\partial c_e(L^-,t) }{\partial x}=D_e^{\mathrm{eff,sep}}(t)\frac{\partial c_e(0^{\mathrm{sep}},t)}{\partial x}$, and $D_e^{\mathrm{eff,sep}}(t)\frac{\partial c_e(L^{\mathrm{sep}},t) }{\partial x}=D_e^{\mathrm{eff},+}(t)\frac{\partial c_e(0^+,t)}{\partial x}$. Substituting the parabolic expressions of $c_e$ into the above boundary conditions yields:
}
\begin{equation}
    \label{equ:ce_flux_balance}
    \frac{L^-D_e^{\mathrm{eff},-}(t)a_e^-(t)}{D_e^{\mathrm{eff,sep}}(t)} + \frac{L^+D_e^{\mathrm{eff},+}(t)a_e^+(t)}{D_e^{\mathrm{eff,sep}}(t)} = 0, a_e^-(t)L^-\left(L^-+ \frac{D_e^{\mathrm{eff},-}(t)L^{\mathrm{sep}}}{D_e^{\mathrm{eff,sep}}(t)} \right) + b_e^-(t) =  a_e^+(t)L^+\left(L^++\frac{D_e^{\mathrm{eff},+}(t) L^{\mathrm{sep}}}{D_e^{\mathrm{eff,sep}}(t)}\right) + b_e^+(t).
\end{equation}

Eqs. (\ref{equ:Qe}) and (\ref{equ:ce_flux_balance}) can be compacted to matrix form:
\begin{equation}
	\label{equ:cecoef}
	\left[ \begin{matrix}
	\frac{D_e^{\mathrm{eff},-}(t)L^-}{D_e^{\mathrm{eff,sep}}(t)}&		0&		\frac{D_e^{\mathrm{eff},+}(t)L^+}{D_e^{\mathrm{eff,sep}}(t)}&		0\\
	L^-\left( L^-+\frac{D_e^{\mathrm{eff},-}(t)L^{\mathrm{sep}}}{D_e^{\mathrm{eff,sep}}(t)} \right)&		1&		-L^+\left( L^++\frac{D_e^{\mathrm{eff},+}(t)L^{\mathrm{sep}}}{D_e^{\mathrm{eff,sep}}(t)} \right)&		-1\\
	\left( L^- \right) ^3/3&		L^-&		0&		0\\
	0&		0&		\left( L^+ \right) ^3/3&		L^+\\
	\end{matrix} \right] \left[ \begin{array}{c}
	a_{e}^{-}(t)\\
	b_{e}^{-}(t)\\
	a_{e}^{+}(t)\\
	b_{e}^{+}(t)\\
	\end{array} \right] =\left[ \begin{array}{c}
	0\\
	0\\
	Q_{e}^{-}(t)/A^-\varepsilon _e^-\\
	Q_{e}^{+}(t)/A^+\varepsilon _e^+\\
	\end{array} \right]
\end{equation}
We denote the matrix on the LHS of Eq. (\ref{equ:cecoef}) by $\boldsymbol{\rm L}(t)$. The quadratic coefficients, $a_e^-(t)$ and $a_e^+(t)$, can then be expressed by:
\begin{equation}
    \label{equ:aepm}
    a_e^-(t) = \frac{\boldsymbol{\rm L}^{-1}_{1,3}(t) Q_e^-(t)}{A^-\varepsilon_e^-} + \frac{\boldsymbol{\rm L}^{-1}_{1,4}(t) Q_e^+(t)}{A^+\varepsilon_e^+},
    a_e^+(t) = \frac{\boldsymbol{\rm L}^{-1}_{3,3}(t) Q_e^-(t)}{A^-\varepsilon_e^-} + \frac{\boldsymbol{\rm L}^{-1}_{3,4}(t) Q_e^+(t)}{A^+\varepsilon_e^+}.
\end{equation}
where $\boldsymbol{\rm L}^{-1}_{1,3}(t)$ represents the element on the 1st row and 3rd column of the matrix $\boldsymbol{\rm L}^{-1}(t)$.

Since only cations (i.e., \ce{Li^+}) are involved in the reaction, the mass conservation of anions always holds. By electro-neutrality, the total quantity of cations in the solution phase, $Q_{e,0}=(\varepsilon_e^- A^- L^-+\varepsilon_e^+ A^+ L^+)c_{e,0}=Q_e^-(t)+Q_e^+(t)$, is constant at any time. Substituting this into Eq. (\ref{equ:Qe_diff}) yields:
\begin{equation}
	\label{equ:qenp_time}
	\begin{split}
    	&\frac{d Q_e^-(t)}{dt} 
    	 = 2A^-L^-D_e^{\mathrm{eff},-}(t) \left( \frac{  \boldsymbol{\rm L}^{-1}_{1,3}(t) }{A^- \varepsilon_e^-} - \frac{  \boldsymbol{\rm L}^{-1}_{1,4}(t) }{A^+ \varepsilon_e^+} \right) Q_e^-(t) 
    	 + 2A^-L^-D_e^{\mathrm{eff},-}(t)  \frac{  \boldsymbol{\rm L}^{-1}_{1,4}(t) Q_{e,0} }{A^+ \varepsilon_e^+}
    	+ \frac{(1-t_+^0)}{F}I(t).	 \\
    	&\frac{d Q_e^+(t)}{dt} 
    	 = 2A^+L^+D_e^{\mathrm{eff},+}(t) \left( \frac{  \boldsymbol{\rm L}^{-1}_{3,4}(t) }{A^+ \varepsilon_e^+} - \frac{  \boldsymbol{\rm L}^{-1}_{3,3}(t) }{A^- \varepsilon_e^-} \right) Q_e^+(t) 
    	 + 2A^+L^+D_e^{\mathrm{eff},+}(t)\frac{  \boldsymbol{\rm L}^{-1}_{3,3}(t) Q_{e,0} }{A^- \varepsilon_e^-}
    	- \frac{(1-t_+^0)}{F}I(t).
	\end{split}
\end{equation}
For notational simplicity, Eq. (\ref{equ:qenp_time}) is written in:
\begin{equation}
    \label{equ:qe_time}
    \tau_e^{\pm}(t) \frac{d Q_e^{\pm}(t)}{dt} = - Q_e^{\pm}(t) + K_{Q_e}^{\pm}(t).
\end{equation}
where $\tau_e^{\pm}(t)$ and $K_{Q_e}^{\pm}(t)$ can be derived from Eq. (\ref{equ:qenp_time}).

The time trajectories of $Q_e^{\pm}$ are modelled by two coupled first-order inertial processes. Once $Q_e^{\pm}$ are obtained, $c_e(x,t)$ at any point in the x-axis can be calculated by Eq. (\ref{equ:cecoef}). To improve the model fidelity, the original diffusion coefficient of the electrolyte solvent is corrected by the Bruggeman relation for porous electrodes, i.e., $D_e^{\mathrm{eff}} = D_e(\varepsilon_e)^p$.


\subsection{Solid-phase diffusion model}\label{sec:soliddiffusion}
\textcolor{blue}{For intercalated active materials, \ce{Li^+} diffuses along the radial direction of active particles according to Eq. (\ref{equ:solid-diffusion}). However, this adds another spatial coordinate, $r$, to the model and increases the complexity; i.e., the solid-phase \ce{Li^+} concentration $c_s$ varies with $x$, $r$ and $t$ synchronously. Actually, only the average \ce{Li^+} concentration and the surface \ce{Li^+} concentration of the active particle are considered in practice because the former determines the remaining charge in the battery, while the latter determines the reaction rate in the battery.
} To this end, many works have proposed different methods to simplify the solid-phase diffusion model as reviewed above, \textcolor{blue}{and their basic idea is similar, i.e., approximate the difference between the average and surface $c_s$ by a series of inertial processes.} However, after attempting existing methods, we found that two challenges still remain. First, for in situ monitoring of the battery, a series of points must be selected along the thickness direction, and the diffusion model must be constructed at every point independently, which means that the model should be reduced to be as simple as possible. Second, we find that existing approximated models whose time constants are derived from the Pad\'{e} approximation or the frequency response optimization are likely to suffer from oscillations. The same situations also arise for the xRA and volume-averaging methods. This problem may be caused by a relatively small time constant obtained from these methods. Thus, a simple first-order inertial process with tuned time constants is used to realize a trade-off between accuracy, stabilization, and simplicity. \textcolor{blue}{In this part, we introduce the derivation of coefficients in the approximated inertial process.}

First, we derive the expression of the average solid-phase concentration. Multiplying the two sides of Eq. (\ref{equ:solid-diffusion}) by $r^2$ and then integrating both sides along the r-axis yields:
\begin{equation}
    \label{equ:cs_integration}
	\frac{1}{4\pi}\frac{\partial \int_{0}^{R_s} 4\pi r^2 c_s(x,r,t) dr }{ \partial t} = D_s(x,t) R_s^2 \frac{\partial c_s(x,r,t)}{\partial r} \bigg|_{r=R_s}.
\end{equation}
\textcolor{blue}{Note that the numerator of the LHS of Eq. (\ref{equ:cs_integration}), $\int_{0}^{R_s} 4\pi r^2 c_s(x,r,t) dr$, equals the total quantity of \ce{Li^+} in the active particle, which can also be represented by the bulk-averaged \ce{Li^+} concentration, denoted by $\bar{c}_s$: $\int_{0}^{R_s} 4 \pi r^2 c_s(x,r,t) dr = \frac{4}{3} \pi R_s^3 \bar{c}_s(x,t)$.} Based on material conservation, the \ce{Li^+} flux at the surface of the active particle is proportional to the pore-wall flux $j_n$, i.e., $\frac{\partial c_s(x,r,t)}{\partial r} \bigg|_{r=R_s} = -\frac{j_n(x,t)}{D_s}.$ Substituting these two terms into Eq. (\ref{equ:cs_integration}) yields:
\begin{equation}
    \label{equ:cs_averaged}
    \frac{\partial \bar{c}_s(x,t)}{\partial t} = - \frac{3}{R_s} j_n(x,t).
\end{equation}

\textcolor{blue}{
To describe the surface solid-phase concentration, we first introduce
an intermediate variable $w$ to depict the difference between the average and surface \ce{Li^+} concentration. By Laplace transformation, the closed-form expression of $w$ in the frequency domain can be derived from Eq. (\ref{equ:solid-diffusion})~\cite{li_physics-based_2017}}: $\frac{w(x,s)}{j_n(x,s)} = \frac{R_s}{D_s}\frac{1}{1-R_s\sqrt{\frac{s}{D_s}}\coth(R_s\sqrt{\frac{s}{D_s}})}+\frac{3}{R_s}\frac{1}{s}$. \textcolor{blue}{Its limitation in the frequency domain at $s=0$ equals} $\lim_{s \to 0} \frac{w(x,s)}{j_n(x,s)} = - \frac{R_s}{5D_s},$, indicating that $w$
gradually approaches $-\frac{R_s j_n}{5D_s}$ in the time domain. \textcolor{blue}{For model simplicity, the transition of $w$ to its steady state is approximated by a first-order inertial process. The physical interpretation of this process is that it takes time for \ce{Li^+} inside the active particle to diffuse to the surface.} Thus, the time constant is set in proportion to the ratio between the radius square and the diffusion coefficient: $\tau_s(x,t) = k_s \frac{R_s^2}{D_s(x,t)}$. The transition equation is expressed by:
\begin{equation}
	\label{equ:inertial_w}
	\tau_s(x,t) \frac{\partial w(x,t)}{\partial t} = -w(x,t) - \frac{R_s j_n(x,t)}{5D_s(x,t)}.
\end{equation}
where $k_s$ is a dimensionless coefficient fitting the approximated process to the actual process. Once $w$ is obtained, the surface solid-phase concentration, denoted by $c_{ss}$, can be calculated directly:
\begin{equation}
	\label{equ:css-cs}
	c_{ss}(x,t) = \bar{c}_s(x,t) + w(x,t).
\end{equation}

\textcolor{blue}{The specific value of $k_s$ varies in different studies,}, e.g., $k_s=\frac{1}{35}$ in the Pad\'{e} approximation, $k_s=\frac{1}{30}$ in the volume-averaging method, $k_s$=0.04356 or 0.03459 in~\cite{han_simplification_2015_1}, and $k_s=0.0214$ in the frequency response optimization (in the frequency band $[10^{-4},10^4]$ Hz). By testing the above settings, we find that the results commonly suffer from oscillation except for~\cite{han_simplification_2015_1}, which indicates that a smaller $k_s$ is likely to bring instability to the model. However, a larger $k_s$ makes the model less accurate, especially under dynamic currents. In this work, $k_s$ is determined by fitting the experimental data to realize a trade-off between accuracy and stabilization.

\textcolor{blue}{The simplified solid-phase model is appropriate for both the negative electrode and positive electrode, so the superscripts $+$ and $-$ are omitted for notation simplicity. Note also that in the following text, the average and surface solid-phase concentrations are sometimes replaced by the average and surface \ce{Li^+} stoichiometry for notational simplicity, denoted by $\theta_s$ and $\theta_{ss}$, respectively. The transformations between them are simple: $\theta_s=\frac{\bar{c}_s}{c_{s,max}}$ and $\theta_{ss}=\frac{c_{ss}}{c_{s,max}}$, where $c_{s,max}$ is the maximum concentration the active particle can store.
}

\subsection{Reaction rate distribution model}\label{subsubsec:jn}
\textcolor{blue}{Generally, the reaction rate $j_n$ is non-uniform along the thickness direction of the battery cell. However, it remains a challenge to express the spatial distribution of $j_n$ because it is determined by Eqs. (\ref{equ:Butler-Volmer})-(\ref{equ:solution-phi}) simultaneously. Thus, different from simplifying (\ref{equ:solution-migration}) and (\ref{equ:solid-diffusion}), as introduced in the above section, we need to consider the coupling between these formulas and design the specific simplification strategy.
}


\subsubsection{Chemical system}
\textcolor{blue}{First, we start from Eq. (\ref{equ:Butler-Volmer}) and simplify the chemical system of the battery. Since Eq. (\ref{equ:Butler-Volmer}) brings non-linearity to the model and increases its complexity,} several approximations have been proposed for simplicity, e.g., the Tafel equation, linear current-potential equation, and hyperbolic sine approximation \cite{noren_clarifying_2005}. However, some methods sacrifice generality to some extent, especially under high currents. \textcolor{blue}{In this work, an adaptive linear approximation method that automatically adjusts the linear coefficients according to the actual applied current is designed. Thus, the simplified expression can be adaptive to critic conditions without too much loss of accuracy.}

For the applied current $I$, we first calculate the average pore-wall flux in the negative electrode and positive electrode: $\bar{j}_n^{\pm}(t)=\mp \frac{I(t)}{a_s^{\pm} F A^{\pm} L^{\pm}}$. Then, we apply the first-order Taylor expansion at $\bar{j}_n^{\pm}$ on the inverse function of Eq. (\ref{equ:Butler-Volmer}):
\begin{equation}
    \label{equ:inverse_BV}
    \eta(x,t) = \frac{2RT(t)}{F}\ln{\left( \frac{F j_n(x,t)+\sqrt{ F^2 j_n^2(x,t) + 4 i_0^2(t) }}{2i_0(t)} \right)} \Rightarrow \eta(x,t) \approx a_{j_n}(t) \left(j_n(x,t) -\bar{j}_n(t) \right) + b_{j_n}(t). 
\end{equation}
The closed-form expression of the linear coefficient $a_{j_n}(t)$ in Eq. (\ref{equ:inverse_BV}) is given by:
\begin{equation}
    \label{equ:ajn}
    a_{j_n}(t) = \frac{RT(t)}{i_0(t)} \frac{\sqrt{\frac{F^2 \bar{j}_n^2(t)}{4i_0^2(t)}+1}+\frac{F\bar{j}_n(t)}{2i_0(t)}}{\frac{F\bar{j}_n(t)}{2i_0(t)}\sqrt{\frac{F^2\bar{j}_n^2(t)}{4i_0^2(t)}+1}+\frac{F^2\bar{j}_n^2(t)}{4i_0^2(t)}+1}.
\end{equation}
\textcolor{blue}{In the formula above, the exchange current densities in the negative electrode and positive electrode are expressed by $i_0^{\pm} = k_r^{\pm} (\bar{c}_e^{\pm})^{\alpha_a}(c_{s,\mathrm{max}}^{\pm}-\bar{c}_{ss}^{\pm})^{\alpha_c} (\bar{c}_{ss}^{\pm})^{\alpha_a}$, where $\bar{c}_e^{\pm}$ and $\bar{c}_{ss}^{\pm}$ refer to the average concentrations across the electrode, respectively. Generally, the anodic and cathodic transfer coefficients, $\alpha_a$ and $\alpha_c$, are set at $0.5$ because the proportions of the anodic and cathodic directions of the total intercalation reaction are assumed to be equal~\cite{thomas_mathematical_2002}.}

\textcolor{blue}{To build the coupling between the chemical equation and electrical equations, we need to introduce variables directly related to the potential rather than using the intermediate variable $\eta$.} By definition, the over-potential $\eta$ in Eq. (\ref{equ:inverse_BV}) also equals $\Phi_{s-e}-U_{\mathrm{OCP}}-F R_f j_n$, where $\Phi_{s-e}$ equals the potential difference between the solid-phase and solution-phase at the surface of the active particle. Substituting this equality into Eq. (\ref{equ:inverse_BV}) and implementing the differential operation yields:
\begin{equation}
    \label{equ:se potential diff1}
     \frac{\partial \Phi_{s-e}(x,t)}{\partial x} = \left( a_{j_n}(t) + F R_f \right)\frac{\partial j_n(x,t)}{\partial x}  +  \frac{\partial U_{\mathrm{OCP}}(x,t)}{\partial x}. 
\end{equation}
\textcolor{blue}{The formula above retains a term to be addressed, i.e., the differential of $U_{\mathrm{OCP}}$. As introduced in the text above, $U_{\mathrm{OCP}}$ is determined by the surface \ce{Li^+} stoichiometry $\theta_{ss}$. Thus, the expression of $U_{\mathrm{OCP}}(x,t)$ can be fitted based on the knowledge of $\theta_{ss}(x,t)$ along the thickness direction. To achieve a balance between the complexity and accuracy, four points evenly distributed in each electrode along the thickness direction are selected as checkpoints; i.e., the coordinates of checkpoints in the negative and positive electrodes are $0^{\pm}$, $\frac{L^{\pm}}{3}$, $\frac{2L^{\pm}}{3}$, and $L^{\pm}$. This helps the proposed model be applicable for in situ monitoring at four checkpoints in each electrode while still simple enough for online control in practical use. Through numerical experiments, we find that using a cubic polynomial can achieve an acceptable performance. The coefficients of the polynomial are fitted on $U_{\mathrm{OCP}}$ at four checkpoints in each electrode.
} Denote the analytical expression of $U_{\mathrm{OCP}}$ by $U_{\mathrm{OCP}}(x,t)=a_{P}(t) x^3 + b_{P}(t) x^2 + c_{P}(t)x + d_{P}(t)$ and substitute it into Eq. (\ref{equ:se potential diff1}):
\begin{equation}
    \label{equ:se potential diff}
     \frac{\partial \Phi_{s-e}(x,t)}{\partial x} \approx \left( a_{j_n}(t) + F R_f \right)\frac{\partial j_n(x,t)}{\partial x}  + 3 a_{P}(t) x^2 + 2 b_{P}(t) x + c_{P}(t). 
\end{equation}

\textcolor{blue}{Notably, Eqs. (\ref{equ:inverse_BV})-(\ref{equ:se potential diff}) are appropriate for both negative electrode and positive electrode, so the superscripts $+$ and $-$ are omitted for notation simplicity.}

\subsubsection{Electrical system}
\textcolor{blue}{By analysing the chemical system, we have obtained the relationship between $\Phi_{s-e}$ and $j_n$. Now, we turn to simplifying the electrical system inside the battery. Note that Eq. (\ref{equ:solid-phi}) depicts the relationship between $\Phi_{s}$ and $j_n$, while Eq. (\ref{equ:solution-phi}) depicts the relationship between $\Phi_{e}$ and $j_n$. Thus, we try to couple these two equations to derive the relationship between $\Phi_{s-e}$ and $j_n$ in the electrical system.
First, Eq. (\ref{equ:solid-phi}) can be decomposed into two equations by introducing a new variable representing the current density in the solid phase, denoted by $i_s$.
}:
\begin{equation}
	\label{equ:solid-phase potential diff}
	\frac{\partial \Phi_s(x,t)}{\partial x}  = -\frac{i_s(x,t)}{\sigma_s^{\mathrm{eff}}},\quad
	\frac{\partial i_s(x,t)}{\partial x} = -a_s F j_n(x,t).
\end{equation}
\textcolor{blue}{
The left formula above is derived based on Ohm's law, and the right formula is derived based on Faraday's law. Similarly, Eq. (\ref{equ:solution-phi}) can be decomposed into two equations by introducing a new variable representing the current density in the solution phase, denoted by $i_e$:
}
\begin{equation}
	\label{equ:solution-phase potential diff}
    \frac{\partial \Phi_e(x,t)}{\partial x}  = -\frac{i_e(x,t)}{\kappa^{\mathrm{eff}}(t)} -  \frac{ \kappa_D^{\mathrm{eff}}(t) \partial \ln c_e(x,t)}{\kappa^{\mathrm{eff}}(t) \partial x}, \quad
    \frac{\partial i_e(x,t)}{\partial x} = a_s F j_n(x,t).
\end{equation}
The second term in the left formula above represents the concentration polarization potential in the electrolyte, and the effective diffusional conductivity $\kappa_D^{\mathrm{eff}}$ is derived from concentrated solution theory, expressed by $\kappa_D^{\mathrm{eff}} = 2\kappa^{\mathrm{eff}}\frac{RT}{F}(t_+^0-1)(1+ \frac{d \ln f_{\pm}}{d \ln c_e})$, where $f_{\pm}$ is the mean molar activity coefficient. Generally, the term $\frac{d \ln f_{\pm}}{d \ln c_e}$ is assumed to be constant~\cite{smith_control_2007}.
\textcolor{blue}{However, in this work, to improve model fidelity, a parabolic polynomial is used to fit the relationship between $\frac{d \ln f_{\pm}}{d \ln c_e}$ and $c_e$ based on the experimental data in~\cite{AutoLion}. The solid-phase and solution-phase conductivities $\sigma_s$ and $\kappa$ are corrected by the Bruggeman correction, i.e., $\sigma_s^{\mathrm{eff}} = \sigma_s \varepsilon_s$, $\kappa^{\mathrm{eff}} = \kappa \varepsilon_e^p$. Additionally, since Eqs. (\ref{equ:solid-phase potential diff})-(\ref{equ:solution-phase potential diff}) are appropriate for both negative and positive electrodes, so the superscripts $+$ and $-$ are omitted for notation simplicity.}

Based on KCL, the boundary conditions of $i_s$ and $i_e$ can be obtained: $i_s(0^-,t)=i_e(L^-,t)=\frac{I(t)}{A^-}$, $i_e(0^+,t)=i_s(L^+,t)=\frac{I(t)}{A^+}$,
$i_e(0^-,t)=i_s(L^-,t)=0$,
$i_s(0^+,t)=i_e(L^+,t)=0$. Thus, expressions of $i_s$ and $i_e$ can be obtained by integrating the RHSs
in Eqs. (\ref{equ:solid-phase potential diff})-(\ref{equ:solution-phase potential diff}):
\begin{equation}
    \label{equ:solid-phase current}
    i_s(x,t)= 
	\begin{cases}
	    -a_s^- F \int_{0^-}^{x} j_n(l,t) dl + I(t)/A^-, x \in [0^-,L^-]; \\
	     a_s^+ F \int_{x}^{L^+} j_n(l,t) dl + I(t)/A^+, x \in [0^+,L^+].
	\end{cases} \quad
	i_e(x,t)= 
	\begin{cases}
	    a_s^- F \int_{0^-}^{x} j_n(l,t) dl , x \in [0^-,L^-]; \\
	    -a_s^+ F \int_{x}^{L^+} j_n(l,t) dl , x \in [0^+,L^+].
	\end{cases}
\end{equation}

Subtracting the left formulas in Eqs. (\ref{equ:solid-phase potential diff})-(\ref{equ:solution-phase potential diff}) and substituting Eq. (\ref{equ:solid-phase current}) yields:
\begin{equation}
	\label{equ:solution-solid potential}
	\frac{\partial \Phi_{s-e}(x,t)}{\partial x} = 
	\begin{cases}
    -\frac{I(t)}{A^-\sigma_s^{\mathrm{eff},-}} + a_s^-F\left(\frac{1}{\sigma_s^{\mathrm{eff},-}}+\frac{1}{\kappa^{\mathrm{eff},-}(t)}\right)\int_{0^-}^{x}j_n(l,t){\rm d}l  +\frac{\kappa_D^{\mathrm{eff},-}(t)}{\kappa^{\mathrm{eff},-}(t)}\frac{\partial \ln\left(c_e(x,t)\right)}{\partial x},  x \in [0^-,L^-]; \\
    -\frac{I(t)}{A^+\sigma_s^{\mathrm{eff},+}} -a_s^+F\left(\frac{1}{\sigma_s^{\mathrm{eff},+}}+\frac{1}{\kappa^{\mathrm{eff},+}(t)}\right)\int_{x}^{L^+}j_n(l,t){\rm d}l  +\frac{\kappa_D^{\mathrm{eff},+}(t)}{\kappa^{\mathrm{eff},+}(t)}\frac{\partial \ln\left(c_e(x,t)\right)}{\partial x}, x \in [0^+,L^+].
	\end{cases}
\end{equation}
\textcolor{blue}{
The last term in (\ref{equ:solution-solid potential}) still makes the derivation of analytical expressions intractable. According to numerical experiments, we find that $\frac{\partial \ln(c_e(x,t))}{\partial x}$ can be approximated by:
}
\begin{equation} 
    \label{equ:lnce}
    \frac{\partial \ln(c_e(x,t))}{\partial x}= 
	\begin{cases}
	    \frac{2a_e^-(t)x}{a_e^-(t)x^2+b_e^-(t)} \approx  \frac{2a_e^-(t)}{b_e^-(t)}x, x \in [0^-,L^-]; \\
	    \frac{2a_e^+(t)\left(x-L^+\right)}{a_e^+(t)\left(x-L^+\right)^2+b_e^+(t)} \approx \frac{2a_e^+(t)\left(x-L^+\right)}{b_e^+(t)}, x \in [0^+,L^+].
	\end{cases}
\end{equation}

\subsubsection{Mathematical representation}
\textcolor{blue}{By analysing the chemical system and electrical system, two independent equations depicting the relationship between $\Phi_{s-e}$ and $j_n$ are obtained. We simultaneously solve them to derive the expression of $j_n$.} We denote the integration of $j_n(x,t)$ over the electrode by $J_n(x,t)$, i.e., $J_n^-(x,t)=\int_{0^-}^{x} j_n^-(l,t)dl$ for the negative electrode and $J_n^+(x,t)=\int_{x}^{L^+} j_n^+(l,t)dl$ for the positive electrode. Combining (\ref{equ:se potential diff}) and (\ref{equ:solution-solid potential}) yields:
\begin{equation}
   \label{equ:solution-solid potential formula}
   \mp k^{\pm}_1(t) J_n^{\pm}(x,t)  \pm k^{\pm}_2(t) \frac{\partial^2 J_n^{\pm}(x,t)}{\partial x^2} + k^{\pm}_3(t)x^2 + k^{\pm}_4(t)x  +   k_5^{\pm}(t)= 0.
\end{equation}
In the formula above, $k_1^{\pm}(t) = a_s^{\pm} F \left(\frac{1}{\sigma_s^{\mathrm{eff},\pm}}+\frac{1}{\kappa^{\mathrm{eff},\pm}(t)} \right)$,
$k_2^{\pm}(t) = a_{j_n}^{\pm}(t) + F R_f^{\pm} $, $k_3^{\pm}(t)=-3a_{P}^{\pm}(t)$,
$k_4^{\pm}(t) = \frac{2a_e^{\pm}(t)\kappa_D^{\mathrm{eff},\pm}(t)}{b_e^{\pm}(t)\kappa^{\mathrm{eff},\pm}(t)}-2b_{P}^{\pm}(t)$,
$k_5^-(t)=-\frac{I(t)}{A^{-}\sigma_s^{\mathrm{eff},-}}-c_{P}^-(t)$,
$k_5^+(t)=-\frac{I(t)}{A^{+}\sigma_s^{\mathrm{eff},+}}-c_{P}^+(t)-2\frac{a_e^{+}(t)\kappa_D^{\mathrm{eff},+}(t)}{b_e^{+}(t)\kappa^{\mathrm{eff},+}(t)}L^+$.
The boundary conditions of $J_n$ are equivalent to $i_s$:
\begin{equation}
    \label{equ:jn_boundary}
        J_n^{-}(0^-,t) = 0, \quad  J_n^{-}(L^-,t) = \frac{I(t)}{a_s^-A^-F}, 
         J_n^{+}(0^+,t) = -\frac{I(t)}{a_s^+A^+F}, \quad  J_n^{+}(L^+,t) = 0.
\end{equation}
The expression of $j_n$ can be obtained by applying the differential operation to $J_n$:
\begin{equation}
    \label{equ:jn}
        j_n^{\pm}(x,t) = \pm m_1^{\pm}(t) \sqrt{\frac{k_1^{\pm}(t)}{k_2^{\pm}(t)}} \exp \left(  - \sqrt{\frac{k_1^{\pm}(t)}{k_2^{\pm}(t)}} x \right)  \mp m_2^{\pm}(t) \sqrt{\frac{k_1^{\pm}(t)}{k_2^{\pm}(t)}} \exp \left(   \sqrt{\frac{k_1^{\pm}(t)}{k_2^{\pm}(t)}} x \right) - \frac{2k_3^{\pm}(t)}{k_1^{\pm}(t)}x - \frac{k_4^{\pm}(t)}{k_1^{\pm}(t)}.
\end{equation}
\textcolor{blue}{where $m_{1,2}^{\pm}(t)$ can be obtained by substituting the boundary conditions into Eq. (\ref{equ:solution-solid potential formula}):
\begin{equation}
\begin{split}
&\left[ \begin{array}{c}
m_{1}^{-}(t)\\
m_{2}^{-}(t)\\
\end{array} \right] =\left[ \begin{matrix}
1&		1\\
\exp \left( -\sqrt{\frac{k_{1}^{-}}{k_{2}^{-}}}L^- \right)&		\exp \left( \sqrt{\frac{k_{1}^{-}}{k_{2}^{-}}}L^- \right)\\
\end{matrix} \right] ^{-1}\left[ \begin{array}{c}
\frac{k_{5}^{-}}{k_{1}^{-}}+\frac{2k_{2}^{-}k_{3}^{-}}{\left( k_{1}^{-} \right) ^2}\\
\frac{k_{5}^{-}}{k_{1}^{-}}+\frac{2k_{2}^{-}k_{3}^{-}}{\left( k_{1}^{-} \right) ^2}+\frac{k_{3}^{-}\left( L^- \right) ^2}{k_{1}^{-}}+\frac{k_{4}^{-}L^-}{k_{1}^{-}}+\frac{I}{FA^-a_{s}^{-}}\\
\end{array} \right] \\
&\left[ \begin{array}{c}
	m_{1}^{+}(t)\\
	m_{2}^{+}(t)\\
\end{array} \right] =-\left[ \begin{matrix}
	1&		1\\
	\exp \left( -\sqrt{\frac{k_{1}^{+}}{k_{2}^{+}}}L^+ \right)&		\exp \left( \sqrt{\frac{k_{1}^{+}}{k_{2}^{+}}}L^+ \right)\\
\end{matrix} \right] ^{-1}\left[ \begin{array}{c}
	\frac{I}{FA^+a_{s}^{+}}+\frac{k_{5}^{+}}{k_{1}^{+}}+\frac{2k_{2}^{+}k_{3}^{+}}{\left( k_{1}^{+} \right) ^2}\\
	\frac{k_{5}^{+}}{k_{1}^{+}}+\frac{2k_{2}^{+}k_{3}^{+}}{\left( k_{1}^{+} \right) ^2}+\frac{k_{3}^{+}\left( L^+ \right) ^2}{k_{1}^{+}}+\frac{k_{4}^{+}L^+}{k_{1}^{+}}\\
\end{array} \right]
\end{split}
\end{equation}
}
\textcolor{blue}{Once the reaction rate across the electrode domain is obtained, the spatial distribution of potentials and current densities in the solid-phase and solution-phase can all be derived through Eqs. (\ref{equ:solid-phase potential diff})-(\ref{equ:solution-phase potential diff}), the in situ monitoring of the battery cell can be realized, providing detailed information to upper-level applications.
}

\subsection{Output model}
The measurable output of the battery cell includes the terminal voltage and surface temperature. \textcolor{blue}{This part introduces the calculation of these two measurable states.
}

\subsubsection{Terminal voltage}
\textcolor{blue}{
Since the solid phase of the electrode is directly connected to the current collector, the terminal voltage equals the potential difference between $\Phi_s(L^+,t)$ and $\Phi_s(0^-,t)$. However, directly calculating $\Phi_s$ through the expression of $j_n$ presented above is impossible since this requires two potential reference points, one for the negative electrode and one for the positive electrodes. However, when viewing the battery as a whole system, only one potential reference point can be selected. To solve this problem, we start from $\Phi_{s-e}$ and $\Phi_e$ to calculate $V_t$ indirectly because $\Phi_s$ is equivalent to $\Phi_{s-e}+\Phi_e$ as well.
}
By denoting the ohmic resistance between the current collector and electrode by $R_c$, $V_t$ is expressed by:
\begin{equation}
    \label{equ:vt_formula}
    V_t(t) = \Phi_{s-e}(L^+,t) + \Phi_e(L^+,t) - \Phi_{s-e}(0^-,t) - \Phi_e(0^-,t) - R_c I(t).
\end{equation}
In the formula above, $\Phi_{s-e}$ at the boundary can be directly calculated according to the B-V equation:
\begin{equation}
    \label{equ:phise-boundary}
    \begin{cases}
    \Phi_{s-e}(L^+,t) = U_{\mathrm{OCP}}^{+}(L^+,t) + F R_f^{+} j_n(L^+,t) +  \frac{2RT(t)}{F} \ln \left( \frac{F j_n(L^+,t)}{2i_0^+(t)}+\sqrt{ \left( \frac{F j_n(L^+,t)}{2i_0^+(t)} \right) ^2+1} \right), \\
    \Phi_{s-e}(0^-,t) = U_{\mathrm{OCP}}^{-}(0^-,t) + F R_f^{-} j_n(0^-,t) +  \frac{2RT(t)}{F} \ln \left( \frac{F j_n(0^-,t)}{2i_0^-(t)}+\sqrt{ \left( \frac{F j_n(0^-,t)}{2i_0^-(t)} \right) ^2+1} \right).
	\end{cases}
\end{equation}

\textcolor{blue}{
We now turn to the potential drop in the solution phase, which is composed of two parts, namely, the ohmic potential drop and polarization potential drop, corresponding to the first and second terms on the RHS of Eq. (\ref{equ:solution-phase potential diff}).
}
The polarization potential drop between $x=L^+$ and $x=0^-$, denoted by $\Delta \Phi_{e,polar}$, can be expressed by the sum of the polarization potential drops in each domain:
\begin{equation}
    \label{equ:phie_ce}
    \Delta \Phi_{e,polar}(t) = -\frac{\kappa_D^{\mathrm{eff,-}}}{\kappa^{\mathrm{eff,-}}}\ln\left(\frac{c_e(L^-,t)}{c_e(0^-,t)}\right) - \frac{\kappa_D^{\mathrm{eff,sep}}}{\kappa^{\mathrm{eff,sep}}}\ln\left(\frac{c_e(L^{\mathrm{sep}},t)}{c_e(0^{\mathrm{sep}},t)}\right) -\frac{\kappa_D^{\mathrm{eff,+}}}{\kappa^{\mathrm{eff,+}}}\ln\left(\frac{c_e(L^+,t)}{c_e(0^+,t)}\right).
\end{equation}
The ohmic potential drop, denoted by $\Delta \Phi_{e,ohm}$, can be obtained by substituting Eq. (\ref{equ:jn}) into Eq. (\ref{equ:solution-phase potential diff}). In the electrode domain, $\Delta \Phi_{e,ohm}$ is expressed by:
\begin{equation}
    \label{equ:phie_ie}
    \begin{split}
        &\Delta \Phi_{e,ohm}^{\pm}(t) = \pm\frac{a_s^{\pm}F}{\kappa^{\mathrm{eff},\pm}(t)}\biggl( -m_1^{\pm}(t) \sqrt{\frac{k_2^{\pm}(t)}{k_1^{\pm}(t)}} \left( \exp \left(  - \sqrt{\frac{k_1^{\pm}(t)}{k_2^{\pm}(t)}} L^{\pm} \right) 
         -1\right)+ m_2^{\pm}(t) \sqrt{\frac{k_2^{\pm}(t)}{k_1^{\pm}(t)}} \left( \exp \left( \sqrt{\frac{k_1^{\pm}(t)}{k_2^{\pm}(t)}} L^{\pm} \right) -1 \right) \\& \pm \frac{k_3^{\pm}(t)}{3k_1^{\pm}(t)} (L^{\pm})^3 
           {\pm} \frac{k_4^{\pm}(t)}{2k_1^{\pm}(t)} (L^{\pm})^2 {\pm} \left(\frac{k_5^{\pm}(t)}{k_1^{\pm}(t)}+ \frac{2k_2^{\pm}(t)k_3^{\pm}(t)}{k_1^{\pm}(t)^2} \right)L^{\pm}\biggr). 
    \end{split}
\end{equation}
In the separator domain, $\Delta \Phi_{e,ohm}^{\mathrm{sep}}(t)=-\frac{L^{\mathrm{sep}}I(t)}{\kappa^{\mathrm{eff,sep}}(t)A^{\mathrm{sep}}}$. Thus, the total solution-phase potential drop between $L^+$ and $0^-$ equals:
\begin{equation}
    \label{equ:phie drop}
    \Phi_e(L^+,t) - \Phi_e(0^-,t) = \Delta \Phi_{e,ohm}^{\mathrm{sep}}(t) + \Delta \Phi_{e,ohm}^{-}(t) + \Delta \Phi_{e,ohm}^{+}(t) + \Delta \Phi_{e,polar}(t).
\end{equation}
By substituting (\ref{equ:phie drop}) into (\ref{equ:vt_formula}), $V_t(t)$ can be obtained.

\subsubsection{Cell temperature}\label{sec:thermal}
The temperature can significantly affect the operating characteristics of the battery. To track the trajectory of battery temperature during its operation, a lumped thermal model is developed to predict the cell temperature based on the following assumptions. First, the temperature distribution is uniform at any instant in time, i.e., the surface temperature is always equal to the core temperature \cite{Rao_1997}. Second, the enthalpy mixing and phase-change heat are neglected \cite{botte_influence_1999}. Third, the reversible entropy change of the reaction is neglected \cite{gu_thermal-electrochemical_2000}. \textcolor{blue}{Many studies have proposed much more sophisticated thermal models than the lumped model. However, they were not adopted in this work for two main reasons. First, we aim to use the proposed model in upper-level applications, e.g., online controlling or operating optimization. Thus, the proposed model is not expected to be very complex, reflecting that the basic properties meet the requirement. Second, in the latest real-world applications, the temperature sensor is quite advanced and can be deployed at the cell level; thus, the predicted temperature can be corrected according to the measurement in real time. Considering the target applications of this work, a lumped thermal model that can approximately track the temperature is acceptable. We care more about depicting those states that cannot be directly measured, such as potentials and concentrations.}

In a lumped thermal model, the energy conservation equation is written as follows:
\begin{equation}
    \label{equ:energy conservation}
    m C_p\frac{d T(t)}{d t} = h_c A_{\mathrm{surf}} \left( T_{\mathrm{amb}}(t)-T(t) \right) + H(t).
\end{equation}
The first term on the RHS of the formula above accounts for the heat transfer rate from the cell to the environment, and the second term refers to the heat generated by the reaction, calculated by:
\begin{equation}
    \label{equ:reaction heat}
    \begin{split}
    H(t) & =  - F\left(A^-\int_{0^-}^{L^-} a_s^-j_n(x,t)U_{\mathrm{OCP}}^-(x,t)dx  + A^+\int_{0^+}^{L^+} a_s^+j_n(x,t)U_{\mathrm{OCP}}^+(x,t)dx\right)  - I(t)V(t).\\
    & \approx \left( \bar{U}_{\mathrm{OCP}}^+(t)-\bar{U}_{\mathrm{OCP}}^-(t)-V(t)\right)I(t).
    \end{split}
\end{equation}
\textcolor{blue}{where $\bar{U}_{\mathrm{OCP}}^{\pm}$ are average values of $U_{\mathrm{OCP}}^{\pm}$ at $0^{\pm}, \frac{L^{\pm}}{3}, \frac{2L^{\pm}}{3}, L^{\pm}$.} For notation simplicity, (\ref{equ:energy conservation}) is represented by:
\begin{equation}
    \label{equ:thermal model}
    \tau_T \frac{d T(t)}{dt} = - T + K_T(t).
\end{equation}
where $\tau_T = m C_p / h_c A_s$ and $K_T(t)=H(t)/h_c A_s + T_{\mathrm{amb}}(t)$.


\textcolor{blue}{
The entire bottom-up approach to construct the simplified model is shown in Fig.~\ref{fig:modelfram}.
}

\begin{figure}[!htb]
    \centering
\begin{minipage}{.53\textwidth}\label{fig:modelfram}
        \centering
        \includegraphics[width=\textwidth]{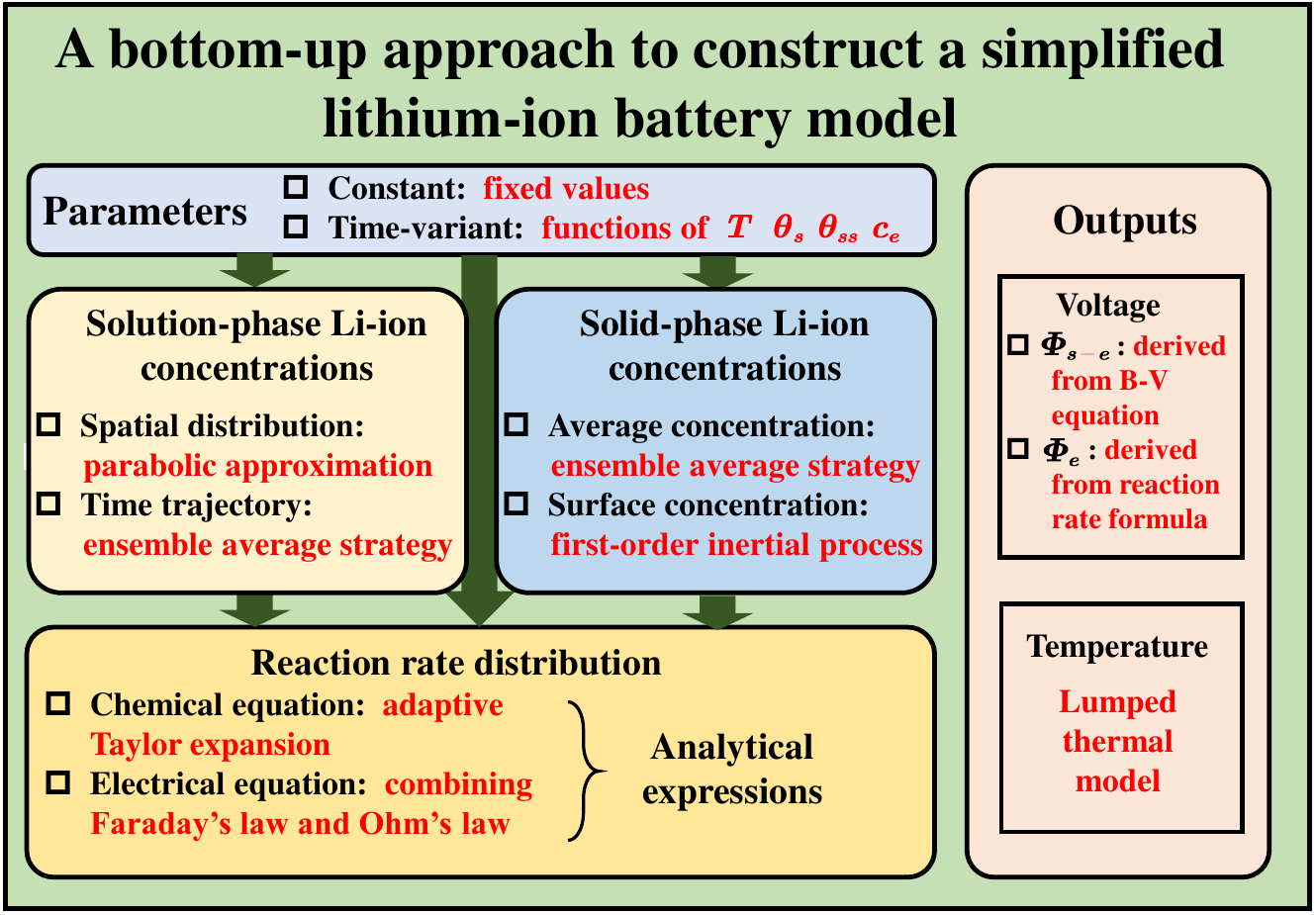}
\subcaption{\textcolor{blue}{Modelling framework. }}\label{fig:modelfram}
\end{minipage}%
\begin{minipage}{0.44\textwidth}\label{fig:flowchart}
        \centering
        \includegraphics[width=\textwidth]{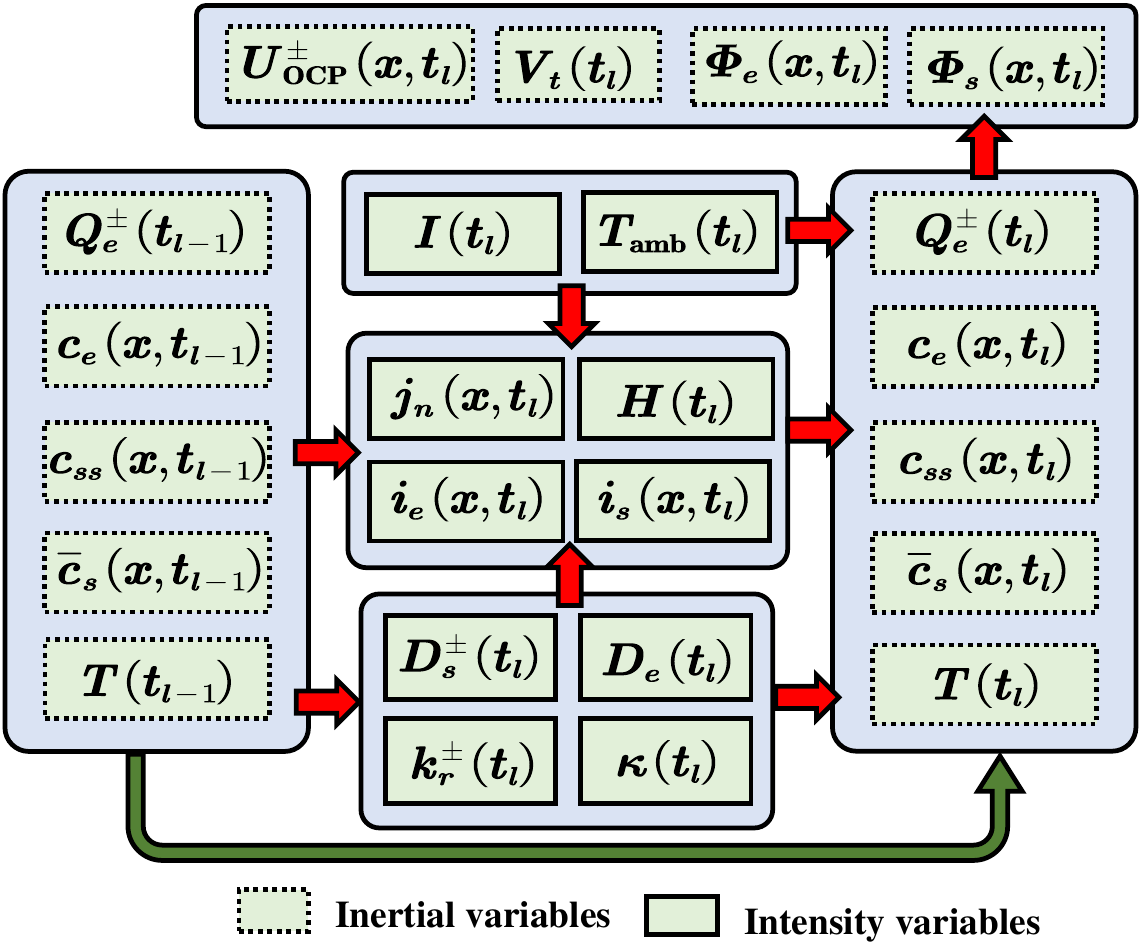}
\subcaption{\textcolor{blue}{The flowchart of updating states within one simulation step}}\label{fig:flowchart}
\end{minipage}
\caption{\textcolor{blue}{Sketches of the modelling approach and the iterative step. }}
\end{figure}

\section{Closed-loop simulation framework} \label{sec:method}
\textcolor{blue}{
After introducing the modelling approach of the lithium-ion battery, the next step is to design the simulation framework so that the proposed work can be applied in practical scenarios such as online control or real-time monitoring.
}

\subsection{Discrete-time state-space realization}\label{sec:discrete}
To enable real-time simulation of the model, a discrete-time state-space representation is necessary. \textcolor{blue}{Since the original simplified model is continuous on the time horizon, before discretization, the following assumptions should be declared. First,} the model inputs, including $I(t)$ and $T_{\mathrm{amb}}(t)$, are treated as intensity variables; i.e., they are updated at the start of every simulation step and remain constant until the end of the current step. \textcolor{blue}{Second, the intensity variables, including $H$, $j_n$, $i_s$ and $i_e$, and the time-variant parameters, including $\kappa$, $D_s$, $D_s$ and $k_r$, are treated similarly to the inputs; they are calculated based on the internal states of the battery at the start of every simulation step and assumed to remain constant until the end of the current step. Third, the inertial variables, including $\bar{c}_s$, $c_{ss}$, $c_e$, $Q_e$ and $T$, are updated based on the intensity variables taking effect in the current step. Their values at the end of the current step are calculated by the discrete state-space equations. Fourth, potential variables or parameters, including $\Phi_s$, $\Phi_e$, $U_{\mathrm{OCP}}$ and $V_t$, are updated at the end of every simulation step.}

We denote the time stamps at the start and end of the $l$-th simulation step by $t_{l-1}$ and $t_{l}$ and denote the current time interval by $\Delta t_{l}=t_{l}-t_{l-1}$. \textcolor{blue}{When the last simulation step stops at $t_{l-1}$, the values of $V_t(t_{l-1})$, $T(t_{l-1})$, $Q_e^{\pm}(t_{l-1})$, $c_e(x,t_{l-1})$, $c_{ss}(x,t_{l-1})$ and $\bar{c}_s(x,t_{l-1})$ are known, plotted by green boxes with boundaries made of dashed-dotted lines, as shown in Fig.~\ref{fig:flowchart}. Based on these values and the latest input $I(t)$ and $T_{\mathrm{amb}}(t)$, time-variant parameters $D_s^{\pm}(t_l)$, $D_e(t_l)$, $\kappa(t_l)$ and $k_r^{\pm}(t_l)$ are updated via Eqs. (\ref{equ:de_kappa})-(\ref{equ:kr}) first. The intensity variables $j_n(t_l)$, $H(t_l)$, $i_s(t_l)$ and $i_e(t_l)$ are updated via Eqs. (\ref{equ:solid-phase current}), (\ref{equ:jn}),and (\ref{equ:reaction heat}) next. They are plotted by green boxes with boundaries made of
solid line, as shown in Fig.~\ref{fig:flowchart}. Once intensity variables and parameters are known, inertial variables at the end of the simulation step can be updated via state-space equations in the discrete-time form, as given below: }
\begin{equation}
    \label{equ:inertial states}
    \begin{split}
        & Q_e^{\pm}(t_l) = Q_e^{\pm}(t_{l-1}) \exp \left( -\frac{\Delta t_{l}}{\tau_e^{\pm}}(t_l) \right) + K_{Q_e}^{\pm}(t_l)  \left( 1-\exp \left( -\frac{\Delta t_{l}}{\tau_e^{\pm}(t_l)} \right) \right). \\
        & \bar{c}_s(x,t_l) = \bar{c}_s(x,t_{l-1}) - \frac{R_s^{\pm}\Delta t_{l}}{3} j_n(x,t_{l}). \\
        & c_{ss}(x,t_l) = \bar{c}_s(x,t_l)+ \left( c_{ss}(x,t_{l-1})-\bar{c}_s(x,t_{l-1}) \right) \exp \left( -\frac{\Delta t_{l}}{\tau_s(x,t_l)} \right) - \frac{R_s^{\pm}j_n(x,t_l)}{5D_s^{\pm}(x,t_l)} \left( 1-\exp \left( -\frac{\Delta t_{l}}{\tau_s(x,t_l)} \right) \right). \\
        & T(t_l) = T(t_{l-1}) \exp \left( -\frac{\Delta t_{l}}{\tau_T(t_{l})} \right) + K_T(t_{l}) \left( 1-\exp \left( -\frac{\Delta t_{l}}{\tau_T(t_{l})} \right) \right).
    \end{split}
\end{equation}
\textcolor{blue}{where $x=0^{\pm},\frac{L^{\pm}}{3},\frac{2L^{\pm}}{3},L^{\pm}$. Finally, the potential variables and parameters $U_{\mathrm{OCP}}^{\pm}(x,t_l)$, $V_t(t_l)$, $\Phi_s(x,t_l)$ and $\Phi_e(x,t_l)$ are updated via Eqs. (\ref{equ:vt_formula})-(\ref{equ:phie drop}). Then, the above steps are repeated for the next interval.}

\subsection{Initializing process}
At the simulation start, the battery initial states should be determined. First, the parameters involved in constructing the battery model should be determined. \textcolor{blue}{Considering the data sources, parameters can be categorized into three types: determined by the material properties, determined by the manufacturing and assumed to fit the battery characteristics. The benchmark and adopted values of all parameters in this paper for simulating the LFPO cell and NCM cell are listed in Table~\ref{tab:params}.}
\begin{table}[p]
    \resizebox{\textwidth}{!}{\begin{threeparttable}
    \caption{Parameter settings of the lithium-ion battery cell used in this work.}\label{tab:params}
    \begin{tabular}{ccc}
    \toprule
    Parameters & Benchmark & Set values \\
    \midrule
    $^{\rm m}$$m$ & - &  \makecell{LFPO: 3.69$\times 10^{-2}$,NCM523: 3.95$\times 10^{-2}$, \\NCM811: 3.85$\times 10^{-2}$} \\
    $^{\rm m}$$L^+,L^-,L^{\mathrm{sep}}$ & - & 7.75$\times 10^{-5}$,8.1$\times 10^{-5}$, 2$\times 10^{-5}$  \\
    $^{\rm m}$$A^+, A^-, A^{\mathrm{sep}}, A_{\mathrm{surf}}$ & - & 6.1$\times 10^{-2}$, 6.41$\times 10^{-2}$, 6.36$\times 10^{-2}$, 4.4$\times 10^{-3}$ \\
    $^{\rm p}$$D_s^+$ & \makecell{LFPO: 1.25$\times 10^{-15}$\cite{xu_fast_2019},\\ NCM: 1-10$\times 10^{-14}$\cite{zhang_electrochemical_2020,gao_implementation_2021,li_electrochemical_2020,li_parameter_2020}}
     & (\ref{equ:ds}) \\
    $^{\rm p}$$D_s^-$ & C: 3.9$\times 10^{-14}$-5.5$\times 10^{-14}$\cite{zou_framework_2016,khaleghi_rahimian_extension_2013,luo_new_2013,torchio_lionsimba_2016,xu_fast_2019,botte_influence_1999} & (\ref{equ:ds}) \\
    $^{\rm p}$$D_e$ & 2.6-7.5$\times 10^{-10}$\cite{zou_framework_2016,han_simplification_2015_1,smith_control_2007,luo_new_2013,li_electrochemical_2020,torchio_lionsimba_2016,botte_influence_1999} & (\ref{equ:de_kappa}) \\
    $^{\rm a}$$\sigma_s^+$ & LFPO: 10.8\cite{xu_fast_2019}, NCM: 1-68\cite{zhang_electrochemical_2020,bi_adaptive_2020,li_electrochemical_2020,yin_new_2019} & LFPO\&NCM523\&NCM811: 3.8 \\
    $^{\rm a}$$\sigma_s^-$ & 100\cite{zhang_electrochemical_2020,zou_framework_2016,khaleghi_rahimian_extension_2013,smith_control_2007,luo_new_2013,torchio_lionsimba_2016,botte_influence_1999} & 100  \\
    $^{\rm p}$$\kappa$ & 3.46\cite{li_reduced-order_2021} & (\ref{equ:kappa}) \\ 
    $^{\rm a}$$t_+^0$ & 0.36-0.4\cite{smith_control_2007,li_electrochemical_2020,torchio_lionsimba_2016,xu_fast_2019,zhao_electrochemical-thermal_2019} & 0.38 \\ 
    $^{\rm a}$$R_c$ & - & 0.0064 \\ 
    $^{\rm a}$$R_f^+$ & 0 & 1.3$\times 10^{-4}$ \\ 
    $^{\rm a}$$R_f^-$ & 0.001-0.1\cite{li_electrochemical_2020,li_parameter_2020,yin_new_2019} & 3.3$\times 10^{-4}$ \\ 
    $^{\rm a}$$R_s^+$ & \makecell{LFPO: 0.2-1.7$\times 10^{-7}$\cite{xu_fast_2019}, \\ NCM: 1-18$\times 10^{-6}$\cite{gao_implementation_2021,li_electrochemical_2020,li_parameter_2020,zhao_electrochemical-thermal_2019}} & LFPO: 5.2$\times 10^{-8}$, NCM523\&NCM811: 5$\times 10^{-6}$ \\ 
    $^{\rm a}$$R_s^-$ & 1-12.5$\times 10^{-6}$\cite{zhang_electrochemical_2020,zou_framework_2016,khaleghi_rahimian_extension_2013,han_simplification_2015_1,luo_new_2013,torchio_lionsimba_2016,xu_fast_2019,botte_influence_1999} & 7.5$\times 10^{-6}$ \\ 
    $^{\rm p}$$M^+$ & \makecell{LFPO:157.7$\times 10^{-3}$, NCM523: 96.5$\times 10^{-3}$,\\NCM811: 97.3$\times 10^{-3}$} & \makecell{LFPO: 157.7$\times 10^{-3}$, NCM523: 96.5$\times 10^{-3}$, \\NCM811: 97.3$\times 10^{-3}$}
    \\ 
    $^{\rm p}$$M^-$ & 72.06$\times 10^{-3}$ & 72.06$\times 10^{-3}$ \\  
    $^{\rm p}$$\rho^+$ & LFPO: 3.6$\times 10^{3}$, NCM523\&NCM811: 4.8$\times 10^{3}$ & LFPO: 3.6$\times 10^{3}$, NCM523\&NCM811: 4.8$\times 10^{3}$ \\ 
    $^{\rm p}$$\rho^-$ & 2.24$\times 10^{3}$ & 2.24$\times 10^{3}$ \\ 
    $^{\rm m}$$\varepsilon_e^+$ & 0.27-0.45\cite{li_electrochemical_2020,li_parameter_2020} & LFPO: 0.4461, NCM523: 0.4401, NCM811:0.5038 \\ 
    $^{\rm m}$$\varepsilon_e^-$ & 0.26-0.5\cite{li_electrochemical_2020,li_parameter_2020} & LFPO: 0.4733, NCM523: 0.4893, NCM811: 0.4893 \\ 
    $^{\rm m}$$\varepsilon_e^{\mathrm{sep}}$ & 0.4-0.55\cite{li_electrochemical_2020,li_parameter_2020} & 0.4 \\ 
    $^{\rm m}$$\varepsilon_s^+$ & 0.35-0.5\cite{li_electrochemical_2020,li_parameter_2020} & LFPO: 0.4928, NCM523: 0.4806, NCM811: 0.4258 \\ 
    $^{\rm m}$$\varepsilon_s^-$ & 0.4-0.5\cite{li_electrochemical_2020,li_parameter_2020} & LFPO: 0.489, NCM523: 0.4742, NCM811: 0.4742 \\ 
    $^{\rm m}$$c_{e,0}$ & 1000-1200\cite{zou_framework_2016,smith_control_2007,khaleghi_rahimian_extension_2013,han_simplification_2015_1,luo_new_2013,torchio_lionsimba_2016} & 1200 \\ 
    $^{\rm a}$$C_p$ & 746-998\cite{wang_lithium-ion_2020,botte_influence_1999} & 1000 \\ 
    $^{\rm a}$$h_c$ & 5-20\cite{zou_framework_2016,wang_lithium-ion_2020} & 20 \\ 
    $^{\rm p}$$k_r^+$ & LFPO: 9.65$\times 10^{-8}$\cite{xu_fast_2019}, NCM: 9.65-96.5$\times 10^{-7}$\cite{li_electrochemical_2020,li_parameter_2020} & (\ref{equ:kr}) \\ 
    $^{\rm p}$$k_r^-$ & 1.7-9.6$\times 10^{-6}$\cite{li_electrochemical_2020,zou_framework_2016,khaleghi_rahimian_extension_2013,luo_new_2013,torchio_lionsimba_2016} & (\ref{equ:kr}) \\
    $^{\rm a}$$p$ & 1.5-4.1\cite{zou_framework_2016,khaleghi_rahimian_extension_2013,han_simplification_2015_1,torchio_lionsimba_2016,xu_fast_2019} & 1.5 \\ 
    \bottomrule
    \end{tabular}
    \begin{tablenotes}
        \item 
        $^{\rm a}$: Assumed.  
        $^{\rm m}$: Manufactured. 
        $^{\rm p}$: Material properties.
    \end{tablenotes}
    \end{threeparttable}}
\end{table}

Second, we acquire the working region of the battery, including the low cut-off and high cut-off voltages $V_{\mathrm{min}}$ and $V_{\mathrm{max}}$. By conducting the full-cycle low-current charge and discharge in the working region, the total capacity of the battery cell $C_Q$ can be obtained. Then, the stoichiometry region of active particles in the positive electrode and negative electrode, denoted by $\theta_{\mathrm{max}}^{\pm}$ and $\theta_{\mathrm{min}}^{\pm}$, can be obtained by solving the two non-linear equations below:
\begin{equation}
    \label{equ:stoichiometry_region}
    \begin{split}
        &U_{\mathrm{OCP}}^+(\theta_{\mathrm{min}}^+)-U_{\mathrm{OCP}}^-(\theta_{\mathrm{max}}^-) = V_{\mathrm{max}}, \quad
        U_{\mathrm{OCP}}^+(\theta_{\mathrm{max}}^+)-U_{\mathrm{OCP}}^-(\theta_{\mathrm{min}}^-) = V_{\mathrm{min}}, \\
        &A^+L^+c_{s,\mathrm{max}}^+\varepsilon_s^+\left(\theta_{\mathrm{max}}^+-\theta_{\mathrm{min}}^+\right) = \frac{3.6C_Q}{F}, \quad
        A^-L^-c_{s,\mathrm{max}}^-\varepsilon_s^-\left(\theta_{\mathrm{max}}^--\theta_{\mathrm{min}}^-\right) = \frac{3.6C_Q}{F}.
    \end{split}
\end{equation}
\textcolor{blue}{The first equation refers to the situation in which the battery is fully charged, when the lithium concentration of the active particles in the positive electrode reaches the upper bound and that in the negative electrode reaches the lower bound. The second equation refers to the situation in which the battery is fully discharged. The third and fourth equations ensure charge conservation. Since both $U_{\mathrm{OCP}}^{\pm}$ are monotonic functions, the above equations have a unique solution.}

Third, the commonly used SOC-OCV curve can be derived once the bounds of stoichiometry in active particles of the negative electrode and positive electrode are known:
\begin{equation}
    \label{equ:soc-ocv}
    \mathrm{OCV} = U_{\mathrm{OCP}}^+\left(\theta_{\mathrm{max}}^+-SOC\left(\theta_{\mathrm{max}}^+-\theta_{\mathrm{min}}^+\right)\right) - U_{\mathrm{OCP}}^-\left(\theta_{\mathrm{max}}^-+SOC\left(\theta_{\mathrm{max}}^--\theta_{\mathrm{min}}^-\right)\right).
\end{equation}

Fourth, the initial values of the inertial states of the battery are determined, including $Q_e^{\pm}$, $c_e$, $c_{ss}$, $c_s$ and $T$ (as shown in Fig.~\ref{fig:flowchart}). By measuring the open circuit voltage of the cell, the initial SOC$_0$ can be obtained by interpolation in the SOC-OCV curve obtained in the third step. Thus, $c_{ss}$ and $c_s$ are initialized at:
\begin{equation}
    \label{equ:init_solid}
    \begin{split}
        &c_{ss}(x,t_0)=\bar{c}_s(x,t_0)=c_{s,\mathrm{max}}^+\left(\theta_{\mathrm{max}}^+ - \mathrm{SOC}_0\left( \theta_{\mathrm{max}}^+-\theta_{\mathrm{min}}^+\right)\right), \quad x = 0^+,\frac{L^+}{3},\frac{2L^+}{3},L^+ \\
        &c_{ss}(x,t_0)=\bar{c}_s(x,t_0)=c_{s,\mathrm{max}}^-\left(\theta_{\mathrm{min}}^- + \mathrm{SOC}_0\left( \theta_{\mathrm{max}}^--\theta_{\mathrm{min}}^-\right)\right), \quad x = 0^-,\frac{L^-}{3},\frac{2L^-}{3},L^-. \\
    \end{split}
\end{equation}
At the start, \ce{Li^+} in the solution phase is assumed to be uniformly distributed along the thickness direction of the cell,
Thus, the $Q_e^{\pm}$ values are initialized at:
\begin{equation}
    \label{equ:init_solution}
    Q_e^{\pm}(t_0)=A^{\pm}L^{\pm}\varepsilon_e^{\pm}c_{e,0}.
\end{equation}
Finally, the cell temperature $T(t_0)$ is initialized at the ambient temperature, i.e., $T(t_0)=T_{\mathrm{amb}}(t_0)$.

\subsection{Stabilizing method}\label{sec:stable}
\textcolor{blue}{
In Section~\ref{sec:discrete}, the reaction rate $j_n$ is modelled as the intensity variable, which is assumed to remain constant within one simulation step (the other three intensity variables, $i_s$, $i_e$ and $H$, are all determined by $j_n$). This assumption is unavoidable when discretizing the system, which nevertheless brings additional error to the model. Through numerical experiments, we find that it is appropriate for most working conditions of different batteries. However, when $\frac{d U_{\mathrm{OCP}} }{d \theta_{ss} } $ is very large (as shown in the red dotted box in Fig.~\ref{fig:Uocp}), this assumption is likely to subject the model to oscillation. Because calculating $j_n$ requires $U_{\mathrm{OCP}}$ according to Eq. (\ref{equ:se potential diff1}), when $\frac{d U_{\mathrm{OCP}} }{d \theta_{ss} } $ is small, we can calculate $j_n$ based on $U_{\mathrm{OCP}}$ at $c_{ss}(x,t_{l-1})$, but when $\frac{d U_{\mathrm{OCP}} }{d \theta_{ss} } $ varies greatly, $U_{\mathrm{OCP}}$ changes significantly within $[t_{l-1},t_l]$ and makes the assumption no longer accurate. Under this kind of circumstance, the model might be unstable.}

\textcolor{blue}{To solve this problem, two measures are taken in this work. First, we reduce the risk of oscillation from the root of modelling. Specifically, we refine the value of $k_s$ in Eq. (\ref{equ:inertial_w}). As mentioned in Section \ref{sec:soliddiffusion}, a smaller $k_s$ is likely to cause oscillation. This is because a smaller $k_s$ leads to a smaller $\tau_s$, and the term $\exp \left( -\frac{\Delta t_{l}}{\tau_s(x,t_l)} \right)$ in Eq. (\ref{equ:inertial states}) is near 0. Then, in every updating step, more weights are allocated to the term $\frac{R_s j_n}{5D_s}$. Since $j_n$ varies significantly between $[t_{l-1},t_l]$ under extreme conditions, the model experiences oscillation under the influence of $k_s$. However, an excessively large $k_s$ can decrease the accuracy because it deviates from the true diffusion characteristics (if not taking the stable problem into account, a perfect $k_s$ should be obtained by frequency response optimization, as mentioned in Section \ref{sec:soliddiffusion}).} Considering the above points, by experiments, $k_s$ is set as $1/28$ for the graphite and NCM active particles, and $1/9$ for the LFPO active particles to realize a trade-off between accuracy and stabilization.

\textcolor{blue}{Nevertheless, after testing the proposed model under various working conditions for different batteries, we still find that oscillation can occur under some extreme working conditions. Thus, to ensure the practicability of this work, we develop a second measure to handle the oscillation,}; i.e., the Savitzky--Golay filter (SGF)~\cite{press1990savitzky} is applied to eliminate the oscillation after it happens. The SGF can smooth the sequence in a moving window with little resolution loss. In addition, the fitted weights of the moving window are calculated in advance, which makes it highly efficient for online implementations. \textcolor{blue}{The hyperparameters of the SGF include the order $N_{\mathrm{SG}}$ and the moving window length $M_{\mathrm{SG}}$}. The framework and formulas for applying the SGF are given below.
\begin{framed}\noindent
\begin{minipage}{0.45\textwidth}
    \label{fig:sgf}
    \centering
	\includegraphics[width=0.9\textwidth]{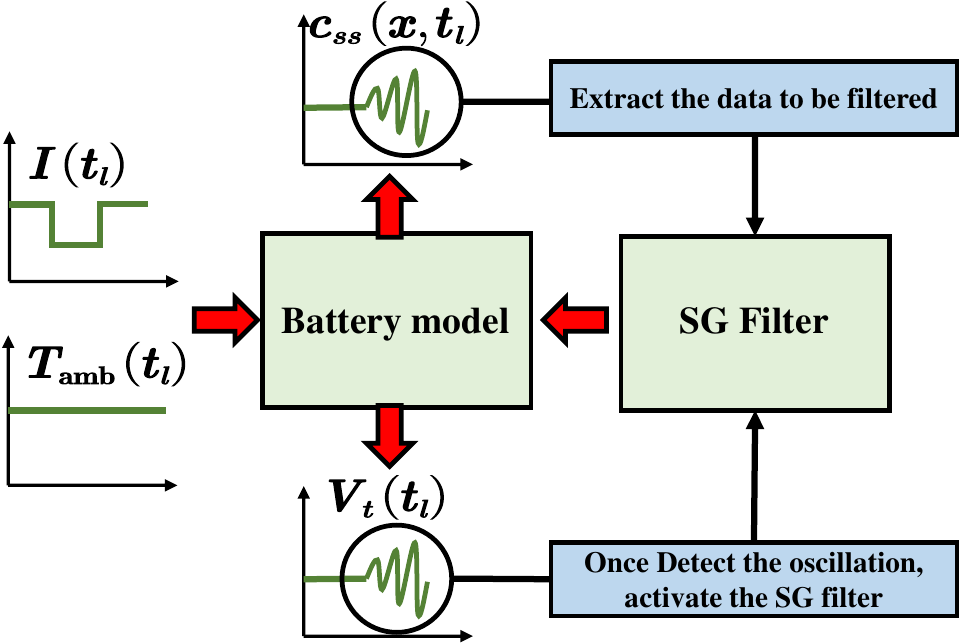}
\captionof{figure}{Framework of applying the SGF to eliminate the oscillation.}
\end{minipage}%
\begin{minipage}{0.5\textwidth}
\begin{equation}\label{equ:sgf}
\begin{split}
&\mathbf{X}=\left[ \begin{matrix}
	\left( \begin{array}{c}
	\frac{1-M_{\mathrm{SG}}}{2}\\
	\frac{3-M_{\mathrm{SG}}}{2}\\
	\vdots\\
	\frac{M_{\mathrm{SG}}-1}{2}\\
\end{array} \right) ^{N_{\mathrm{SG}}}&		\left( \begin{array}{c}
	\frac{1-M_{\mathrm{SG}}}{2}\\
	\frac{3-M_{\mathrm{SG}}}{2}\\
	\vdots\\
	\frac{M_{\mathrm{SG}}-1}{2}\\
\end{array} \right) ^{N_{\mathrm{SG}}-1}&		\cdots&		\left( \begin{array}{c}
	\frac{1-M_{\mathrm{SG}}}{2}\\
	\frac{3-M_{\mathrm{SG}}}{2}\\
	\vdots\\
	\frac{M_{\mathrm{SG}}-1}{2}\\
\end{array} \right) ^0\\
\end{matrix} \right] 
\\
&\mathbf{B}= \mathbf{X}\left( \mathbf{X}^{\mathrm{T}}\mathbf{X} \right) ^{-1}\mathbf{X}^{\mathrm{T}}  
\\
&\left[ \begin{array}{c}
	\hat{c}_{ss}\left( x,t_{l-M_{\mathrm{SG}}+1} \right)\\
	\hat{c}_{ss}\left( x,t_{l-M_{\mathrm{SG}}+2} \right)\\
	\vdots\\
	\hat{c}_{ss}\left( x,t_l \right)\\
\end{array} \right] =\mathbf{B}\left[ \begin{array}{c}
	c_{ss}\left( x,t_{l-M_{\mathrm{SG}}+1} \right)\\
	c_{ss}\left( x,t_{l-M_{\mathrm{SG}}+2} \right)\\
	\vdots\\
	c_{ss}\left( x,t_l \right)\\
\end{array} \right] 
\end{split}
\end{equation}
\end{minipage}
\end{framed}
\textcolor{blue}{
$\mathbf{B}$ can be calculated and stored in advance to reduce the computation cost. During the simulation, once the oscillation is detected, only the newest $M_{\mathrm{SG}}$ data points are filtered. Notably, only $c_{ss}$ is selected as the state to be filtered because the ill approximation of $j_n$ affects $c_{ss}$ first according to the analysis above. By numerical experiments, we find that filtering only $c_{ss}$ can eliminate the oscillation effectively.}

\subsection{Closed-loop correction scheme}
\textcolor{blue}{Due to incorrect initialization or error accumulation, the model accuracy cannot always remain high during long-term continuous simulation. Thus, a real-time closed-loop correction scheme is designed in this work, which can adaptively correct the battery states based on the measurable output.} In previous research, Kalman filters were widely adopted to handle the problem. \textcolor{blue}{However, upon attempting different variants of Kalman filters, we found that they are not appropriate for the proposed model because there are nearly 30 states internal to the battery, which places a huge burden on calculation, e.g., computing the Jacob matrix in an extended Kalman filter or the square root of the sigma point matrix in an unscented Kalman filter, making the model impractical. Conventional Kalman filters ignore the high correlation between different states, which can be utilized to simplify the correction complexity of our model. By conducting numerical experiments in open-loop, we found that once the model is initialized at the correct SOC, it can track the true trajectories of other states accurately on the long time-scale (as presented in Section~\ref{sec:insitumonitoring}-\ref{sec:outputprediction}), inspiring us that the key to correct our model is the SOC, i.e., the average \ce{Li^+} concentration $\bar{c}_s$. Based on this idea,} a heuristic correction method is proposed.

We denote the time stamp at the start of correction by $t_l$ and the measured terminal voltage by $\hat{V}_t(t_l)$. Since we aim to find the appropriate SOC for the current time, other terms in the terminal voltage are eliminated, and only the equilibrium potential that is directly related to the SOC remains:
\begin{equation}
    \label{equ:ocv_corr}
    \hat{U}_{\mathrm{OCV}}(t_l) =  \widehat{V}(t_l) - \Phi_e(L^+,t_l) + \Phi_e(0^-,t_l) + R_c I(t_l) -\eta(t_l).  
\end{equation}
where $\Phi_e(0^-,t_l)- \Phi_e(L^+,t_l)$ is calculated by Eq. (\ref{equ:phie drop}) and $\eta(t_l)$ is calculated by:
\begin{equation}
    \label{equ:eta}
    \begin{split}
       \eta(t_l) = & F R_f^{+} j_n(L^+,t_l) - F R_f^{-} j_n(0^-,t_l)  + \frac{2RT(t_l)}{F} \ln \left( \frac{F j_n(L^+,t_l)}{2i_0^+(t_l)}+\sqrt{ \left( \frac{F j_n(L^+,t_k)}{2i_0^+(t_l)} \right) ^2+1} \right)  \\&- \frac{2RT(t_l)}{F} \ln \left( \frac{F j_n(0^-,t_l)}{2i_0^-(t_l)}+\sqrt{ \left( \frac{F j_n(0^-,t_l)}{2i_0^-(t_l)} \right) ^2+1} \right). 
    \end{split}
\end{equation}
\textcolor{blue}{
Eq. (\ref{equ:ocv_corr}) demonstrates that to let the predicted $V_t$ approach the true value $\hat{V}_t$, the open circuit voltage of the battery should approach $\hat{U}_{\mathrm{OCV}}$. Similar to constructing the simplified solid-phase diffusion and solution-phase migration models, an ensemble average strategy is adopted here. To match $\hat{U}_{\mathrm{OCV}}$, the solid-phase stoichiometry in both the negative electrode and the positive electrode needs modification. We denote the ensemble average correction quantity in the negative electrode by $\Delta\theta^-(t_l)$, which means $\theta_s(x,t_l)$ and $\theta_{ss}(x,t_l)$ ($x=0^-,\frac{L^-}{3},\frac{2L^-}{3},L^-$) should incorporate $\Delta\theta^-(t_l)$ after correction. Similarly, in the positive electrode, we have $\Delta\theta^+(t_l)$. The correction quantities should satisfy two conditions. First, the total \ce{Li} in the two electrodes should remain unchanged. Second, the open circuit voltage should be equal to $\hat{U}_{\mathrm{OCV}}$. Thus, $\Delta\theta^-(t_l)$ and $\Delta\theta^+(t_l)$ can be obtained by solving the non-linear equations below:}
\begin{equation}
    \label{equ:y_corr}
    \begin{split}
    & A^+L^+\varepsilon_s^+c_{s,\mathrm{max}}^+\Delta \theta^+(t_l) + A^-L^-\varepsilon_s^-c_{s,\mathrm{max}}^-\Delta \theta^-(t_l)= 0, \\
    & U_{\mathrm{OCP}}^+\left(\theta^+_{ss}(L^+,t_l)+\Delta \theta^+(t_l)\right) - U_{\mathrm{OCP}}^-\left(\theta^-_{ss}(0^-,t_l)+\Delta \theta^-(t_l)\right) = \hat{U}_{\mathrm{OCV}}(t_l).
    \end{split}
\end{equation}
Since both $U_{\mathrm{OCP}}^{\pm}$ are monotonic functions, the above equations have a unique solution. Then, the correction quantities of solid-phase concentrations can be computed by $\Delta c_s^{\pm}(t_l) = c_{s,\mathrm{max}}^{\pm} \Delta \theta^{\pm}(t_l) $. \textcolor{blue}{After obtaining the correction terms at $t_l$, there still remains a problem to solve. Mathematically, directly adding these terms to the current $\bar{c}_s$ and $c_{ss}$ is equivalent to adding an instantaneous process to the solid-phase diffusion model. However, as analysed in Section~\ref{sec:stable}, this introduces instability into the model and leads to oscillation. To maintain model stability, when the ideal correction quantities, $\Delta c_s^{\pm}$, are obtained at $t_l$, the actual correction quantities, denoted by $\Delta \hat{c}_s^{\pm}$, are determined by the historical actual correction quantities and the latest ideal correction quantities together via a first-order inertial process:}
\begin{equation}
    \label{equ:dcs}
    \Delta \hat{c}_s^{\pm}(t_l) = \exp(-\frac{\Delta t_l}{\tau_{\Delta}^{\pm}}) \Delta \hat{c}_s^{\pm}(t_{l-1}) + \left(1-\exp(-\frac{\Delta t_l}{\tau_{\Delta}^{\pm}})\right) \Delta c_s^{\pm}(t_l).
\end{equation}
where $\tau_{\Delta}^{\pm}$ are appropriate
time constants that control the stability of the correction scheme. In this work, $\tau_{\Delta}$ for the positive electrode is set as 0.2, and for the negative electrode, it is set as 60. Then, the solid-phase concentrations are corrected by the actual correction quantities:
\begin{equation}
    \label{equ:cs_corr}
    \hat{\bar{c}}_s(x,t_l) = \bar{c}_s(x,t_l) + \Delta \hat{c}_s^{\pm}(t_l), \quad \hat{c}_{ss}(x,t_l) = c_{ss}(x,t_l) + \Delta \hat{c}_s^{\pm}(t_l),\quad  x=0^{\pm},\frac{L^{\pm}}{3},\frac{2L^{\pm}}{3},L^{\pm}.
\end{equation}

\textcolor{blue}{It is also noteworthy that since solving the correction terms also requires computing resources, it is recommended to activate the correction step when the predicted voltage error exceeds a given threshold, denoted by $V_{\mathrm{error}}$, which is set as 0.02 V in this work.} The steps of the entire simulation framework are given below.

\begin{framed}\noindent
\begin{minipage}{0.5\textwidth}
\begin{algorithm}[H]\small
\caption{Simulation steps}
    \label{algo:1}
\begin{algorithmic}[1] 
\REQUIRE Manufacturing and material information of the battery. 
    \STATE Set values of parameters in Table~\ref{tab:params}.
    \STATE Set cut-off voltages $V_{\mathrm{min}}$ and $V_{\mathrm{max}}$, and solve the stoichiometry regions of electrodes $\theta_{\mathrm{min}}^{\pm}$ and $\theta_{\mathrm{max}}^{\pm}$ via Eq.~(\ref{equ:stoichiometry_region}).
    \STATE Calculate the SOC-OCV curve via Eq.~(\ref{equ:soc-ocv}).
    \STATE Measure the open circuit voltage $U_{\mathrm{OCV}}(t_0)$ and ambient temperature $T_{\mathrm{amb}}(t_0)$.
    \STATE Calculate SOC$_0$ by interpolation in the SOC-OCV curve.
    \STATE Initialize $c_{ss}^{\pm}(x,t_0)$ and $\bar{c}_s^{\pm}(x,t_0)$ via Eq.~(\ref{equ:init_solid}), initialize $Q_e^{\pm}(t_0)$ via Eq.~(\ref{equ:init_solution}), initialize $T(t_0)$ at  $T_{\mathrm{amb}}(t_0)$, initialize $\Delta \hat{c}_s^{\pm}(t_0)$ at 0, and initialize $V_t(t_0)$ at $U_{\mathrm{OCV}}(t_0)$.
\FOR{each $t_l = t_1, t_2,\cdots t_N$}
        \STATE Acquire $I(t_l)$, $T_{\mathrm{amb}}(t_l)$ and $\Delta t_l$ for the current simulation step,
        \STATE Update time-variant parameters $D_e^{\pm,\mathrm{sep}}(t_l)$, $\kappa^{\pm,\mathrm{sep}}(t_l)$, $\kappa_D^{\pm,\mathrm{sep}}(t_l)$, $D_s^{\pm}(x,t_l)$, $k_r^{\pm}(t_l)$ and $U_{\mathrm{OCP}}^{\pm}(x,t_l)$ based on $c_e(x,t_{l-1})$, $\bar{c}_s(x,t_{l-1})$, $c_{ss}(x,t_{l-1})$ and $T(t_{l-1})$ via Eqs.~(\ref{equ:de_kappa})-(\ref{equ:kr}) and Fig.~\ref{fig:Uocp}.
        \STATE Calculate intensity states $j_n(x,t_l)$ and $H(t_l)$ based on $c_e(x,t_{l-1})$, $c_{ss}(x,t_{l-1})$, $T(t_{l-1})$, $V_t(t_{l-1})$ and $I(t_l)$ via Eq.~(\ref{equ:jn}) and Eq.~(\ref{equ:reaction heat}).
        \STATE Calculate inertial states $Q_e^{\pm}(t_l)$, $c_e(x,t_l)$, $\bar{c}_s(x,t_l)$, $c_{ss}(x,t_l)$ and $T(t_l)$ based on their values at the end of last step and intensity state values in the current step via Eq.~(\ref{equ:inertial states}).
        \STATE Calculate the terminal voltage at the end of the current step $V_t(t_l)$.
        \STATE Measure the battery voltage at the end of the current step $\hat{V}_t(t_l)$.
        \IF{ $\left| V_t(t_l)-\hat{V}_t(t_l) \right| > V_{\mathrm{error}}$ }
            \STATE Calculate the ideal correction term $\Delta c_s^{\pm}(t_l)$ based on $c_{ss}(x,t_l)$ and $\hat{V}_t(t_l)$ via Eq.~(\ref{equ:y_corr}).
\ELSE
            \STATE Set the ideal correction term $\Delta c_s^{\pm}(t_l)=0$.
        \ENDIF
        \STATE Calculate the actual correction term $\Delta \hat{c}_s^{\pm}(t_l)$ based on $\Delta c_s^{\pm}(t_l)$ and $\Delta \hat{c}_s^{\pm}(t_{l-1})$ via Eq.~(\ref{equ:dcs}).
        \STATE Correct $\bar{c}_s(x,t_l)$ and $c_{ss}(x,t_l)$ based on $\Delta \hat{c}_s^{\pm}(t_l)$ via Eq.~(\ref{equ:cs_corr}).
        \IF{Oscillation is detected and $l \ge M_{\mathrm{SG}}$}
            \STATE Update $c_{ss}(x,t_{l-M_{\mathrm{SG}}+1})\sim c_{ss}(x,t_{l})$ via Eq.~(\ref{equ:sgf}).
        \ENDIF
\ENDFOR
\end{algorithmic}
\end{algorithm}
\end{minipage}%
\begin{minipage}{0.4\textwidth}
    \label{fig:entire}
    \centering
	\includegraphics[width=0.9\textwidth]{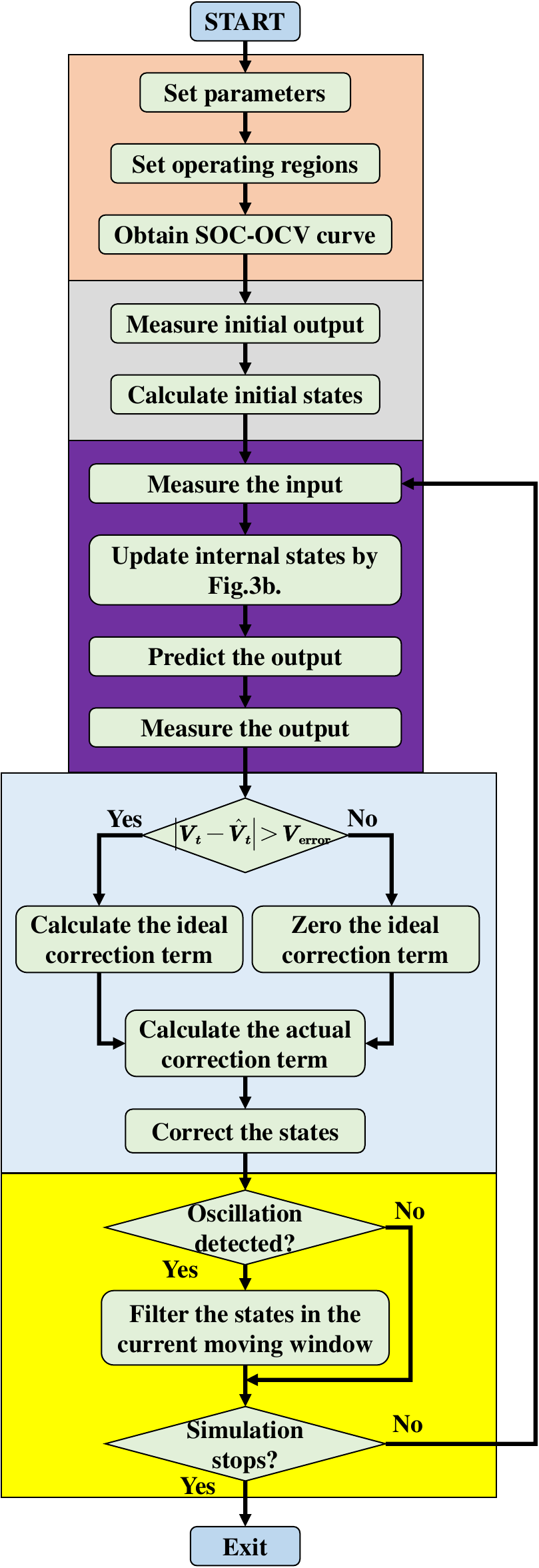}
\captionof{figure}{Complete simulation framework.}
\end{minipage}
\end{framed}

\section{Numerical experiments}\label{sec:case}
To evaluate the performance of this work, numerical experiments are designed and conducted for validation. The proposed model is compared against two highly cited simplified models: a classic ESP model and a recently proposed advanced ESP model~\cite{han_simplification_2015_1,han_simplification_2015_2}. The benchmark is a full-order P2D model that contains 51 elements on the x-axis in each electrode, 11 elements on the x-axis in the separator, 18 elements in each active particle along the r-axis, and \textcolor{blue}{1847 elements in total. All the models and simulation programs are written and run on the MATLAB R2021A platform. The hardware for computation is a 2.11 GHz Intel Core i5-10210U processor with 16 GB of RAM. Note that the proposed model contains only basic operators and that it is convenient to write the model in other programming languages, such as Python and Java.
}

\subsection{Designs}
\textcolor{blue}{As mentioned above, the proposed model contains constant parameters and time-variant parameters expressed by functions. The values of the constant parameters are listed in Table~\ref{tab:params}. Coefficients of functions depicting time-variant parameters are fit to material experiment data~\cite{AutoLion} and listed in Table~\ref{table:params_coefficients}.}

\begin{table}[width=.8\linewidth,cols=4,pos=h]
\caption{Fitted coefficients in expressions of $D_s$, $k_r$ and $\frac{d \ln f_{\pm}}{d \ln c_e}$.}\label{table:params_coefficients}
\begin{tabular*}{\tblwidth}{@{} CCCCC@{} }
\toprule
 & LFPO(+) & NCM523(+) & NCM811(+) & Graphite(-) \\
\midrule
$E_{A,k_{D_s}}$ & 0      & -7349    & -7330     & 19626 \\
$E_{A,b_{D_s}}$ & 30011  & -313     & -309      & 19626 \\
$k_{D_s}^{\mathrm{ref}}$   & 0      & -2.05e-14& -2.05e-14 & -2.4e-14 \\
$b_{D_s}^{\mathrm{ref}}$   & 8e-18  & 2.65e-14 & 2.65e-14  & 2.9e-14 \\
$E_{A,k_r}$     & 31997  & 51997    & 51997     & 67995 \\
$k_{r}^{\mathrm{ref}}$     & 5.3e-6 & 2.3e-6   & 2.6e-6    & 2.3e-5 \\
$\frac{d \ln f_{\pm}}{d \ln c_e}$ & \multicolumn{4}{c}{ $0.55(c_e/1000)^2+1.08(c_e/1000)-0.44$  } \\
\bottomrule
\end{tabular*}
\end{table}

\begin{figure}[!htb]
    \centering
\begin{minipage}{.32\textwidth}
        \centering
        \includegraphics[width=\textwidth]{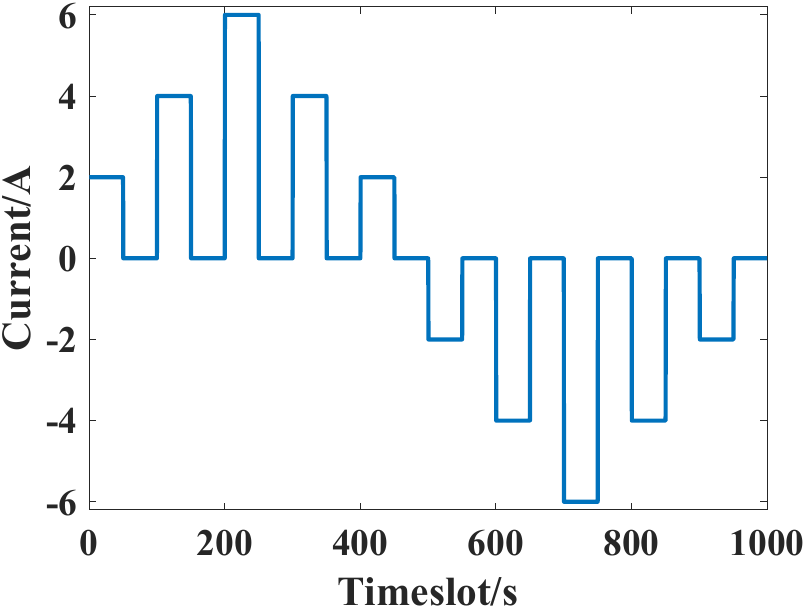}
\subcaption{\textcolor{blue}{ACC. }}\label{fig:currentACC}
\end{minipage}%
\begin{minipage}{0.32\textwidth}
        \centering
        \includegraphics[width=\textwidth]{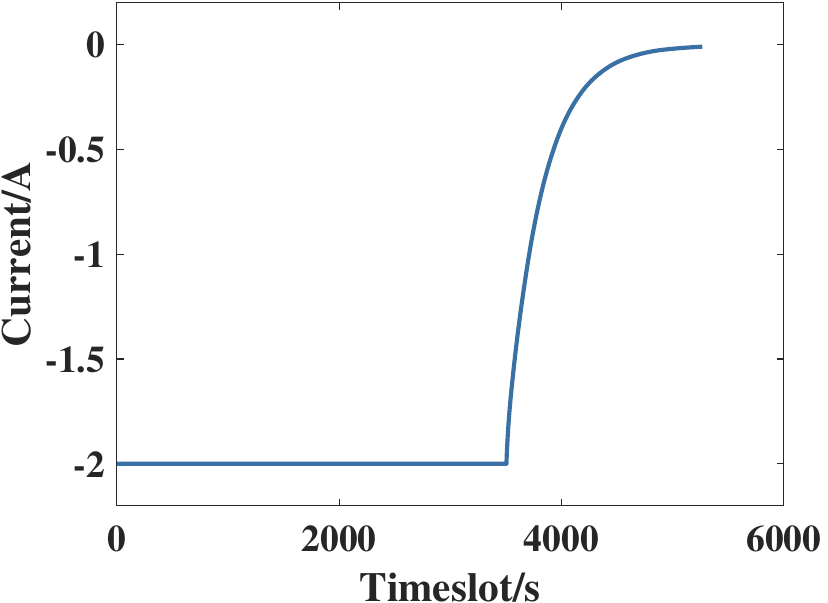}
\subcaption{\textcolor{blue}{CCCV. }}\label{fig:currentCCCV}
\end{minipage}
\begin{minipage}{0.32\textwidth}
        \centering
        \includegraphics[width=\textwidth]{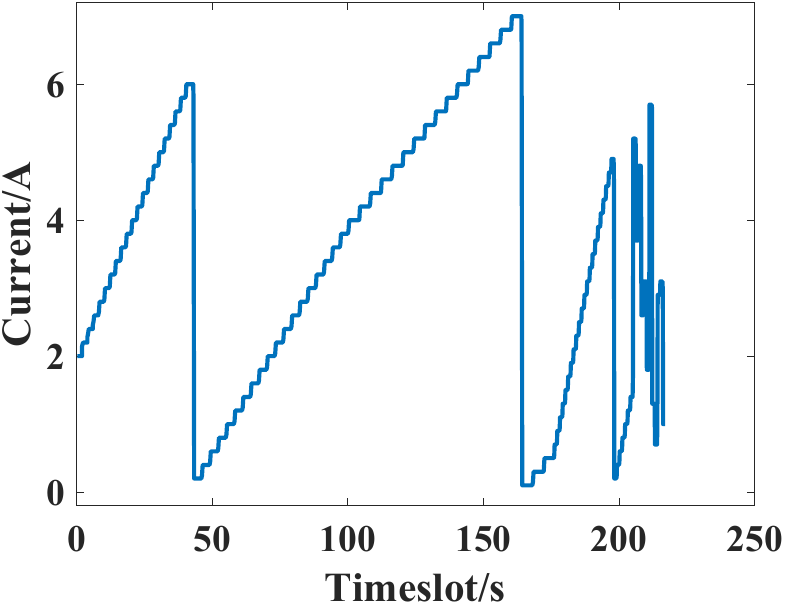}
\subcaption{\textcolor{blue}{RC. }}\label{fig:currentRC}
\end{minipage}
\caption{\textcolor{blue}{Working profiles of protocols with time-variant currents. }}
\end{figure}

\textcolor{blue}{
To test the proposed model comprehensively, the working conditions to simulate should consider three points. First, they should cover a wide range of current amplitudes and ambient temperatures. Second, they should contain various working profiles, including galvanostatic and dynamic currents. Third, they should start at different initial points.} Considering the above points, eight scenarios were designed for each type of battery to test, as listed in Table~\ref{table:sceniros}. \textcolor{blue}{Scenario nos. 1-3 and nos. 7-8 test the model under the galvanostatic discharging protocol, and the difference is the current amplitude or ambient temperature. Scenario no. 4 tests the model under the standard constant-current constant-voltage charging protocol (CCCV). Scenario no. 5 tests the model under the alternate charging and discharging protocol (ACC), and the current amplitude also varies during the switch of the current direction. Scenario no. 6 tests the model under the random discharging current protocol (RC). The dynamic current profiles of the latter three scenarios are shown in Figs.~\ref{fig:currentACC}-\ref{fig:currentRC}.}

To evaluate the model performance in simulating internal states and output, the mean absolute error (MAE), root mean squared error (RMSE) and R-squared (R$^2$) are applied. Taking the voltage as an example, for time steps from $t_1,\cdots,t_l$, the above three metrics are calculated by:
\begin{equation}
    \label{equ:index}
    R^2 = 1-\frac{\sum\nolimits_{i=1}^l \left( V_t(t_i) - \hat{V}_t(t_i)  \right)^2}{\sum\nolimits_{i=1}^l \left( V_t(t_i) - \bar{V_t}  \right)^2}, \quad MAE = (\frac{1}{l} \sum\limits_{i=1}^l \left| V_t(t_i) - \hat{V}_t(t_i)  \right|), \quad RMSE = \sqrt{\frac{1}{l} \sum\limits_{i=1}^l \left( V_t(t_i) - \hat{V}_t(t_i)   \right)^2}.
\end{equation}
where $V_t(t_i)$ is the true value and $\hat{V}_t(t_i)$ is the predicted value.

\begin{table}[width=1\textwidth, pos=h]
\caption{Simulation scenarios for testing.}\label{table:sceniros}
\begin{tabular}[width=1.5\linewidth]{CCCCCCC}
\toprule
No. & Protocol & Current amplitude & Ambient temperature & LFPO & NCM523 & NCM811 \\
\midrule
1 & galvanostatic & 1C-rate & 298 K & SOC$_0$=1  & SOC$_0$=1    & SOC$_0$=1     \\
2 & galvanostatic & 2C-rate & 298 K & SOC$_0$=1  & SOC$_0$=1    & SOC$_0$=1      \\
3 & galvanostatic & 4C-rate & 298 K & SOC$_0$=1  & SOC$_0$=1    & SOC$_0$=1 \\
4 & CCCV & -1C-rate\textasciitilde0C-rate & 298 K & SOC$_0$=0  & SOC$_0$=0    & SOC$_0$=0  \\
5 & ACC & -5C-rate\textasciitilde5C-rate  & 298 K & SOC$_0$=1  & SOC$_0$=0.7  & SOC$_0$=0.7     \\
6 & RC & 0C-rate\textasciitilde5C-rate & 298 K    & SOC$_0$=1  & SOC$_0$=1    & SOC$_0$=1   \\
7 & galvanostatic & 1C-rate & 273 K & SOC$_0$=1  & SOC$_0$=1    & SOC$_0$=1     \\
8 & galvanostatic & 1C-rate & 313 K & SOC$_0$=1  & SOC$_0$=1    & SOC$_0$=1    \\
\bottomrule
\end{tabular}
\end{table}

First, the running time of the simplified model in different working scenarios is listed in Table~\ref{table:run_time}. \textcolor{blue}{The computational efficiency of the simplified model is significantly increased as expected, ensuring that the proposed model is practical in real-world applications. Although the operating time of batteries in dynamic current scenarios (e.g., ACC and RC) is shorter than that in galvanostatic scenarios, sometimes their simulation process takes a longer time. This is because under dynamic current, we raise the sampling frequency, resulting in more total simulation steps than under galvanostatic currents.}

\begin{table*}[width=2.1\linewidth,cols=3,pos=h]
\caption{Run time (s) for simulations of different cells in eight scenarios.}\label{table:run_time}
\begin{tabular}{ccccccccc}
\toprule
LFPO & No. 1             & No. 2             & No. 3             & No. 4             & No. 5             & No. 6             & No. 7             & No. 8               \\ \hline
Operating time & 3465          & 1665.5        & 768           & 4194.7        & 1000          & 217           & 3321          & 3508          \\
P2D model              & 1157.04       & 634.79        & 483.24        & 1393.88       & 603.45        & 239.04        & 585.47        & 2176.61       \\
Simplified model          & \textbf{1.31} & \textbf{1.25} & \textbf{0.96} & \textbf{1.46} & \textbf{1.73} & \textbf{0.72} & \textbf{1.25} & \textbf{1.32} \\ \hline
NCM523            & No. 1             & No. 2             & No. 3             & No. 4             & No. 5             & No. 6             & No. 7             & No. 8              \\
Operating time & 3472          & 1653.5        & 762           & 5261.5        & 1000          & 217           & 3245          & 3519          \\
P2D model              & 337.98        & 328.02        & 256.73        & 341.10        & 418.89        & 162.14        & 307.37        & 335.26        \\
Simplified model          & \textbf{1.33} & \textbf{1.27} & \textbf{0.98} & \textbf{1.37} & \textbf{1.63} & \textbf{0.65} & \textbf{1.27} & \textbf{1.37} \\ \hline
NCM811            & No. 1             & No. 2             & No. 3             & No. 4             & No. 5             & No. 6             & No. 7             & No. 8              \\
Operating time & 3483          & 1661.5        & 767.7         & 5247.8        & 1000          & 217           & 3277          & 3528          \\
P2D model           & 343.35        & 329.85        & 263.01        & 344.77        & 417.47        & 162.59        & 330.28        & 366.18        \\
Simplified model          & \textbf{1.36} & \textbf{1.29} & \textbf{1.00} & \textbf{1.39} & \textbf{1.64} & \textbf{0.65} & \textbf{1.29} & \textbf{1.39} \\  \bottomrule
\end{tabular}
\end{table*}

\subsection{State monitoring}\label{sec:insitumonitoring}
\textcolor{blue}{The proposed model can provide information on internal chemical states in the battery, including $c_e$, $\bar{c}_s$, $c_{ss}$ and $j_n$, where the latter three are key states that can significantly affect the operating characteristics of the battery. Specifically, $\bar{c}_s$ determines the remaining charge of a battery (SOC), $c_{ss}$ determines the extreme instantaneous power the battery can provide or absorb (SOP), and $j_n$ determines the heat generation and degradation process inside the battery (SOH). All four states together determine the electrical variables inside and outside the battery, e.g., $\i_s$, $i_e$, $\Phi_s$, $\Phi_e$, $\Phi_{s-e}$, and $V_t$. The accuracy metrics listed in the tables below are averaged over the eight scenarios.}

\begin{figure}[!htb]
    \centering
\begin{minipage}{.24\textwidth}
        \centering
        \includegraphics[width=\textwidth]{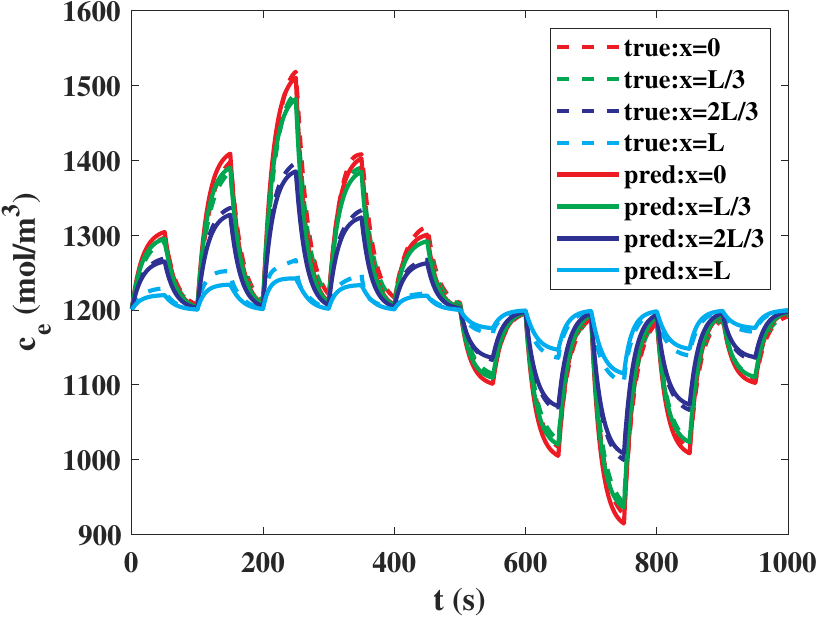}
\subcaption{NCM811, ACC, $c_e$ }\label{fig:ceACC}
\end{minipage}%
\begin{minipage}{0.24\textwidth}
        \centering
        \includegraphics[width=\textwidth]{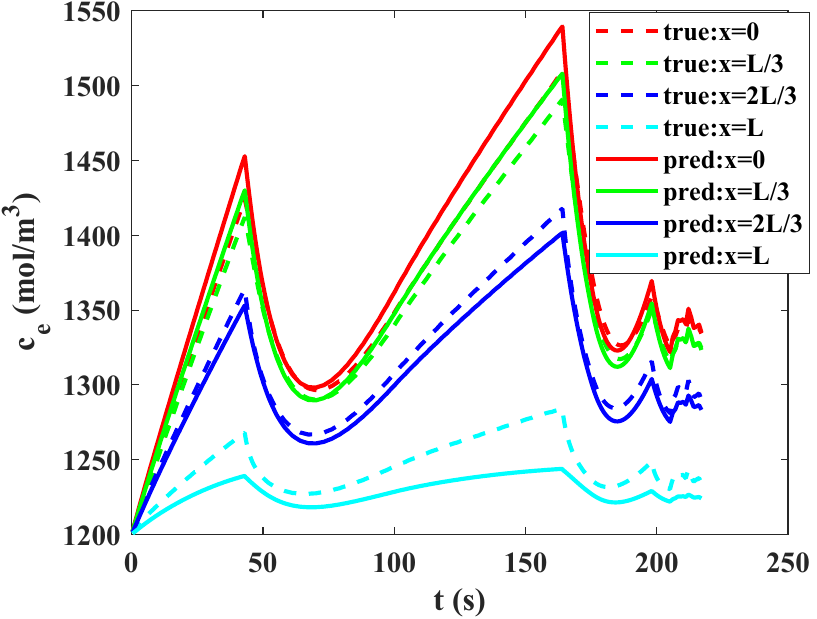}
\subcaption{NCM811, RC, $c_e$ }\label{fig:ceRC}
\end{minipage}
\begin{minipage}{.24\textwidth}
        \centering
        \includegraphics[width=\textwidth]{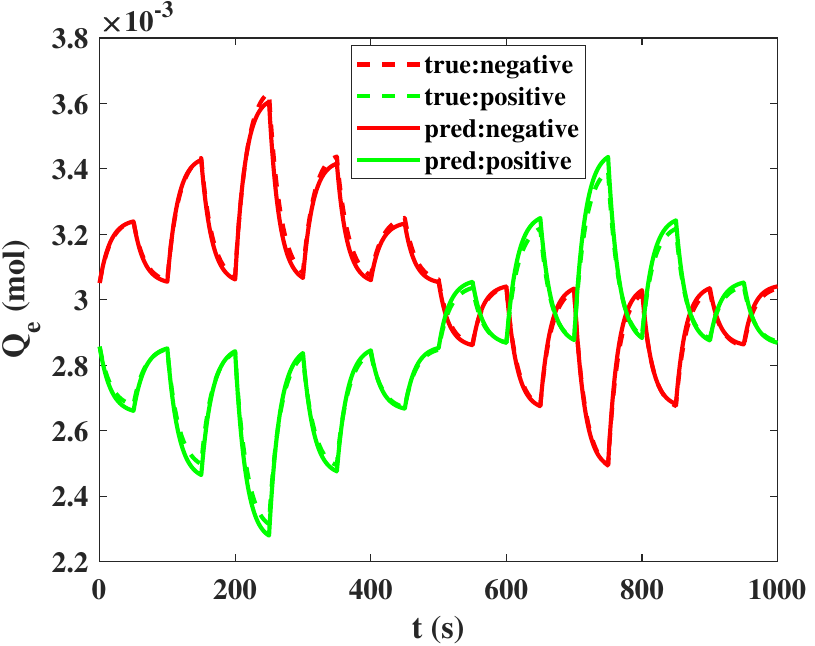}
\subcaption{\textcolor{blue}{NCM811, ACC, $Q_e$. }}\label{fig:QeACC}
\end{minipage}%
\begin{minipage}{0.24\textwidth}
        \centering
        \includegraphics[width=\textwidth]{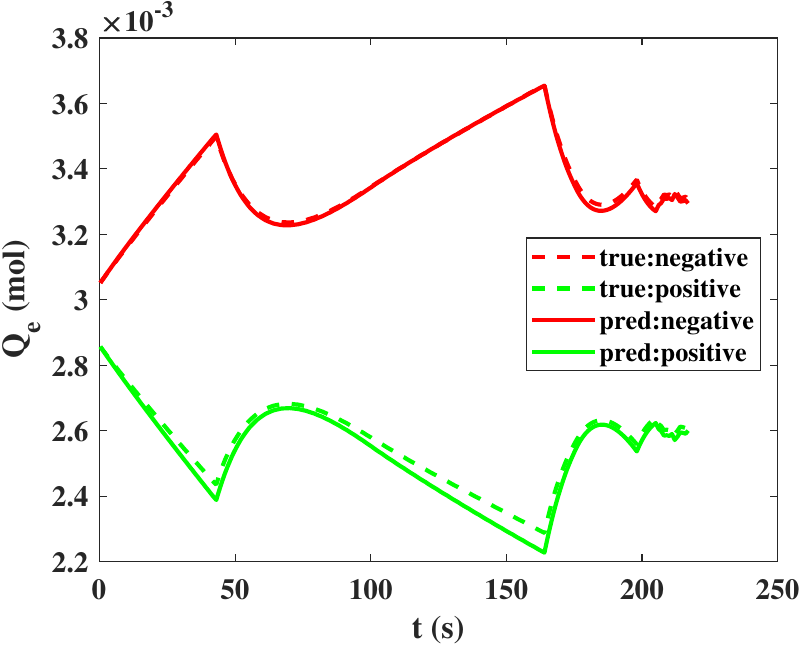}
\subcaption{\textcolor{blue}{NCM811, RC, $Q_e$ }}\label{fig:QeRC}
\end{minipage}
\caption{\textcolor{blue}{Trajectories of $c_e$ along the thickness of the negative electrode and $Q_e$ in the NCM811 battery under ACC and RC protocols. }}
\end{figure}

The prediction accuracy of $c_e$ is shown in Table~\ref{table:ce_accuracy}. \textcolor{blue}{Here, the results of the other two ESP models are not given because we assume a parabolic distribution of $c_e$ across the thickness direction.} Both the RMSE and the MAE are smaller than $100$; since the baseline value of $c_e$ is $1200$ mol/m$^3$, such a level of error is acceptable. \textcolor{blue}{To clearly demonstrate the distribution characteristics of $c_e$, we plot $c_e$ along the thickness direction in the negative electrode of NCM811 cell under dynamic current protocols (scenario nos. 5-6), as shown in Figs.~\ref{fig:ceACC}-\ref{fig:ceRC}. At $x=0^-,\frac{L^-}{3},\frac{2L^-}{3}$, the predicted $c_e$ fits well to the true value. However, at $L^-$, i.e., the point at the boundary between the electrode and separator, the error is slightly larger, especially under the RC protocol, indicating that we should be cautious when using the model to predict $c_e$ near the separator. Actually, the main contribution of this work is not using the parabolic polynomial to depict $c_e$ but proposing the concept of $Q_e$ based on the ensemble average idea and deriving the trajectories of $Q_e$ on the time horizon. Figs. \ref{fig:QeACC}-\ref{fig:QeRC} show the change of $Q_e$ against the time. Our model can accurately track the time-variant change of $Q_e$ under dynamic currents.}

\begin{table*}[width=2.1\linewidth,cols=3,pos=h]
\caption{Prediction accuracy of $c_e$ for LFPO, NCM523 and NCM811 cells.}\label{table:ce_accuracy}
\begin{tabular}{cccccccccccc}
\toprule
& \multicolumn{4}{c}{Negative Electrode}        & \multicolumn{4}{c}{Positive Electrode}        & \multicolumn{3}{c}{Separator}     \\
\midrule
LFPO   & $0^-$ & $L^-/3$ & $2L^-/3$ & $L^-$ & $0^+$ & $L^+/3$ & $2L^+/3$ & $L^+$ & $0^{\mathrm{sep}}$ & $L^{\mathrm{sep}}/2$ & $L^{\mathrm{sep}}$       \\
RMSE   & 19.294   & 16.816  & 12.113  & 12.182  & 66.354   & 36.795  & 34.220  & 39.391  & 13.833   & 12.908   & 11.744  \\
MAE    & 16.124   & 14.097  & 10.078  & 9.884   & 63.037   & 34.388  & 27.890  & 33.160  & 11.158   & 10.257   & 9.268   \\ \hline
NCM523 & $0^-$ & $L^-/3$ & $2L^-/3$ & $L^-$ & $0^+$ & $L^+/3$ & $2L^+/3$ & $L^+$ & $0^{\mathrm{sep}}$ & $L^{\mathrm{sep}}/2$ & $L^{\mathrm{sep}}$       \\
RMSE   & 14.537   & 10.660  & 4.767   & 9.147   & 82.830   & 46.466  & 29.666  & 14.579  & 10.499   & 9.655    & 8.561   \\
MAE    & 11.905   & 8.822   & 4.343   & 8.088   & 79.458   & 44.162  & 27.835  & 12.840  & 9.459    & 8.552    & 7.311   \\ \hline
NCM811 & $0^-$ & $L^-/3$ & $2L^-/3$ & $L^-$ & $0^+$ & $L^+/3$ & $2L^+/3$ & $L^+$ & $0^{\mathrm{sep}}$ & $L^{\mathrm{sep}}/2$ & $L^{\mathrm{sep}}$       \\
RMSE   & 15.169   & 11.217  & 4.405   & 8.476   & 82.993   & 37.941  & 23.754  & 11.111  & 9.758    & 8.896    & 7.849   \\
MAE    & 12.568   & 9.400   & 4.039   & 7.463   & 79.656   & 35.973  & 22.170  & 9.511   & 8.782    & 7.854    & 6.674   \\
\bottomrule
\end{tabular}
\end{table*}

\begin{figure}[!htb]
    \centering
\begin{minipage}{.24\textwidth}
        \centering
        \includegraphics[width=\textwidth]{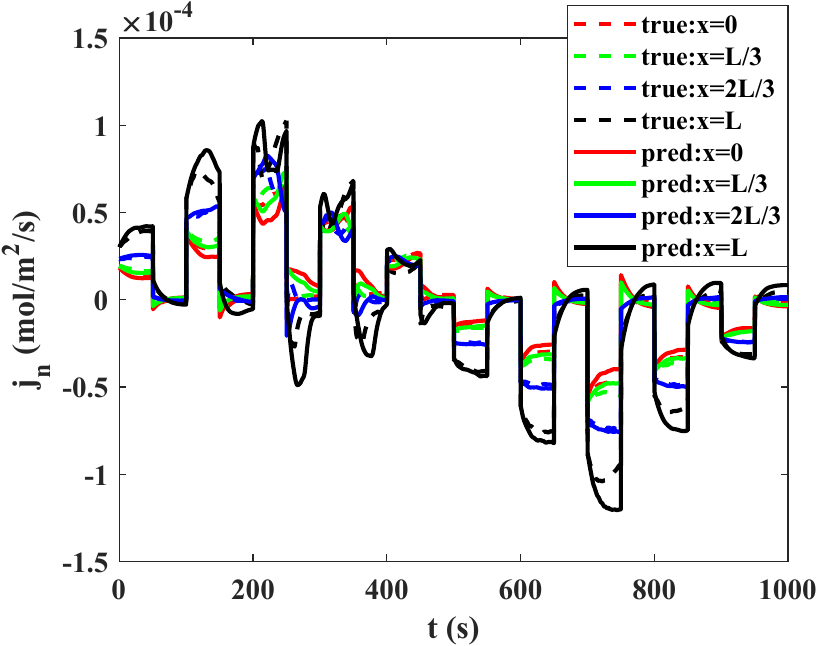}
\subcaption{$j_n$ at points in the negative electrode of NCM523, ACC. }\label{fig:jnACCn}
\end{minipage}%
\begin{minipage}{0.24\textwidth}
        \centering
        \includegraphics[width=\textwidth]{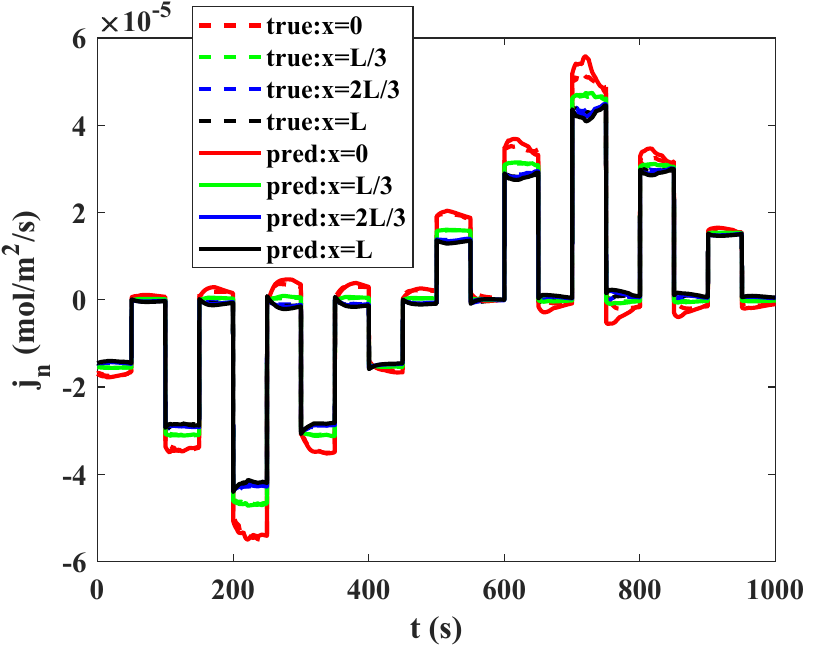}
\subcaption{$j_n$ at points in the positive electrode of NCM523, ACC. }\label{fig:jnACCp}
\end{minipage}
\begin{minipage}{.24\textwidth}
        \centering
        \includegraphics[width=\textwidth]{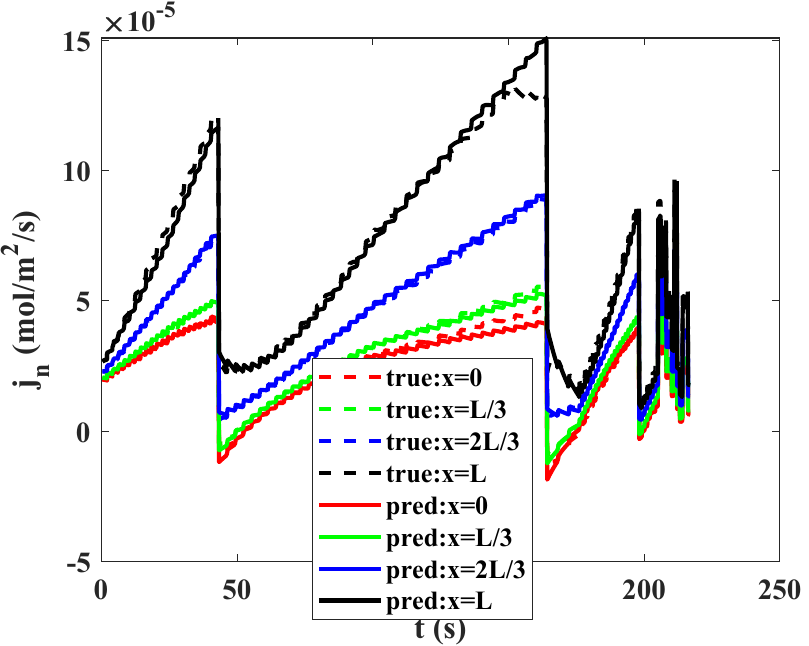}
\subcaption{$j_n$ at points in the negative electrode of NCM523, RC. }\label{fig:jnRCn}
\end{minipage}%
\begin{minipage}{0.24\textwidth}
        \centering
        \includegraphics[width=\textwidth]{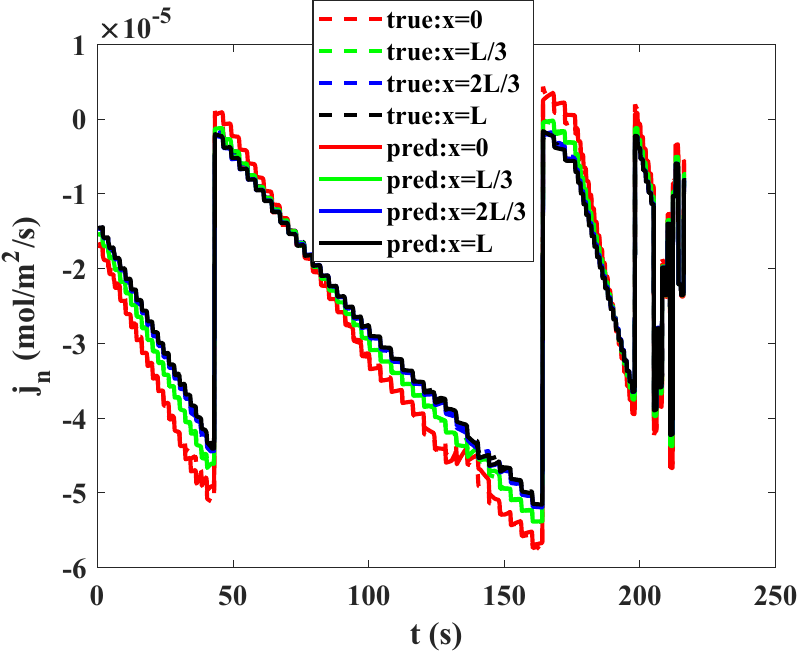}
\subcaption{$j_n$ at points in the positive electrode of NCM523, RC. }\label{fig:jnRCp}
\end{minipage}
\begin{minipage}{.24\textwidth}
        \centering
        \includegraphics[width=\textwidth]{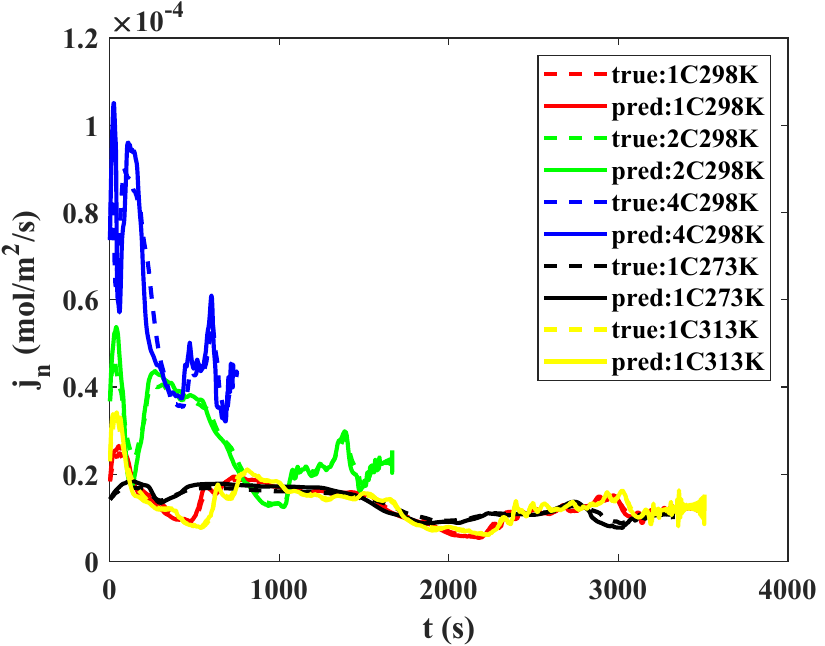}
\subcaption{\textcolor{blue}{$j_n$ at the negative electrode boundary of LFPO, galvanostatic. }}\label{fig:jnCC_LFPO}
\end{minipage}%
\begin{minipage}{0.24\textwidth}
        \centering
        \includegraphics[width=\textwidth]{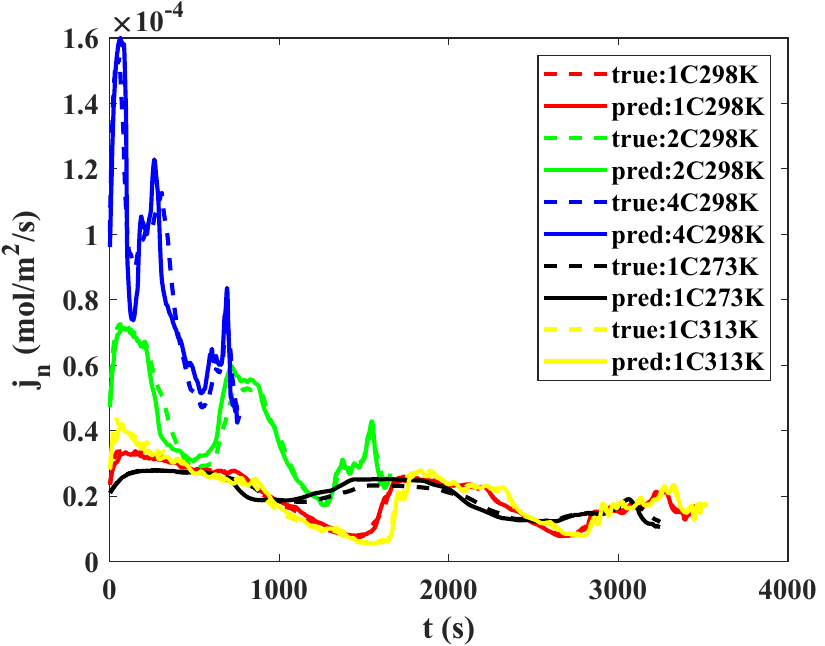}
\subcaption{\textcolor{blue}{$j_n$ at the negative electrode boundary of NCM523, galvanostatic. }}\label{fig:jnCC_NCM523}
\end{minipage}
\begin{minipage}{.24\textwidth}
        \centering
        \includegraphics[width=\textwidth]{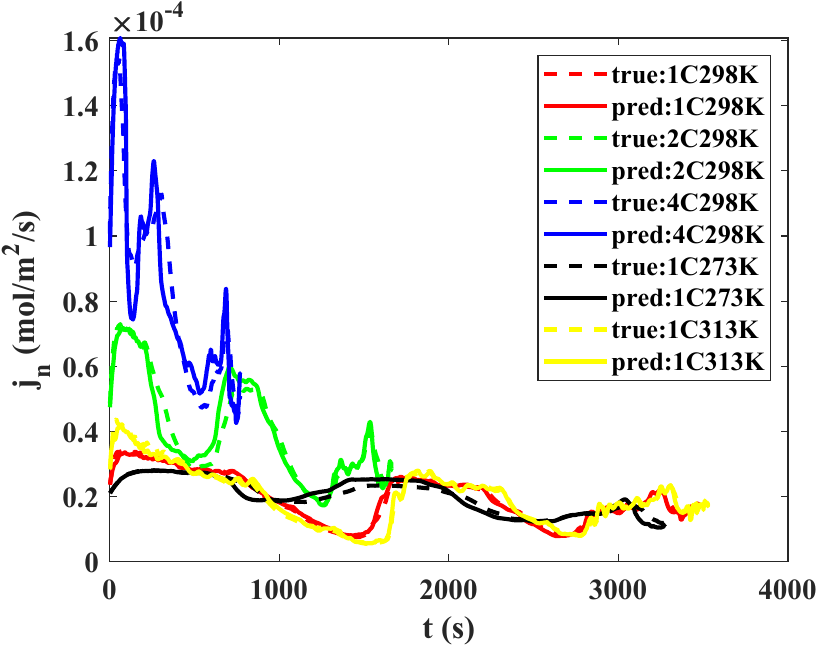}
\subcaption{\textcolor{blue}{$j_n$ at the negative electrode boundary of NCM811, galvanostatic. }}\label{fig:jnCC_NCM811}
\end{minipage}%
\begin{minipage}{0.26\textwidth}
        \centering
        \includegraphics[width=\textwidth]{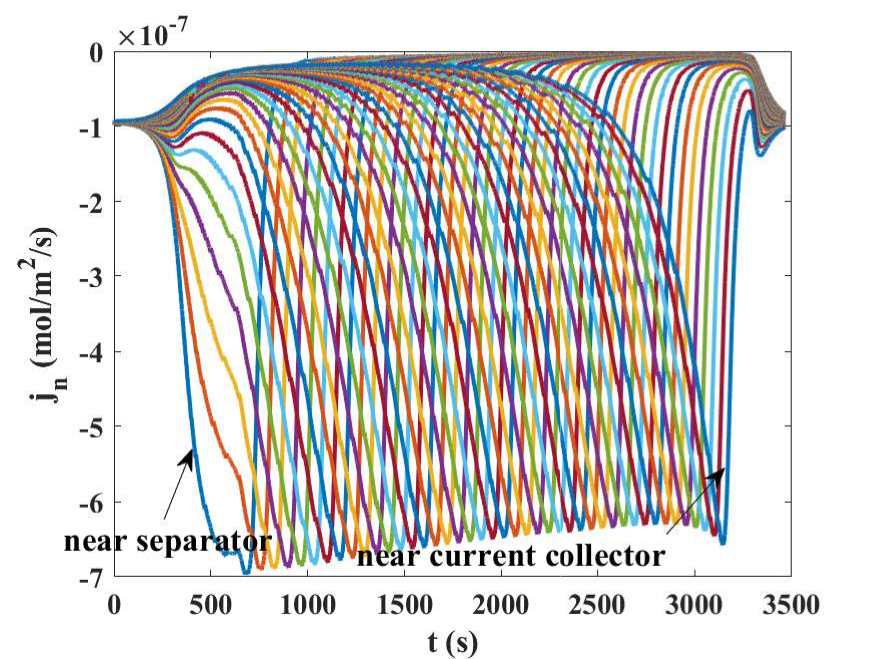}
\subcaption{\textcolor{blue}{$j_n$ at points along the thickness of the positive electrode in LFPO, galvanostatic. }}\label{fig:jnCC_LFPO_p}
\end{minipage}
\caption{\textcolor{blue}{Trajectories of $j_n$ along the thickness of positive and negative electrodes in LFPO, NCM523 and NCM811 cells under different protocols. }}
\end{figure}

The prediction accuracy of $j_n$ is shown in Table~\ref{table:jn_accuracy}. \textcolor{blue}{Note that $j_n$ is the most important state inside the battery. This is not only because it couples the chemical system and electrical system and determines the trajectories of $\bar{c}_s$, $c_{ss}$, $\Phi_s$, $\Phi_e$, etc., on the short time-scale but also because it reflects the degradation pressure that affects the battery status over long time scales. The most important contribution of this work is that we find a simple way to approximate $j_n$, avoiding vast computational costs. We compare the proposed model with an advanced ESP \cite{han_simplification_2015_1,han_simplification_2015_2} that also considers the spatial distribution of $j_n$ and a classic ESP that assumes a uniform distribution of $j_n$. The table shows that at different points along the thickness direction, our model performs better than the advanced ESP. Moreover, the low accuracy of the classic ESP proves the necessity of considering a non-uniform distribution. To clearly demonstrate the distribution characteristics of $j_n$, we plot $j_n$ along the thickness direction in the positive and negative electrodes of the NCM523 cell under dynamic current protocols (scenario nos. 5-6), as shown in Figs.~\ref{fig:jnACCn}-\ref{fig:jnRCp}. Generally, under dynamic currents, the model can accurately predict $j_n$ along the thickness. However, under a very large current, the prediction error of $j_n$ at the interface between the negative electrode and separator is larger, indicating that we should be cautious when applying the model to estimate $j_n$ at $L^-$ under extreme currents. Since the reaction at $L^-$ is the most violent compared with other locations, accurate monitoring of $j_n$ at $L^-$ is more meaningful for analysing the degradation inside the battery. Thus, we plot the results of estimating $j_n(L^-,t)$ under galvanostatic protocols for LFPO, NCM523, and NCM811 cells in Figs.~\ref{fig:jnCC_LFPO}-\ref{fig:jnCC_NCM811}. As the ambient temperatures vary from 273 K to 313 K, the current rates vary from 1C-rate to 4C-rate, and the proposed model can always give accurate results, proving its effectiveness and importance. According to existing research~\cite{han_review_2019}, the degradation of LIBs mainly occurs in graphite-containing negative electrodes; thus, we mainly focus on $j_n$ in the negative electrode above. Now turning to the positive electrode, the estimation of $j_n$ for NCM523 and NCM811 cells is evaluated in Table~\ref{table:jn_accuracy}, and its accuracy is even higher than that of the graphite electrode. Note that the \ce{LiFePO_4} electrode is not simulated as other electrodes. This is because the particle radius of \ce{LiFePO_4} is approximately 100 times smaller than those of NCM and graphite, which makes the peak of the reaction rates across the electrode very narrow and high, as shown in Fig.~\ref{fig:jnCC_LFPO_p}. Since we select only 4 points along the x-axis for simplicity, such granularity is inapplicable for capturing the extremely uneven distribution characteristics of $j_n$ for the \ce{LiFePO_4} electrode, and the same is true for the advanced ESP~\cite{han_simplification_2015_1,han_simplification_2015_2}. However, we can observe from Fig.~\ref{fig:jnCC_LFPO_p} that within a full-cycle operation, the peak of $j_n$ moves steadily from the separator to the current collector, and the $\int_{t_0}^{t_l} j_n(x,t)dt$ values for every point on the x-axis are almost the same. When we conduct the degradation analysis, we focus more on the integration of $j_n(x,t)$ within a specific time period than on its instantaneous values. Thus, the uniform distribution of $j_n$ is adopted for the \ce{LiFePO_4} electrode in this work.}

\begin{table*}[width=1.8\linewidth,cols=3,pos=h]
\caption{Prediction accuracy of $j_n$ in electrodes of LFPO, NCM523, NCM811 cells.}\label{table:jn_accuracy}
\resizebox{\textwidth}{!}{\begin{tabular}{ccccccccccccc}
\toprule
& \multicolumn{4}{c}{R$^2$}                                            & \multicolumn{4}{c}{RMSE($\times10^{-7}$)}                       & \multicolumn{4}{c}{MAE($\times10^{-7}$)}                        \\ \hline
Graphite(LFPO) & $0^-$ & $L^-/3$ & $2L^-/3$ & $L^-$ & $0^-$ & $L^-/3$ & $2L^-/3$ & $L^-$ & $0^-$ & $L^-/3$ & $2L^-/3$ & $L^-$  \\
Proposed         & \textbf{0.920} & \textbf{0.932} & \textbf{0.877} & \textbf{0.940} & \textbf{23.0} & \textbf{16.2} & \textbf{15.5} & \textbf{38.4} & \textbf{14.9} & \textbf{10.3} & \textbf{10.5} & \textbf{25.1} \\
Advanced ESP{\cite{han_simplification_2015_1,han_simplification_2015_2}}     & 0.712          & 0.576          & 0.128          & 0.805          & 41.5          & 32.9          & 34.1          & 63.8          & 23.5          & 15.7       & 15.8          & 36.8          \\
Classic ESP          & 0.191          & 0.255          & 0.349          & 0.181          & 58.3          & 42.4          & 26.9          & 116.4         & 47.5          & 34.2          & 22.1          & 93.4          \\ \hline
Graphite (NCM523) & $0^-$ & $L^-/3$ & $2L^-/3$ & $L^-$ & $0^-$ & $L^-/3$ & $2L^-/3$ & $L^-$ & $0^-$ & $L^-/3$ & $2L^-/3$ & $L^-$             \\
Proposed         & \textbf{0.914} & \textbf{0.931} & \textbf{0.866} & \textbf{0.950} & \textbf{29.2} & \textbf{19.2} & \textbf{22.7} & \textbf{42.4} & \textbf{20.2} & \textbf{13.2} & \textbf{15.4} & \textbf{31.2} \\
Advanced ESP{\cite{han_simplification_2015_1,han_simplification_2015_2}}         & 0.890          & 0.920          & 0.831          & 0.938          & 35.0          & 22.0          & 26.5          & 47.2          & 25.9          & 15.6          & 19.2          & 33.1          \\
Classic ESP          & 0.177          & 0.271          & 0.342          & 0.179          & 80.1          & 56.4          & 42.7          & 147.3         & 66.6          & 47.0          & 36.5          & 123.7         \\ \hline
Graphite (NCM811) & $0^-$ & $L^-/3$ & $2L^-/3$ & $L^-$ & $0^-$ & $L^-/3$ & $2L^-/3$ & $L^-$ & $0^-$ & $L^-/3$ & $2L^-/3$ & $L^-$             \\
Proposed         & \textbf{0.900} & \textbf{0.921} & \textbf{0.844} & \textbf{0.947} & \textbf{31.0} & \textbf{20.2} & \textbf{23.9} & \textbf{43.1} & \textbf{21.8} & \textbf{14.0} & \textbf{16.4} & \textbf{31.8} \\
Advanced ESP{\cite{han_simplification_2015_1,han_simplification_2015_2}}         & 0.871          & 0.904          & 0.799          & 0.932          & 37.2          & 23.3          & 27.9          & 48.7          & 27.8          & 16.7          & 20.4          & 34.5          \\
Classic ESP          & 0.179          & 0.273          & 0.343          & 0.181          & 80.3          & 56.5          & 42.7          & 147.5         & 66.8          & 47.1          & 36.6          & 124.0         \\ \hline
NCM523  & $0^+$ & $L^+/3$ & $2L^+/3$ & $L^+$ & $0^+$ & $L^+/3$ & $2L^+/3$ & $L^+$ & $0^+$ & $L^+/3$ & $2L^+/3$ & $L^+$             \\
Proposed         & \textbf{0.940} & \textbf{0.772} & \textbf{0.916} & \textbf{0.922} & \textbf{5.70} & \textbf{1.99} & \textbf{2.38} & \textbf{3.00} & \textbf{3.91} & \textbf{1.22} & \textbf{1.55} & \textbf{1.90} \\
Advanced ESP{\cite{han_simplification_2015_1,han_simplification_2015_2}}         & 0.938          & 0.757          & 0.915          & 0.916          & 6.01          & 2.10          & 2.46          & 3.20          & 4.10          & 1.34          & 1.63          & 2.08          \\
Classic ESP          & 0.346          & 0.361          & 0.357          & 0.355          & 19.82         & 3.74          & 6.90          & 9.29          & 15.8          & 2.95          & 5.49          & 7.36          \\ \hline
NCM811  & $0^+$ & $L^+/3$ & $2L^+/3$ & $L^+$ & $0^+$ & $L^+/3$ & $2L^+/3$ & $L^+$ & $0^+$ & $L^+/3$ & $2L^+/3$ & $L^+$              \\
Proposed         & \textbf{0.936} & \textbf{0.684} & \textbf{0.916} & \textbf{0.926} & \textbf{4.44} & \textbf{1.64} & \textbf{1.86} & \textbf{2.23} & \textbf{2.77} & \textbf{0.87} & \textbf{1.08} & \textbf{1.36} \\
Advanced ESP{\cite{han_simplification_2015_1,han_simplification_2015_2}}         & 0.928          & 0.673          & 0.908          & 0.917          & 4.85          & 1.75          & 2.00          & 2.45          & 3.05          & 0.97          & 1.18          & 1.52          \\
Classic ESP          & 0.360          & 0.367          & 0.365          & 0.364          & 15.97         & 2.82          & 5.54          & 7.39          & 11.17         & 1.94          & 3.86          & 5.11          \\ 
\toprule
\end{tabular}}
\end{table*}

\begin{figure}[!htb]
    \centering
\begin{minipage}{.24\textwidth}
        \centering
        \includegraphics[width=\textwidth]{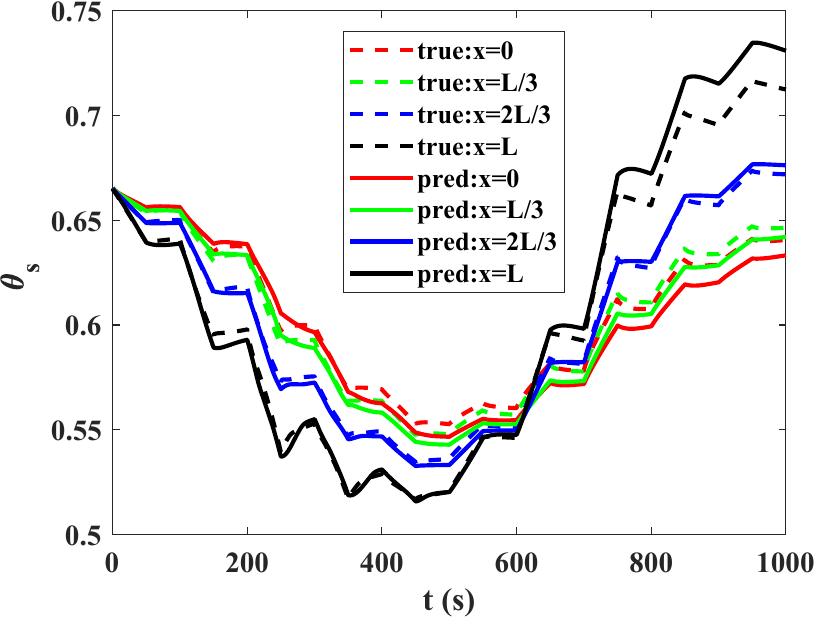}
\subcaption{\textcolor{blue}{$\theta_s$ at points in the negative electrode of NCM811, ACC. }}\label{fig:xsACCn}
\end{minipage}%
\begin{minipage}{0.24\textwidth}
        \centering
        \includegraphics[width=\textwidth]{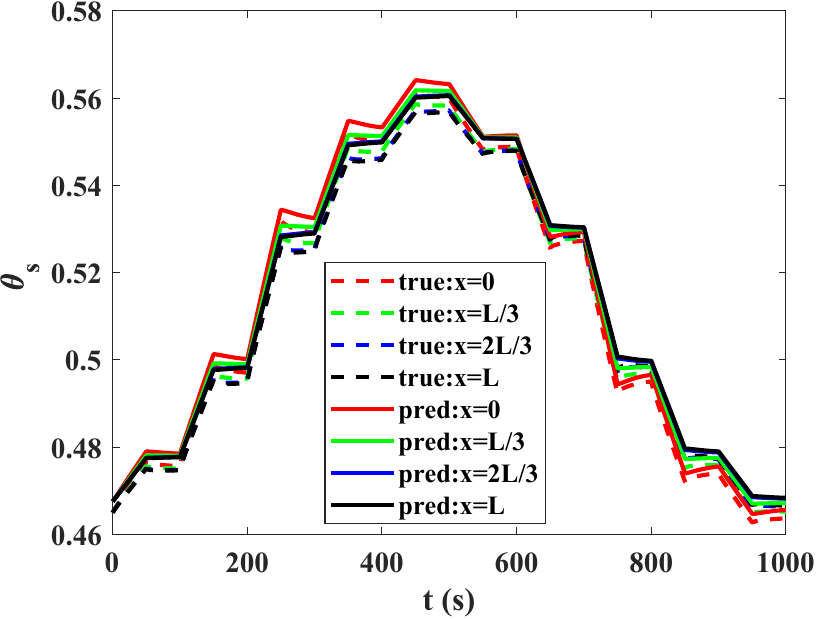}
\subcaption{\textcolor{blue}{$\theta_s$ at points in the positive electrode of NCM811, ACC. }}\label{fig:xsACCp}
\end{minipage}
\begin{minipage}{.24\textwidth}
        \centering
        \includegraphics[width=\textwidth]{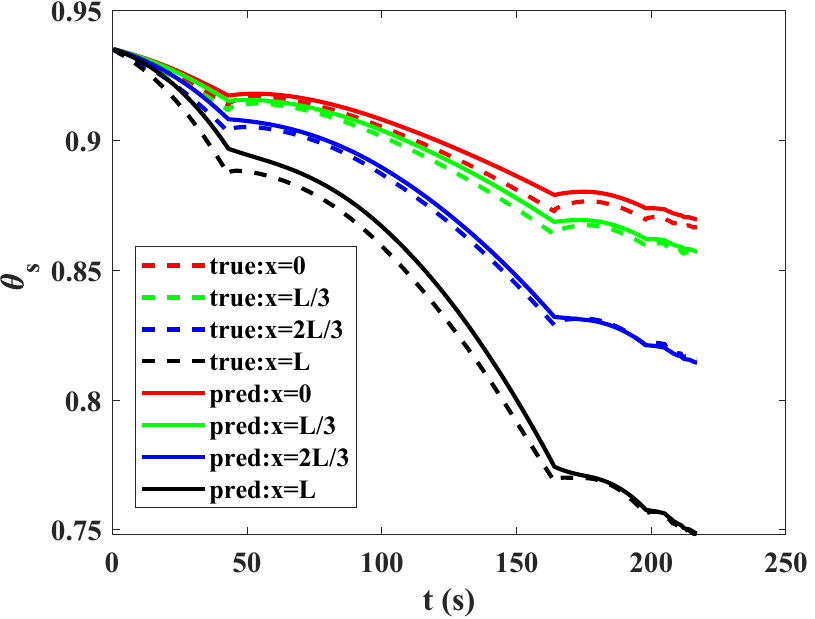}
\subcaption{\textcolor{blue}{$\theta_s$ at points in the negative electrode of NCM523, RC. }}\label{fig:xsRCn}
\end{minipage}%
\begin{minipage}{0.24\textwidth}
        \centering
        \includegraphics[width=\textwidth]{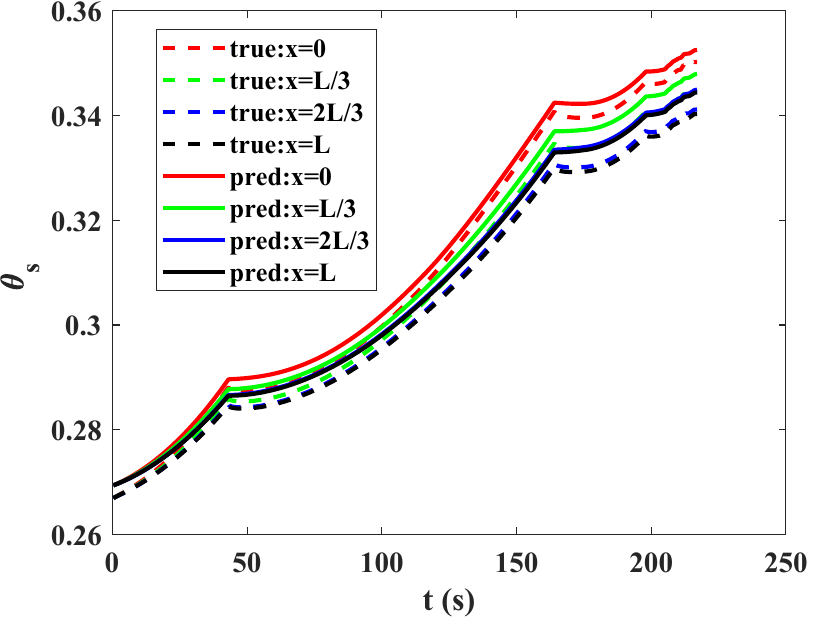}
\subcaption{\textcolor{blue}{$\theta_s$ at points in the positive electrode of NCM523, RC. }}\label{fig:xsRCp}
\end{minipage}
\begin{minipage}{.24\textwidth}
        \centering
        \includegraphics[width=\textwidth]{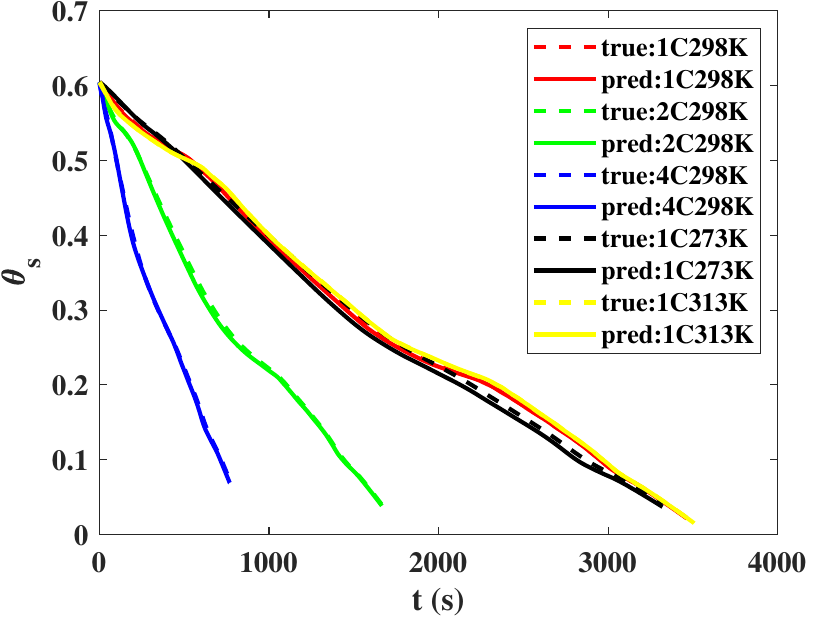}
\subcaption{\textcolor{blue}{$\theta_s$ at the negative electrode boundary of LFPO, galvanostatic. }}\label{fig:xsCC_LFPO}
\end{minipage}%
\begin{minipage}{0.24\textwidth}
        \centering
        \includegraphics[width=\textwidth]{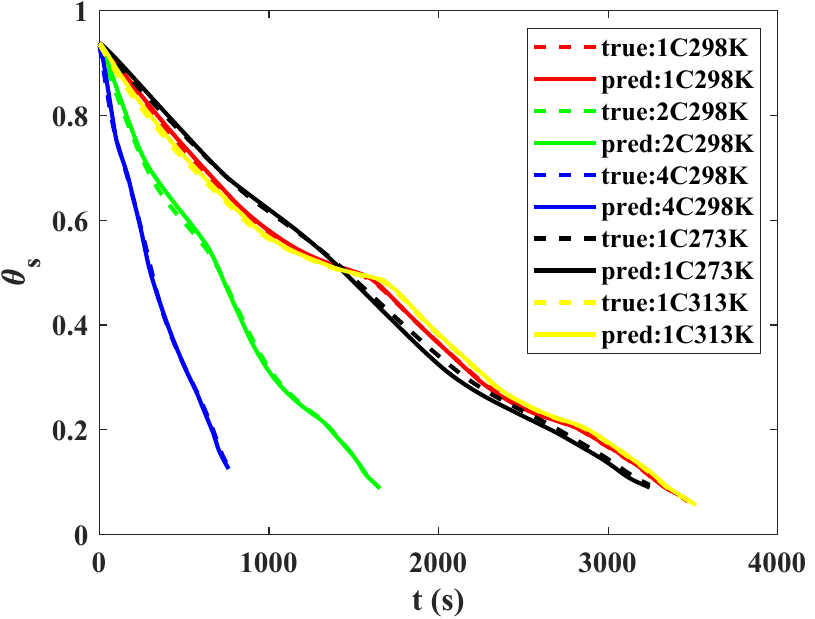}
\subcaption{\textcolor{blue}{$\theta_s$ at the negative electrode boundary of NCM523, galvanostatic. }}\label{fig:xsCC_NCM523}
\end{minipage}
\begin{minipage}{.24\textwidth}
        \centering
        \includegraphics[width=\textwidth]{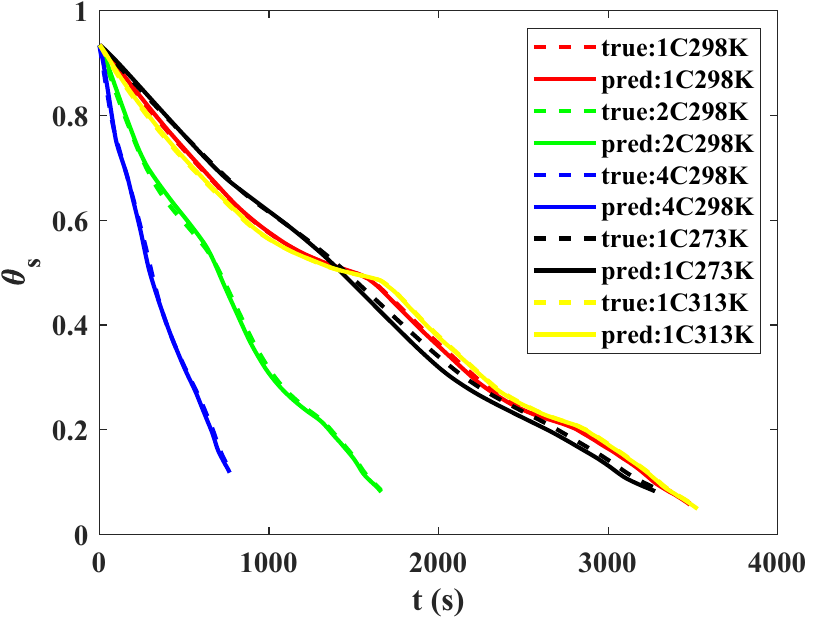}
\subcaption{\textcolor{blue}{$\theta_s$ at the negative electrode boundary of NCM811, galvanostatic. }}\label{fig:xsCC_NCM811}
\end{minipage}%
\begin{minipage}{0.24\textwidth}
        \centering
        \includegraphics[width=\textwidth]{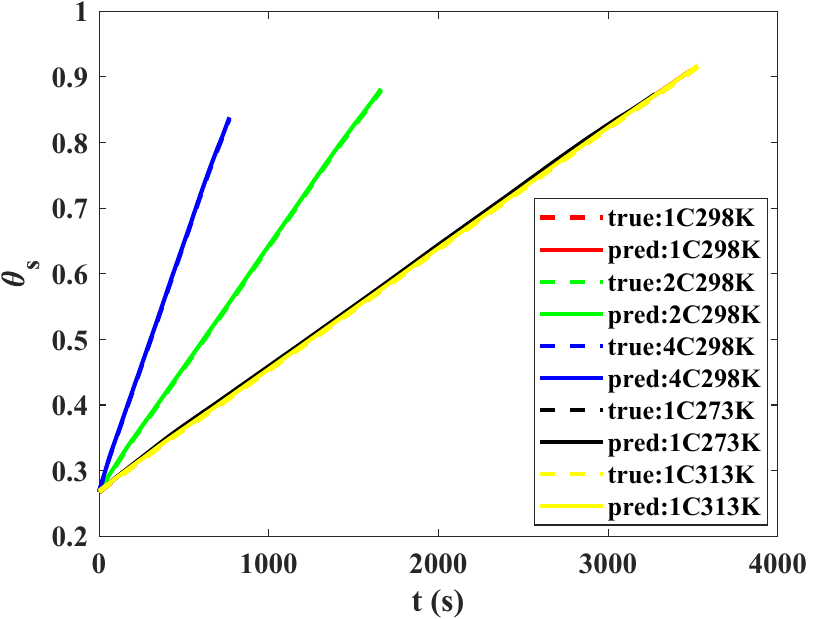}
\subcaption{\textcolor{blue}{$\theta_s$ at the positive electrode boundary of NCM811, galvanostatic. }}\label{fig:xsCC_NCM811_p}
\end{minipage}
\caption{\textcolor{blue}{Trajectories of $\theta_s$ along the thickness of positive and negative electrodes in LFPO, NCM523 and NCM811 cells under different protocols. }}
\end{figure}

The prediction accuracy of $\bar{c}_s$ is shown in Table~\ref{table:xs_accuracy}. For ease of comparison, it is replaced by the normalized value $\theta_s$, i.e., the average stoichiometry in the solid phase. Generally, our model performs better at more points along the thickness direction for the three types of cells. \textcolor{blue}{However, we also notice that at some points, e.g., at $\frac{L^+}{3}$ of the NCM523 positive electrode, $\frac{2L^-}{3}$ of the NCM523 negative electrode and $L^-$ of the NCM811 negative electrode, the accuracy of our model is slightly lower than that of the advanced ESP or classic ESP. There might be two reasons for this phenomenon. First, only one first-order inertial process is used to approximate the diffusion process of \ce{Li^+} in the solid phase for simplicity, while two independent first-order inertial processes are used in the advanced ESP model. Second, as mentioned in the text above, to reduce the potential of being trapped in the oscillations, we made a trade-off between accuracy and stabilization when determining the fitted coefficient $k_s$. Actually, before finally setting $k_s=1/9$ for the LFPO electrode and $k_s=1/28$ for other electrodes, we tried $k_s$ derived from the xRA method, Pad\'{e} approximations and frequency response optimization. Although these methods can generally perform better under dynamic protocols (e.g., ACC/RC), they were found to be trapped in oscillations in galvanostatic protocols during very low or high SOCs. Thus, to ensure the applicability of the proposed model when the battery is cycled in the full range of SOC, we sacrifice the accuracy to some extent and selected the proper $k_s$ to achieve better stabilization. Moreover, considering the high requirements of simplicity in real-world applications, we retain only one first-order inertial process to avoid unnecessary model complexity. The results support our choice because although our model did not perform best everywhere, its absolute accuracy is acceptable. To clearly demonstrate the performance of the model at predicting $\theta_s$, its trajectories along the thickness direction of positive and negative electrodes in three types of cells are plotted in Figs.~\ref{fig:xsACCn}-\ref{fig:xsCC_NCM811_p}. Figs.~\ref{fig:xsACCn}-\ref{fig:xsRCp} show that under dynamic current protocols, the accuracy of $\theta_s$ at points in the middle of the electrode is higher than that at the boundaries for the negative electrode. For the positive electrode, the difference between points at different locations is not very prominent. This indicates that when estimating the SOC of the battery under dynamic loads based on the information of $\theta_s$, it would be better to select points in the middle of the electrode. Since we have observed that the error of $\theta_s$ at the interface between the electrode and separator is higher, we plot their trajectories under galvanostatic protocols in Figs.~\ref{fig:xsCC_LFPO}-\ref{fig:xsCC_NCM811_p}. Under galvanostatic current, when the ambient temperatures vary from 273 K to 313 K, the current rates vary from 1C to 4C, and the model can always provide accurate results in both negative and positive electrodes of all three types of cells.}

\begin{table}[width=2\linewidth,cols=2,pos=h]
\caption{Prediction accuracy of $\theta_s$ in electrodes of LFPO, NCM523, and NCM811 cells.}\label{table:xs_accuracy}
\resizebox{\textwidth}{!}{\begin{tabular}{ccccccccccccc}
\toprule
& \multicolumn{4}{c}{R$^2$} & \multicolumn{4}{c}{RMSE($\times10^{-4}$)}  & \multicolumn{4}{c}{MAE($\times10^{-4}$)}                            \\ \hline
\makecell{Graphite (LFPO)}  & $0^-$ & $L^-/3$ & $2L^-/3$ & $L^-$ & $0^-$ & $L^-/3$ & $2L^-/3$ & $L^-$ & $0^-$ & $L^-/3$ & $2L^-/3$ & $L^-$ \\
Proposed         & \textbf{0.9952} & \textbf{0.9975} & \textbf{0.9992} & \textbf{0.9973} & \textbf{23.10} & \textbf{17.48} & \textbf{17.24} & \textbf{35.24} & \textbf{18.42} & \textbf{14.23} & \textbf{14.48} & \textbf{27.94} \\
Advanced ESP{\cite{han_simplification_2015_1,han_simplification_2015_2}}         & 0.9905          & 0.9959          & 0.9988          & 0.9954          & 35.50          & 25.52          & 27.72          & 53.07          & 27.61          & 19.22          & 18.58          & 39.82          \\
Classic ESP          & 0.9086          & 0.9553          & 0.9915          & 0.8503          & 212.5         & 146.9         & 77.12          & 381.2         & 183.8         & 126.1         & 66.88          & 330.7         \\ \hline
\makecell{Graphite (NCM523)} & $0^-$ & $L^-/3$ & $2L^-/3$ & $L^-$ & $0^-$ & $L^-/3$ & $2L^-/3$ & $L^-$ & $0^-$ & $L^-/3$ & $2L^-/3$ & $L^-$           \\
Proposed         & \textbf{0.9879} & \textbf{0.9936} & 0.9963          & 0.9901          & \textbf{50.90} & \textbf{38.90} & \textbf{41.37} & 85.63          & \textbf{43.23} & \textbf{33.53} & 34.82          & 68.37          \\
Advanced ESP{\cite{han_simplification_2015_1,han_simplification_2015_2}}         & 0.9712          & 0.9874          & \textbf{0.9969} & \textbf{0.9923} & 59.72          & 43.93          & 41.96          & \textbf{71.77} & 49.48          & 37.52          & \textbf{34.49} & \textbf{55.55} \\
Classic ESP          & 0.7948          & 0.9256          & 0.9752          & 0.7514          & 396.7         & 262.6         & 166.8         & 693.2         & 354.4        & 232.4         & 153.3         & 632.4         \\ \hline
\makecell{Graphite (NCM811)} & $0^-$ & $L^-/3$ & $2L^-/3$ & $L^-$ & $0^-$ & $L^-/3$ & $2L^-/3$ & $L^-$ & $0^-$ & $L^-/3$ & $2L^-/3$ & $L^-$            \\
Proposed         & \textbf{0.9913} & \textbf{0.9957} & \textbf{0.9990} & 0.9944          & \textbf{67.88} & 53.19          & \textbf{36.34} & \textbf{64.34} & \textbf{57.67} & 45.76          & \textbf{31.17} & \textbf{51.25} \\
Advanced ESP{\cite{han_simplification_2015_1,han_simplification_2015_2}}         & 0.9813          & 0.9929          & 0.9986          & \textbf{0.9947} & 69.15          & \textbf{51.73} & 43.50          & 66.10          & 58.65          & \textbf{44.90} & 36.59          & 51.43          \\
Classic ESP          & 0.7485          & 0.9026          & 0.9844          & 0.7719          & 417.3         & 282.5         & 144.5         & 671.3         & 375.7        & 253.2         & 129.0         & 607.1         \\ \hline
NCM523           & $0^+$ & $L^+/3$ & $2L^+/3$ & $L^+$ & $0^+$ & $L^+/3$ & $2L^+/3$ & $L^+$ & $0^+$ & $L^+/3$ & $2L^+/3$ & $L^+$            \\
Proposed         & 0.9978          & 0.9974          & \textbf{0.9972} & \textbf{0.9971} & 26.78          & 25.98          & \textbf{24.73} & \textbf{25.40} & 26.24          & 25.64          & \textbf{24.38} & \textbf{24.96} \\
Advanced ESP{\cite{han_simplification_2015_1,han_simplification_2015_2}}         & \textbf{0.9983} & 0.9974          & 0.9971          & 0.9969          & \textbf{21.13} & 25.82          & 25.96          & 26.64          & \textbf{20.83} & 25.50          & 25.58          & 26.17          \\
Classic ESP          & 0.9957          & \textbf{0.9983} & 0.9935          & 0.9914          & 67.24          & \textbf{17.48} & 49.42          & 59.67          & 57.40          & \textbf{16.67} & 46.64          & 55.60          \\ \hline
NCM811          & $0^+$ & $L^+/3$ & $2L^+/3$ & $L^+$ & $0^+$ & $L^+/3$ & $2L^+/3$ & $L^+$ & $0^+$ & $L^+/3$ & $2L^+/3$ & $L^+$              \\
Proposed         & 0.9979          & 0.9972          & \textbf{0.9968} & \textbf{0.9964} & 27.57          & 29.12          & \textbf{30.07} & \textbf{31.48} & 27.32          & 28.88          & \textbf{29.75} & \textbf{31.09} \\
Advanced ESP{\cite{han_simplification_2015_1,han_simplification_2015_2}}         & \textbf{0.9982} & 0.9972          & 0.9967          & 0.9963          & \textbf{24.56} & 29.01          & 30.71          & 32.12          & \textbf{24.35} & 28.77          & 30.37          & 31.71          \\
Classic ESP          & 0.9969          & \textbf{0.9982} & 0.9931          & 0.9911          & 38.65          & \textbf{22.31} & 44.94          & 50.70          & 35.79          & \textbf{22.09} & 43.72          & 48.91          \\ 
\toprule
\end{tabular}}
\end{table}

\begin{figure}[!htb]
    \centering
\begin{minipage}{.24\textwidth}
        \centering
        \includegraphics[width=\textwidth]{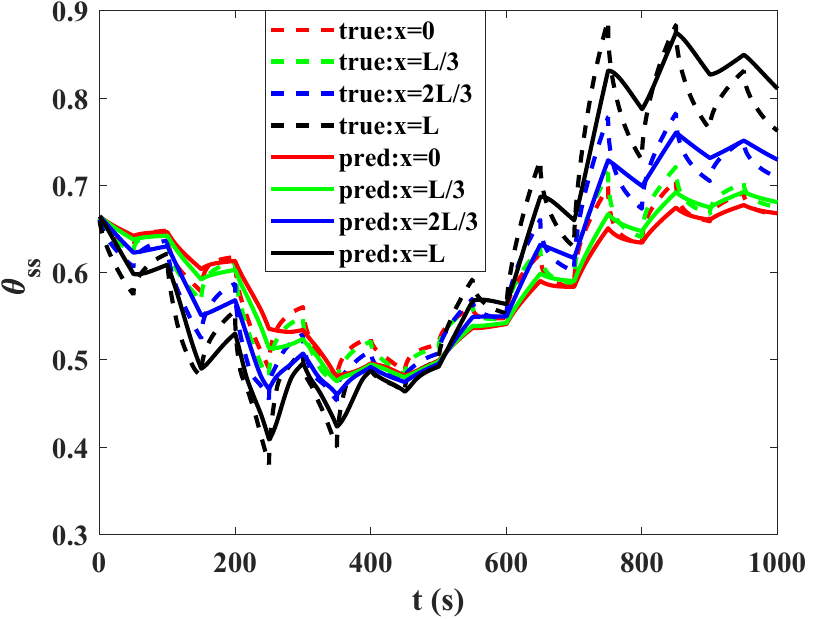}
\subcaption{\textcolor{blue}{$\theta_{ss}$ at points in the negative electrode of NCM811, ACC. }}\label{fig:xssACCn}
\end{minipage}%
\begin{minipage}{0.24\textwidth}
        \centering
        \includegraphics[width=\textwidth]{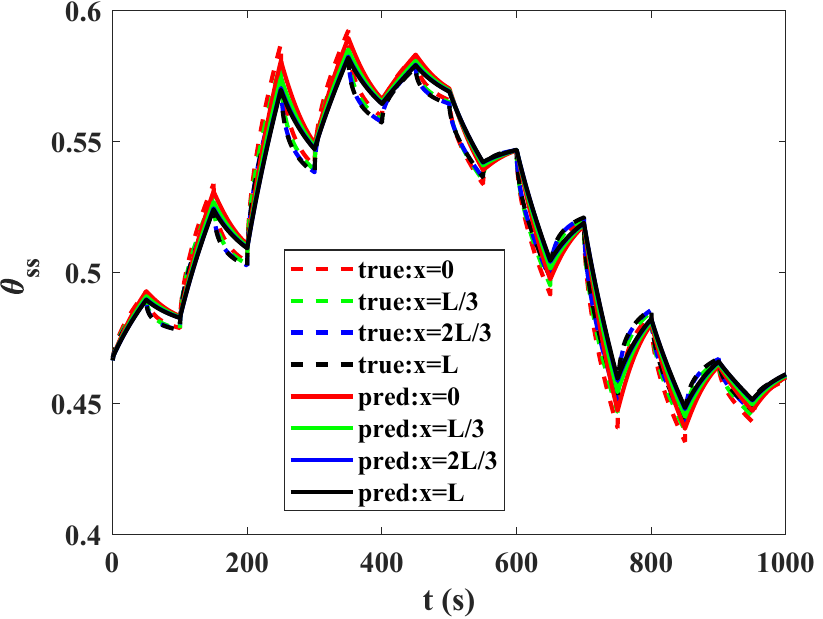}
\subcaption{\textcolor{blue}{$\theta_{ss}$ at points in the positive electrode of NCM811, ACC. }}\label{fig:xssACCp}
\end{minipage}
\begin{minipage}{.24\textwidth}
        \centering
        \includegraphics[width=\textwidth]{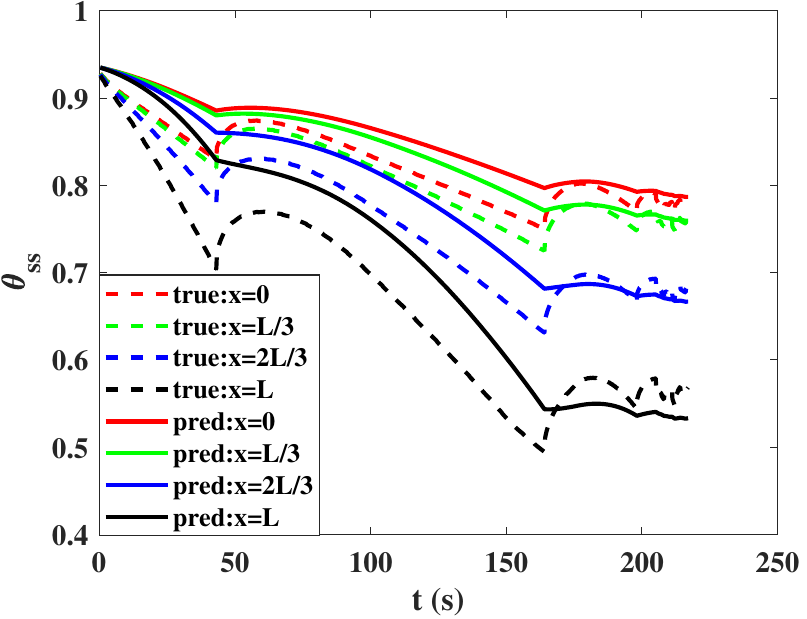}
\subcaption{\textcolor{blue}{$\theta_{ss}$ at points in the negative electrode of NCM523, RC. }}\label{fig:xssRCn}
\end{minipage}%
\begin{minipage}{0.24\textwidth}
        \centering
        \includegraphics[width=\textwidth]{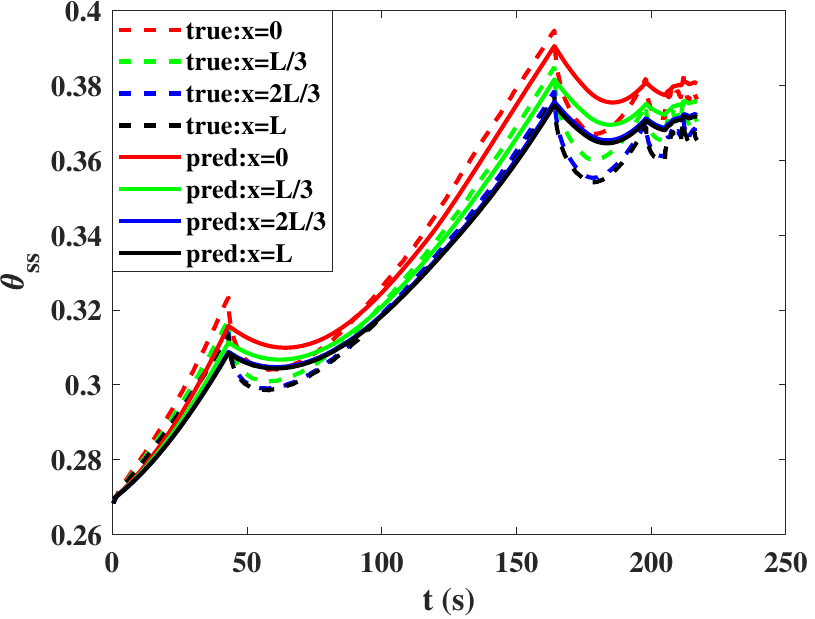}
\subcaption{\textcolor{blue}{$\theta_{ss}$ at points in the positive electrode of NCM523, RC. }}\label{fig:xssRCp}
\end{minipage}
\begin{minipage}{.24\textwidth}
        \centering
        \includegraphics[width=\textwidth]{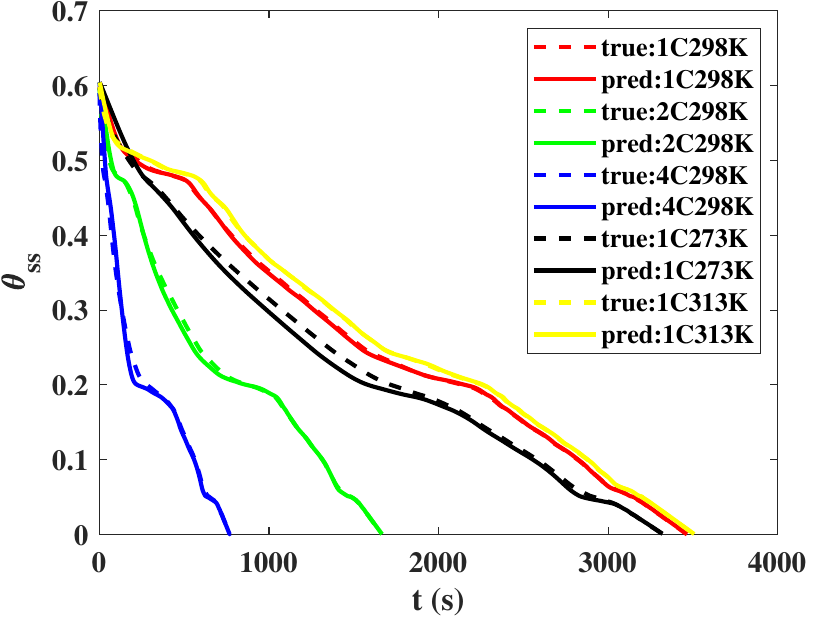}
\subcaption{\textcolor{blue}{$\theta_{ss}$ at the negative electrode boundary of LFPO, galvanostatic. }}\label{fig:xssCC_LFPO}
\end{minipage}%
\begin{minipage}{0.24\textwidth}
        \centering
        \includegraphics[width=\textwidth]{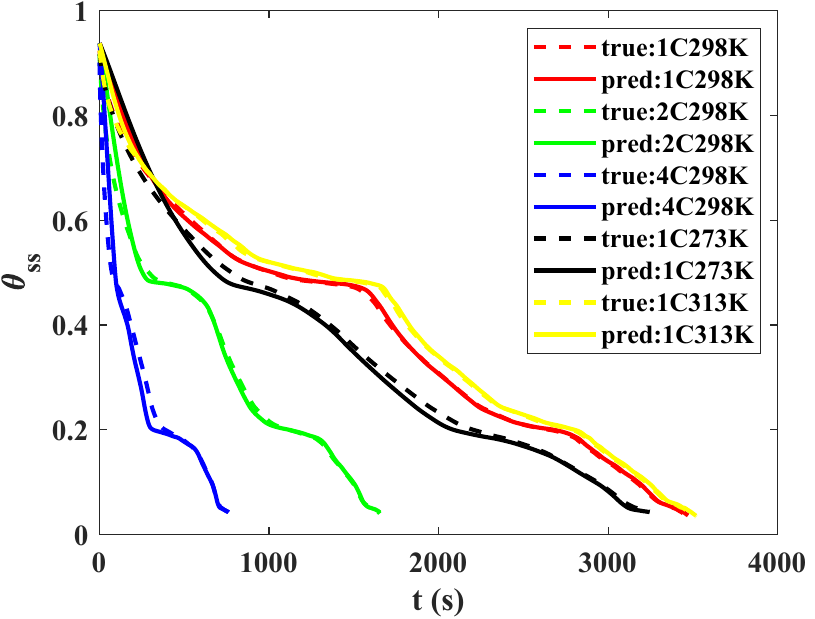}
\subcaption{\textcolor{blue}{$\theta_{ss}$ at the negative electrode boundary of NCM523, galvanostatic. }}\label{fig:xssCC_NCM523}
\end{minipage}
\begin{minipage}{.24\textwidth}
        \centering
        \includegraphics[width=\textwidth]{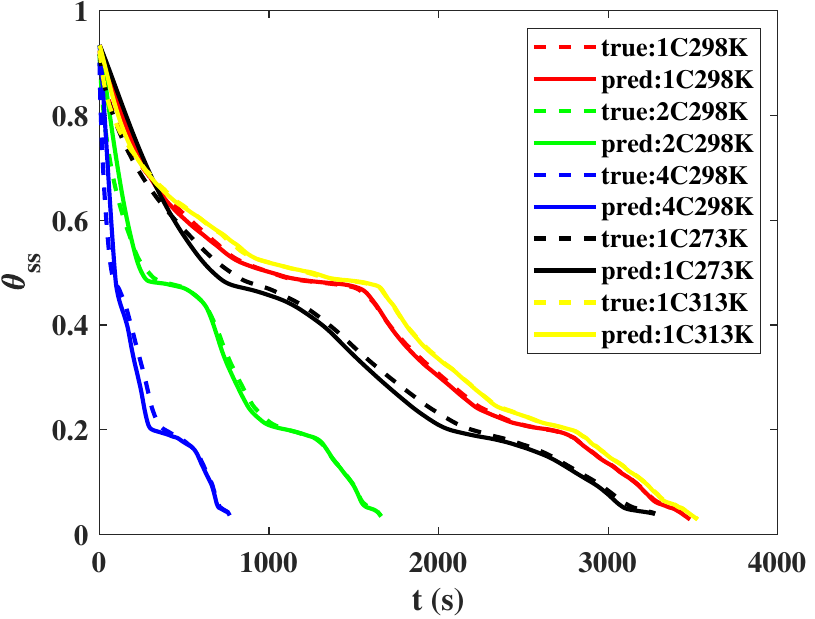}
\subcaption{\textcolor{blue}{$\theta_{ss}$ at the negative electrode boundary of NCM811, galvanostatic. }}\label{fig:xssCC_NCM811}
\end{minipage}%
\begin{minipage}{0.24\textwidth}
        \centering
        \includegraphics[width=\textwidth]{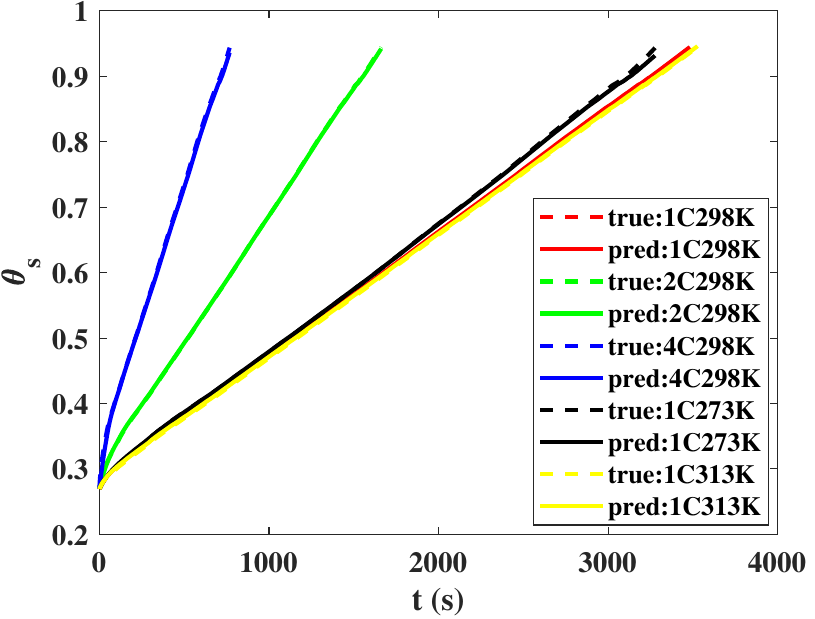}
\subcaption{\textcolor{blue}{$\theta_{ss}$ at the positive electrode boundary of NCM811, galvanostatic. }}\label{fig:xssCC_NCM811_p}
\end{minipage}
\caption{\textcolor{blue}{Trajectories of $\theta_{ss}$ along the thickness of positive and negative electrodes in LFPO, NCM523 and NCM811 cells under different protocols. }}
\end{figure}

The prediction accuracy of $c_{ss}$ is shown in Table~\ref{table:xss_accuracy}. Similarly, the surface stoichiometry $\theta_{ss}$ is used to represent $c_{ss}$ for ease of comparison. The proposed model performs better than the other two models at all points along the thickness direction, especially in the negative electrode. \textcolor{blue}{As mentioned above, side reactions such as SEI generation and lithium plating mainly occur in negative electrodes~\cite{han_review_2019}. Thus, a higher accuracy of estimating $\theta_{ss}$ in the negative electrode is pivotal and meaningful for predicting the SOP or conducting degradation analysis of the battery. To clearly demonstrate the performance of the model at predicting $\theta_{ss}$, its trajectories along the thickness direction of positive and negative electrodes in three types of cells are plotted in Figs.~\ref{fig:xssACCn}-\ref{fig:xssCC_NCM811_p}. Figs.~\ref{fig:xssACCn}-\ref{fig:xssRCp} show that under dynamic current protocols, the accuracy of $\theta_{ss}$ in the negative electrode at the side of the current collector ($x=0^-$) is higher than points at the side of the separator ($x=L^-$). This is reasonable since we estimate $j_n$ better at $x=0^-$. For the positive electrode, the difference between points at different locations is not very prominent, which is similar to the case of $\theta_s$. This indicates that when estimating the SOP of the battery under dynamic loads based on the information of $\theta_{ss}$ in the negative electrode, it would be better to select points at the side of the current collector. The trajectories of $\theta_{ss}$ under galvanostatic protocols are plotted in Figs.~\ref{fig:xssCC_LFPO}-\ref{fig:xssCC_NCM811_p}. Under galvanostatic current, when the ambient temperatures vary from 273 K to 313 K, the current rates vary from 1C-rate to 4C-rate, and the model can always give accurate results in both negative and positive electrodes of all three types of cells.}

\begin{table*}[width=2.1\linewidth,cols=3,pos=h]
\caption{Prediction accuracy of $\theta_{ss}$ in electrodes of LFPO, NCM523, and NCM811 cells.}\label{table:xss_accuracy}
\resizebox{\textwidth}{!}{\begin{tabular}{ccccccccccccc}
\toprule
                 & \multicolumn{4}{c}{R$^2$}                                                & \multicolumn{4}{c}{RMSE ($\times10^{-4}$)}                           & \multicolumn{4}{c}{MAE ($\times10^{-4}$)}                            \\ \hline
\makecell{Graphite (LFPO)}   & $0^-$ & $L^-/3$ & $2L^-/3$ & $L^-$ & $0^-$ & $L^-/3$ & $2L^-/3$ & $L^-$ & $0^-$ & $L^-/3$ & $2L^-/3$ & $L^-$             \\
Proposed         & \textbf{0.9688} & \textbf{0.9731} & \textbf{0.9777} & \textbf{0.9761} & \textbf{80.05} & \textbf{75.01} & \textbf{79.75} & \textbf{120.2} & \textbf{55.22} & \textbf{52.41} & \textbf{59.37} & \textbf{91.10} \\
Advanced ESP{\cite{han_simplification_2015_1,han_simplification_2015_2}}         & 0.9520          & 0.9631          & 0.9748          & 0.9733          & 105.2          & 93.79          & 100.6          & 134.4          & 70.83          & 62.06          & 64.69          & 101.9          \\
Classic ESP          & 0.3102          & 0.4429          & 0.7719          & 0.7696          & 420.5          & 354.2          & 263.92         & 518.6          & 338.9          & 280.1          & 212.4          & 421.7          \\ \hline
\makecell{Graphite(NCM523)} & $0^-$ & $L^-/3$ & $2L^-/3$ & $L^-$ & $0^-$ & $L^-/3$ & $2L^-/3$ & $L^-$ & $0^-$ & $L^-/3$ & $2L^-/3$ & $L^-$        \\
Proposed         & \textbf{0.9319} & \textbf{0.9454} & \textbf{0.9587} & \textbf{0.9493} & \textbf{165.4} & \textbf{160.1} & \textbf{177.1} & \textbf{253.8} & \textbf{124.9} & \textbf{118.8} & \textbf{121.5} & \textbf{158.6} \\
Advanced ESP{\cite{han_simplification_2015_1,han_simplification_2015_2}}         & 0.8803          & 0.9168          & 0.9529          & 0.9458          & 197.0          & 183.9          & 187.9          & 263.5          & 152.2          & 137.8          & 126.8          & 171.6          \\
Classic ESP          & -2.1964         & -1.1228         & 0.3186          & 0.5626          & 949.8          & 830.5          & 653.3          & 926.6          & 737.7          & 607.4          & 481.4          & 773.7          \\ \hline
\makecell{Graphite (NCM811)} & $0^-$ & $L^-/3$ & $2L^-/3$ & $L^-$ & $0^-$ & $L^-/3$ & $2L^-/3$ & $L^-$ & $0^-$ & $L^-/3$ & $2L^-/3$ & $L^-$   \\
Proposed         & \textbf{0.9392} & \textbf{0.9510} & \textbf{0.9626} & \textbf{0.9533} & \textbf{170.8} & \textbf{162.7} & \textbf{172.2} & \textbf{240.9} & \textbf{130.6} & \textbf{121.6} & \textbf{120.6} & \textbf{153.7} \\
Advanced ESP{\cite{han_simplification_2015_1,han_simplification_2015_2}}         & 0.8930          & 0.9265          & 0.9576          & 0.9493          & 197.2          & 181.9          & 182.3          & 253.8          & 155.1          & 138.3          & 127.1          & 172.6          \\
Classic ESP          & -2.1919         & -1.1150         & 0.3289          & 0.5756          & 952.4          & 831.5          & 646.5          & 913.6          & 746.5          & 614.5          & 468.6          & 757.7          \\ \hline
NCM523           & $0^+$ & $L^+/3$ & $2L^+/3$ & $L^+$ & $0^+$ & $L^+/3$ & $2L^+/3$ & $L^+$ & $0^+$ & $L^+/3$ & $2L^+/3$ & $L^+$         \\
Proposed         & \textbf{0.9932} & \textbf{0.9934} & \textbf{0.9935} & \textbf{0.9934} & \textbf{29.27} & \textbf{29.28} & \textbf{30.20} & \textbf{30.39} & \textbf{24.00} & \textbf{24.80} & \textbf{25.60} & \textbf{25.68} \\
Advanced ESP{\cite{han_simplification_2015_1,han_simplification_2015_2}}         & 0.9928          & 0.9928          & 0.9927          & 0.9927          & 32.12          & 31.04          & 31.76          & 31.94          & 27.33          & 26.22          & 26.79          & 26.82          \\
Classic ESP          & 0.9561          & 0.9544          & 0.9461          & 0.9432          & 114.7          & 61.57          & 84.00          & 93.55          & 96.22          & 47.95          & 73.13          & 82.58          \\ \hline
NCM811           & $0^+$ & $L^+/3$ & $2L^+/3$ & $L^+$ & $0^+$ & $L^+/3$ & $2L^+/3$ & $L^+$ & $0^+$ & $L^+/3$ & $2L^+/3$ & $L^+$    \\
Proposed         & \textbf{0.9943} & \textbf{0.9940} & \textbf{0.9936} & \textbf{0.9934} & \textbf{32.46} & \textbf{32.79} & \textbf{33.25} & \textbf{33.57} & \textbf{27.10} & \textbf{27.83} & \textbf{28.33} & \textbf{28.59} \\
Advanced ESP{\cite{han_simplification_2015_1,han_simplification_2015_2}}         & 0.9938          & 0.9934          & 0.9930          & 0.9927          & 35.05          & 34.61          & 34.96          & 35.20          & 29.54          & 29.34          & 29.69          & 29.87          \\
Classic ESP          & 0.9627          & 0.9608          & 0.9527          & 0.9500          & 82.19          & 63.23          & 79.76          & 85.04          & 69.82          & 48.14          & 68.89          & 74.22          \\ 
\toprule
\end{tabular}}
\end{table*}

\subsection{Output prediction}\label{sec:outputprediction}
\begin{figure}[!htb]
    \centering
\begin{minipage}{.24\textwidth}
        \centering
        \includegraphics[width=\textwidth]{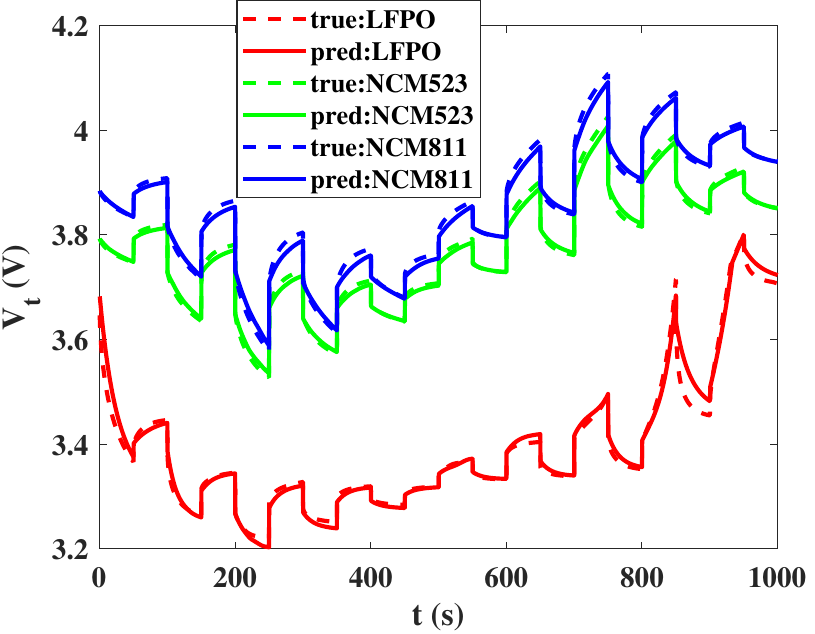}
\subcaption{\textcolor{blue}{$V_t$ of LFPO, NCM523, NCM811 cells, ACC. }}\label{fig:VtACC}
\end{minipage}%
\begin{minipage}{0.24\textwidth}
        \centering
        \includegraphics[width=\textwidth]{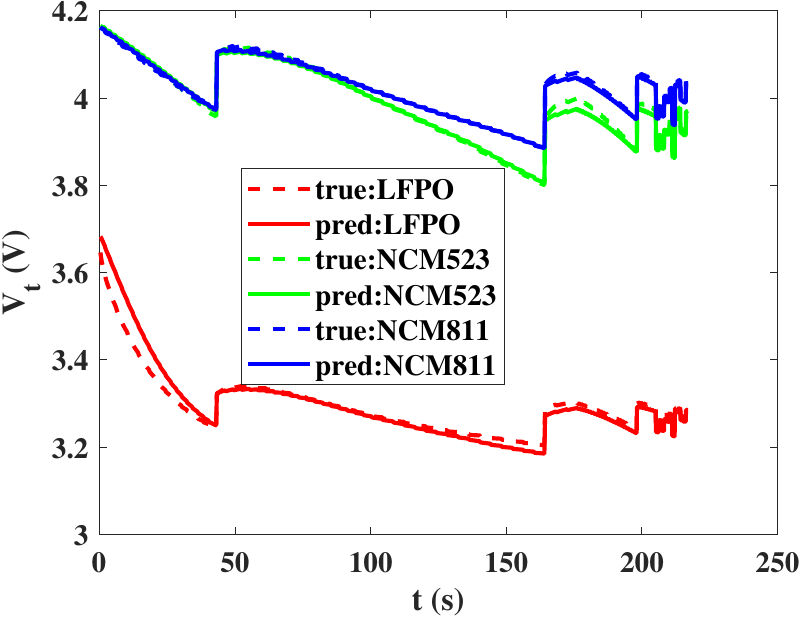}
\subcaption{\textcolor{blue}{$V_t$ of LFPO, NCM523, NCM811 cells, RC. }}\label{fig:VtRC}
\end{minipage}
\begin{minipage}{.24\textwidth}
        \centering
        \includegraphics[width=\textwidth]{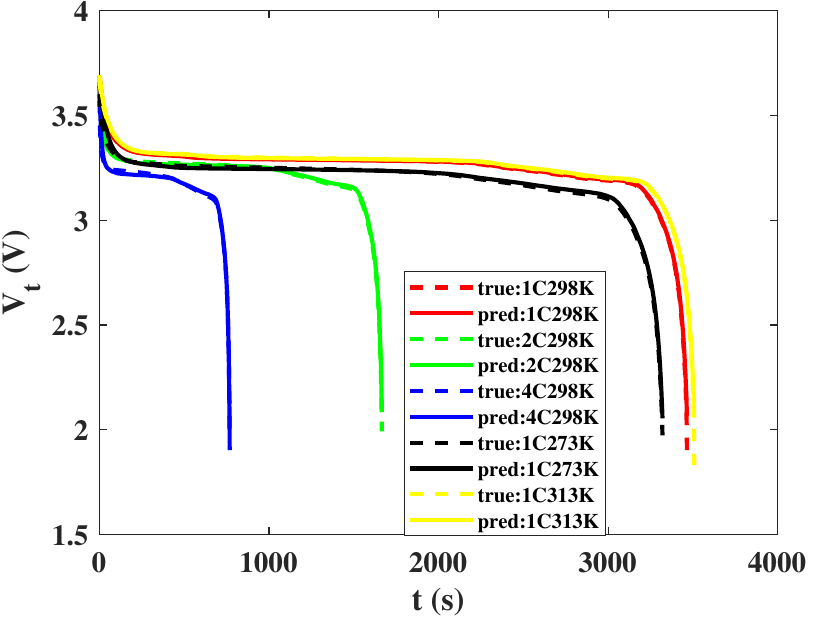}
\subcaption{\textcolor{blue}{$V_t$ of LFPO under different galvanostatic protocols. }}\label{fig:Vt_LFPO}
\end{minipage}%
\begin{minipage}{0.24\textwidth}
        \centering
        \includegraphics[width=\textwidth]{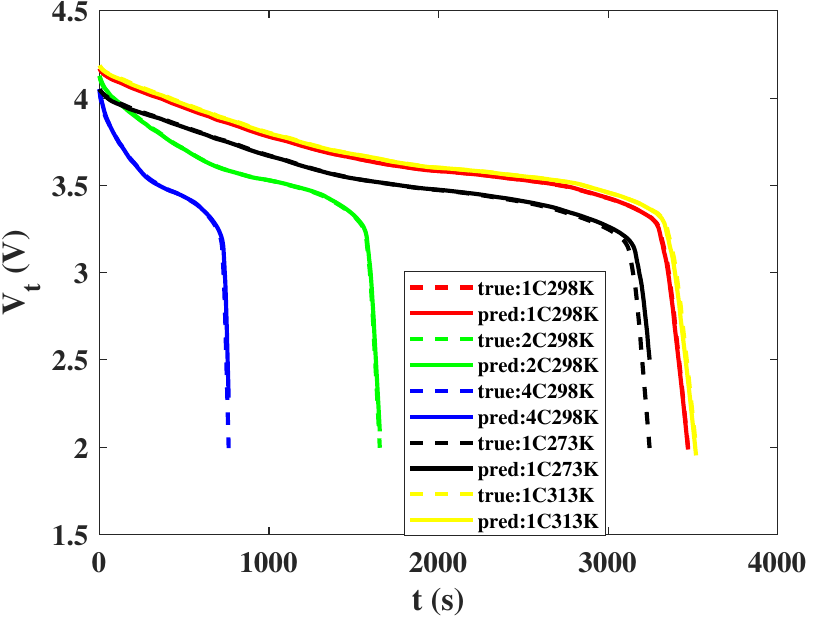}
\subcaption{\textcolor{blue}{$V_t$ of NCM523 under different galvanostatic protocols. }}\label{fig:Vt_NCM523}
\end{minipage}
\begin{minipage}{.24\textwidth}
        \centering
        \includegraphics[width=\textwidth]{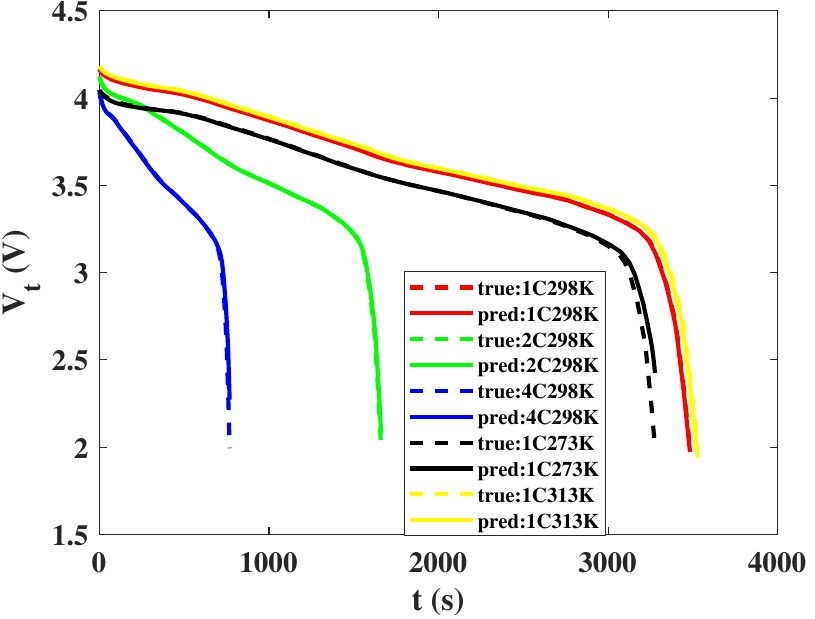}
\subcaption{\textcolor{blue}{$V_t$ of NCM811 under different galvanostatic protocols. }}\label{fig:Vt_NCM811}
\end{minipage}%
\begin{minipage}{0.24\textwidth}
        \centering
        \includegraphics[width=\textwidth]{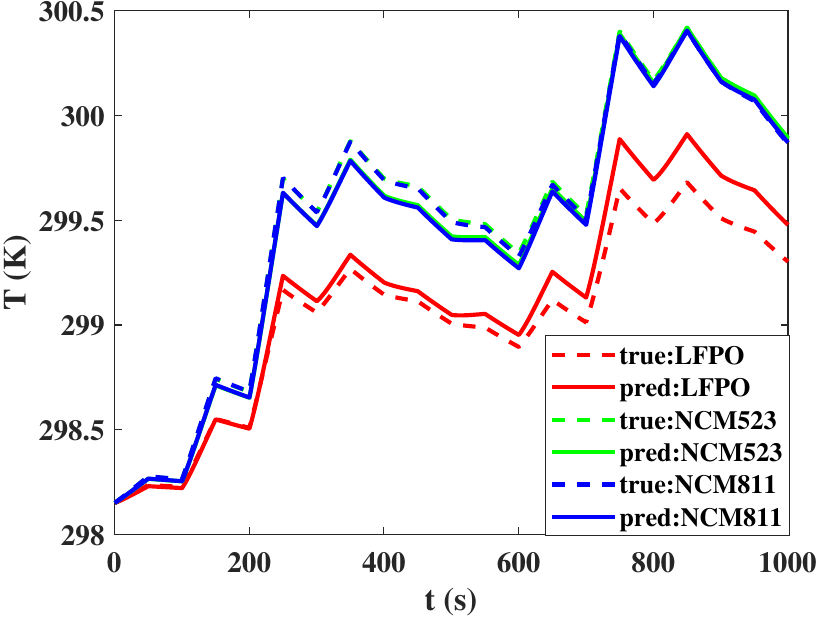}
\subcaption{\textcolor{blue}{$T$ of LFPO, NCM523, NCM811 cells, ACC. }}\label{fig:TtACC}
\end{minipage}
\begin{minipage}{.24\textwidth}
        \centering
        \includegraphics[width=\textwidth]{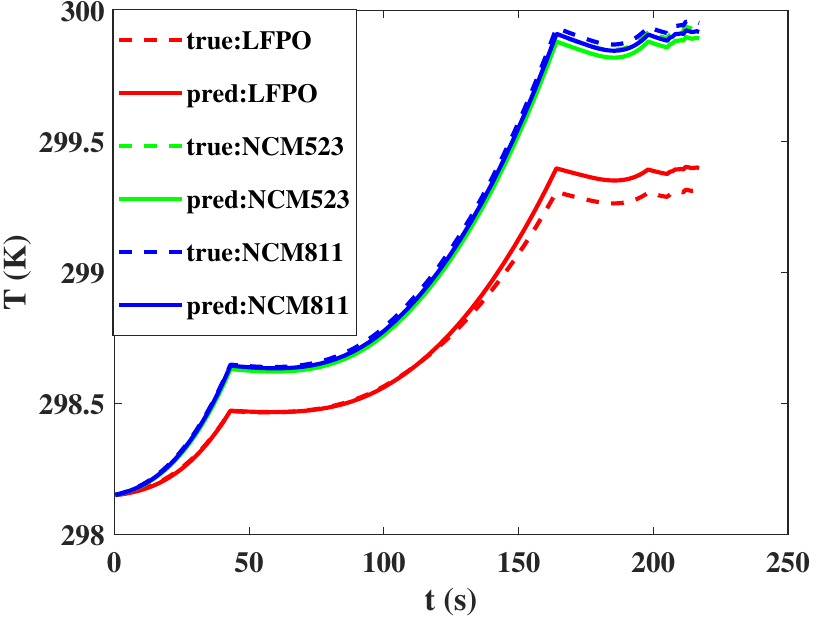}
\subcaption{\textcolor{blue}{$T$ of LFPO, NCM523, NCM811 cells, RC. }}\label{fig:TtRC}
\end{minipage}%
\begin{minipage}{0.24\textwidth}
        \centering
        \includegraphics[width=\textwidth]{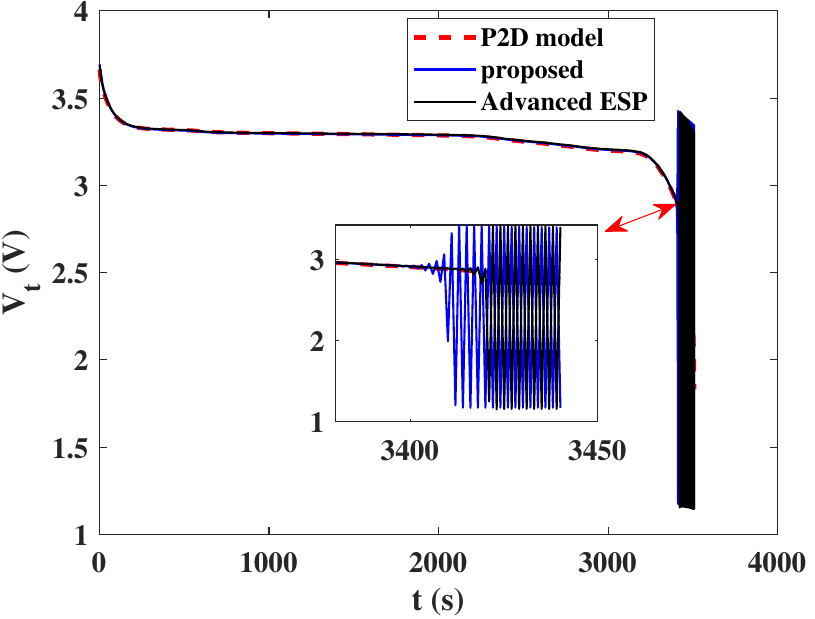}
\subcaption{\textcolor{blue}{$Vt$ of LFPO under the 313 K galvanostatic protocol without SGF. }}\label{fig:Vt_LFPO_313K}
\end{minipage}
\caption{\textcolor{blue}{Trajectories of outputs (including $V_t$ and $T$) of LFPO, NCM523 and NCM811 cells under different protocols. }}
\end{figure}

\textcolor{blue}{
After reviewing the model prediction performance on different internal states, we analyse the output prediction performance in this part.
The outputs, including $V_t$ and $T$, are both predicted in the model. Between them, we care more about the accurate prediction of $V_t$ than that of $T$ for two reasons. First, as explained in Section \ref{sec:thermal}, a lumped thermal model is developed to predict $T$ for simplicity, which can give only approximate predictions. Thus, we only expect the trajectory track of $T$ to meet the basic requirements. Second, the $V_t$ signal can directly help us conduct parameter identification and develop online control strategies; thus, it was often considered in previous research. Actually, when we review existing models of LIBs, regardless of whether they are electro-chemical models or EC models, mapping between the input current and output voltage is always the key problem to discuss. Thus, in this paper, we continue this rule and focus more on $V_t$. The prediction accuracy of $V_t$ is given in Table~\ref{table:v_accuracy}. Generally, our model performs better for all types of cells. Considering the MAE, the prediction accuracy of LFPO is approximately 25\% higher than that of the advanced ESP, and that of NCM cells is approximately 100\% higher than that of the advanced ESP. The $V_t$ values of the three cells under dynamic current protocols and galvanostatic protocols are plotted in Figs. \ref{fig:VtACC}-\ref{fig:VtRC} and Figs. \ref{fig:Vt_LFPO}-\ref{fig:Vt_NCM811}. High agreement can be observed for three cells in different scenarios. Near the end of discharge, the accuracy decreases somewhat, especially at low ambient temperature (273 K), indicating that caution should be taken when using the model under low temperature and that it would be better not to overdischarge the battery. In Figs. \ref{fig:TtACC}-\ref{fig:TtRC}, the predicted $T$ is plotted, and the maximum error under dynamic current protocols is approximately 0.2 K, which basically meets the requirement of practical use. In Fig.~\ref{fig:Vt_LFPO_313K}, we plot the case in which oscillation occurs. Actually, this is the only case we find that exhibits oscillation. As analysed above, the oscillation is caused by the assumption that $j_n$ remains constant within a simulation step and is no longer correct. When the LFPO is near the end of discharge, $\frac{U_{\mathrm{OCP}}}{\theta_{ss}}$ is very large. Additionally, under high ambient temperature (313 K), the reaction is very active. Thus, the variation of $j_n$ can be very substantial in a short time interval and ultimately results in oscillation. Both our model and the advanced ESP exhibit oscillation. Although the possibility of this situation is not high, it is still necessary to deploy a suitable stabilizer to ensure the reliability of the model. In this work, we set the hyperparameters of the SGF, $N_{\mathrm{SG}}=2$ and $M_{\mathrm{SG}}=49$. The solid yellow line in Fig.~\ref{fig:Vt_LFPO} shows $V_t$ after filtering; the oscillation is eliminated effectively.}

\begin{table*}[width=2.1\linewidth,cols=3,pos=h]
\caption{Prediction accuracy of $V_t$ of LFPO, NCM523, and NCM811 cells.}\label{table:v_accuracy}
\resizebox{\textwidth}{!}{\begin{tabular}{cccccccccc}
\toprule
     & \multicolumn{3}{c}{LFPO}                    & \multicolumn{3}{c}{NCM523}            & \multicolumn{3}{c}{NCM811}            \\
     & Proposed         & {\begin{tabular}[c]{@{}c@{}}Advanced\\ ESP{\cite{han_simplification_2015_1,han_simplification_2015_2}}\end{tabular}}
             & {\begin{tabular}[c]{@{}c@{}}Classic\\ ESP\end{tabular}}     & Proposed         & {\begin{tabular}[c]{@{}c@{}}Advanced\\ ESP{\cite{han_simplification_2015_1,han_simplification_2015_2}}\end{tabular}} & {\begin{tabular}[c]{@{}c@{}}Classic\\ ESP\end{tabular}}     & Proposed         & {\begin{tabular}[c]{@{}c@{}}Advanced\\ ESP{\cite{han_simplification_2015_1,han_simplification_2015_2}}\end{tabular}} & {\begin{tabular}[c]{@{}c@{}}Classic\\ ESP\end{tabular}}      \\ \hline
R2   & 0.979            & \textbf{0.983} & 0.953   & \textbf{0.983}   & 0.974    & 0.949   & \textbf{0.992}   & 0.984    & 0.955   \\
RMSE & \textbf{0.01371} & 0.01536        & 0.02143 & \textbf{0.02459} & 0.03023  & 0.02636 & \textbf{0.01995} & 0.02499  & 0.02323 \\
MAE  & \textbf{0.00774} & 0.00927        & 0.01205 & \textbf{0.00843} & 0.01617  & 0.01859 & \textbf{0.00771} & 0.01456  & 0.01629 \\ \bottomrule
\end{tabular}}
\end{table*}

\subsection{Closed-loop correction}
\begin{figure}[!htb]
    \centering
\begin{minipage}{.24\textwidth}
        \centering
        \includegraphics[width=\textwidth]{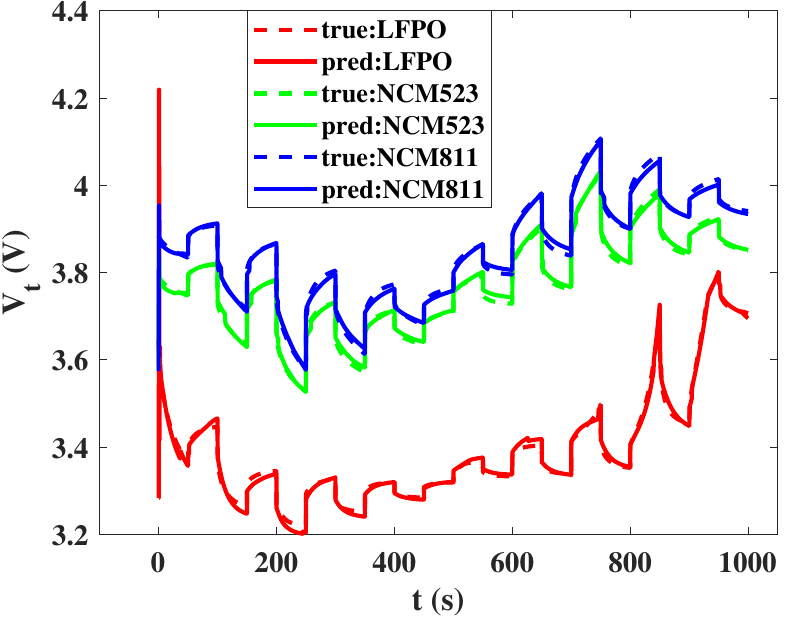}
\subcaption{\textcolor{blue}{$V_t$ of LFPO, NCM523, NCM811 cells, ACC, wrong initialization. }}\label{fig:VtclosedACC}
\end{minipage}%
\begin{minipage}{0.24\textwidth}
        \centering
        \includegraphics[width=\textwidth]{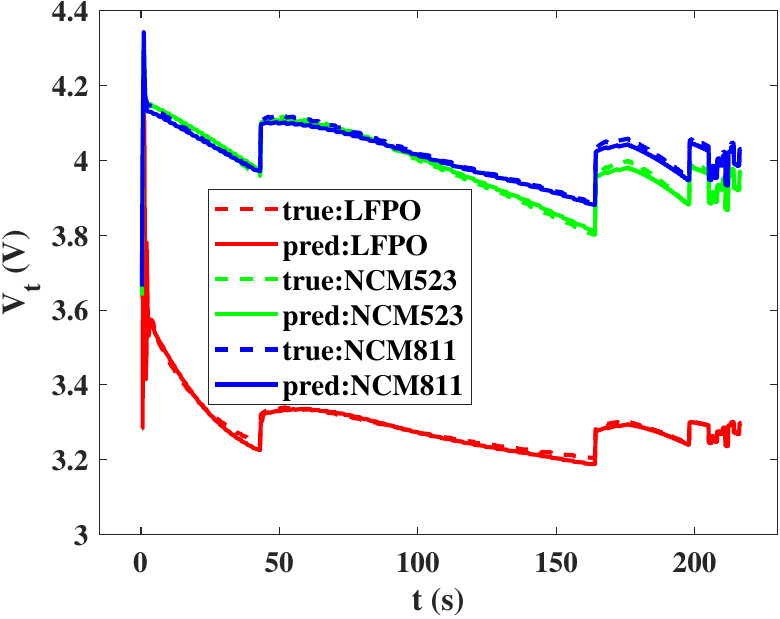}
\subcaption{\textcolor{blue}{$V_t$ of LFPO, NCM523, NCM811 cells, RC, wrong initialization. }}\label{fig:VtclosedRC}
\end{minipage}
\begin{minipage}{.24\textwidth}
        \centering
        \includegraphics[width=\textwidth]{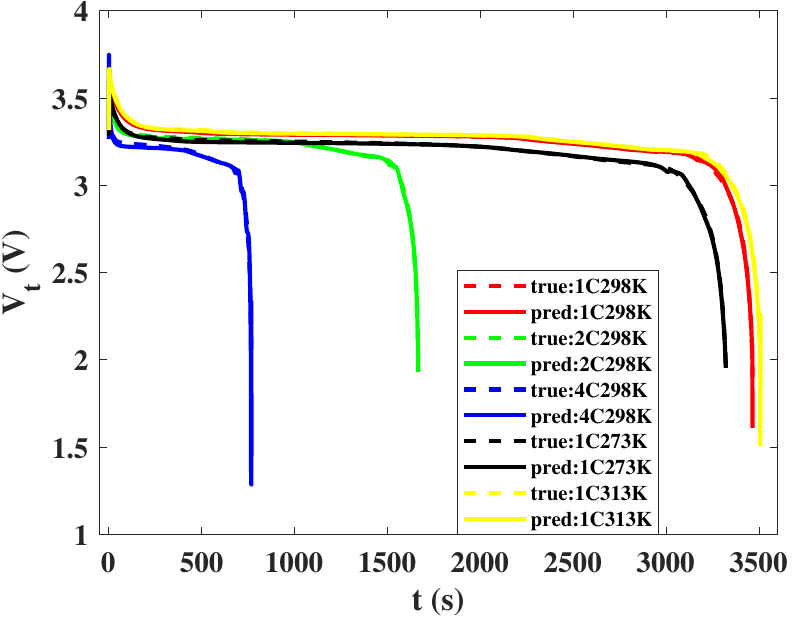}
\subcaption{\textcolor{blue}{$V_t$ of LFPO under different galvanostatic protocols, wrong initialization. }}\label{fig:Vt_closedLFPO}
\end{minipage}%
\begin{minipage}{0.24\textwidth}
        \centering
        \includegraphics[width=\textwidth]{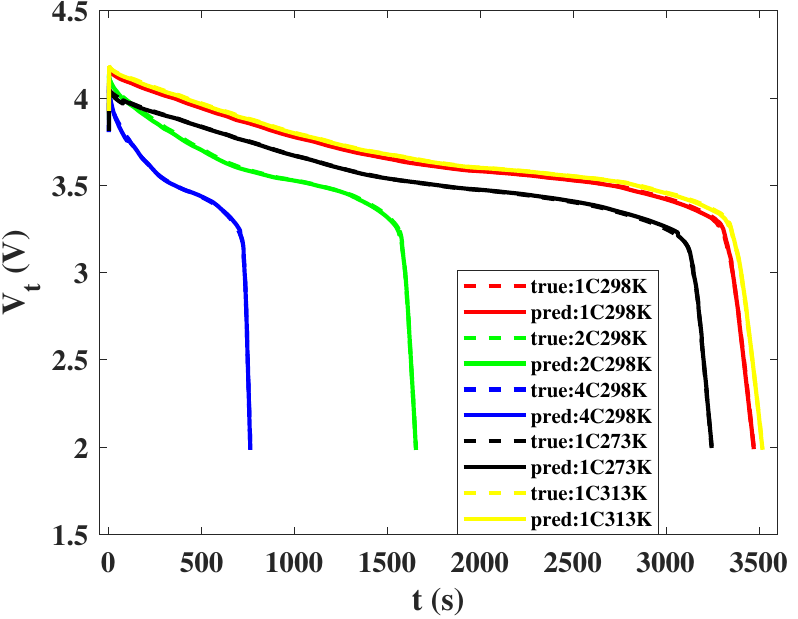}
\subcaption{\textcolor{blue}{$V_t$ of NCM523 under different galvanostatic protocols, wrong initialization. }}\label{fig:Vt_closedNCM523}
\end{minipage}
\begin{minipage}{.24\textwidth}
        \centering
        \includegraphics[width=\textwidth]{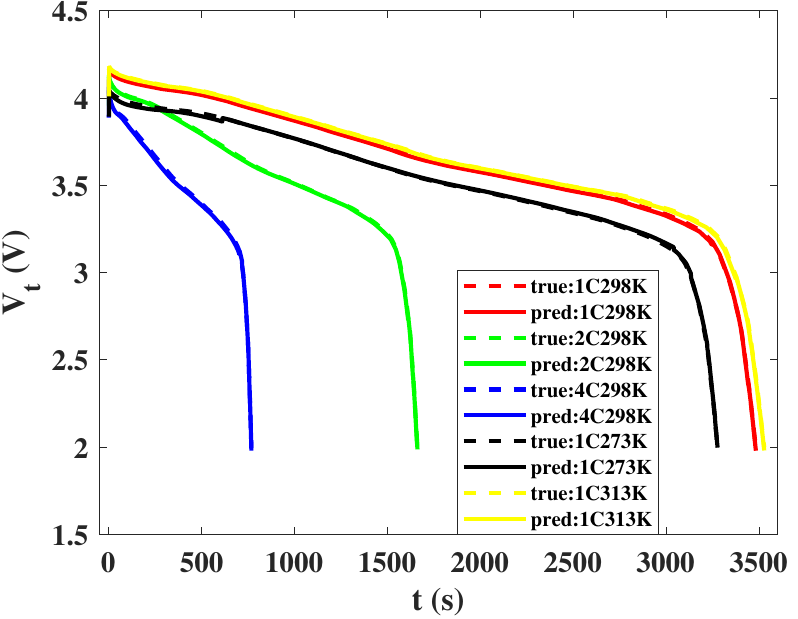}
\subcaption{\textcolor{blue}{$V_t$ of NCM811 under different galvanostatic protocols, wrong initialization. }}\label{fig:Vt_closedNCM811}
\end{minipage}%
\begin{minipage}{0.24\textwidth}
        \centering
        \includegraphics[width=\textwidth]{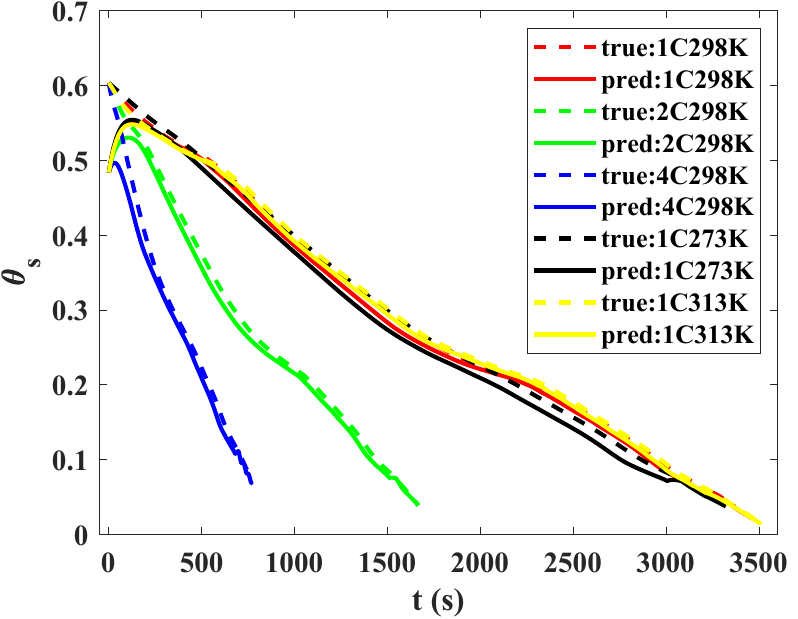}
\subcaption{\textcolor{blue}{$\theta_s$ at the negative electrode boundary of LFPO under different galvanostatic protocols, wrong initialization. }}\label{fig:xsCC_close_LFPO}
\end{minipage}
\begin{minipage}{.24\textwidth}
        \centering
        \includegraphics[width=\textwidth]{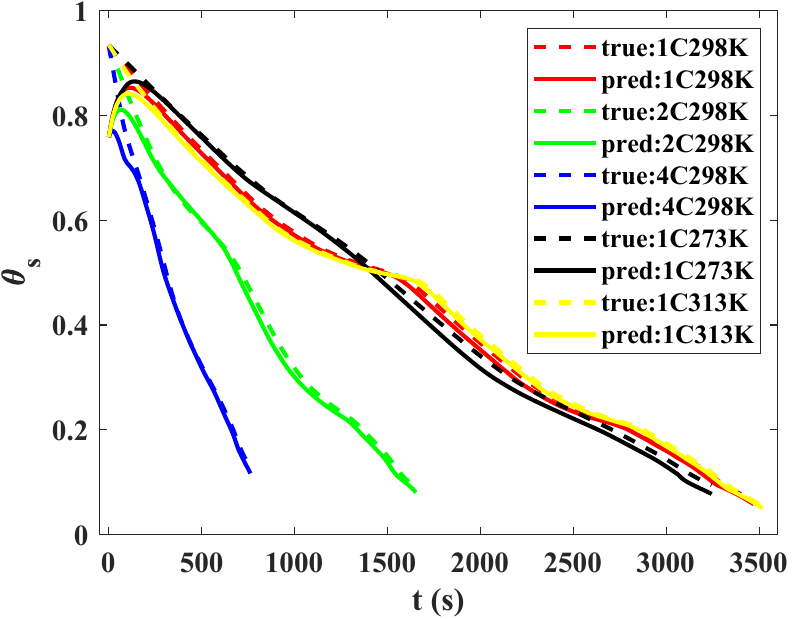}
\subcaption{\textcolor{blue}{$\theta_s$ at the negative electrode boundary of NCM523 under different galvanostatic protocols, wrong initialization. }}\label{fig:xsCC_close_NCM523}
\end{minipage}%
\begin{minipage}{0.24\textwidth}
        \centering
        \includegraphics[width=\textwidth]{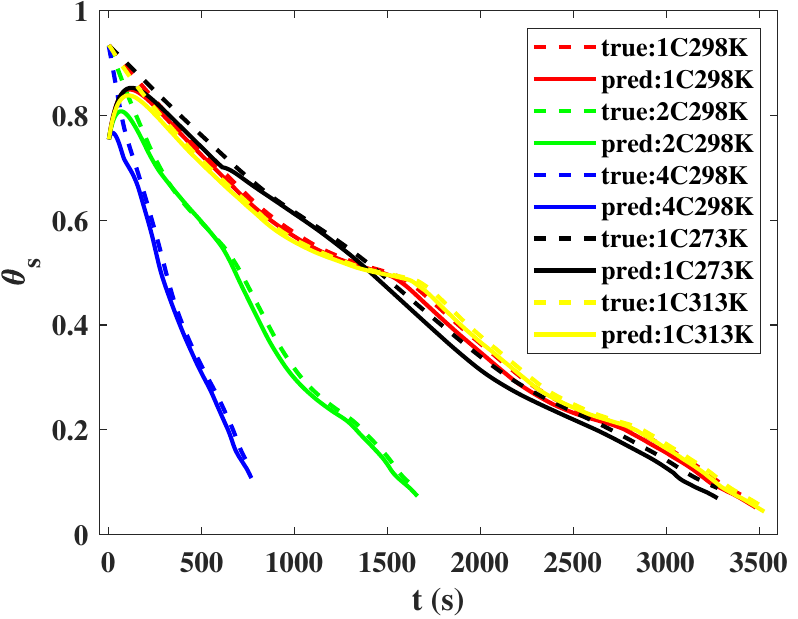}
\subcaption{\textcolor{blue}{$\theta_s$ at the negative electrode boundary of NCM811 under different galvanostatic protocols, wrong initialization. }}\label{fig:xsCC_close_NCM811}
\end{minipage}
\caption{\textcolor{blue}{Trajectories of $V_t$ and $\theta_s$ of LFPO, NCM523 and NCM811 cells under different protocols when the battery is wrongly initialized. }}
\end{figure}

Although the state monitoring and output prediction agree well with those of the full-order P2D model, a closed-loop framework is necessary for real-world applications since the initialization error and the model error accumulate in continuous simulation. \textcolor{blue}{To evaluate the reliability of the proposed closed-loop correction scheme, we set the initial SOC of the battery cell to the wrong values. Specifically, the initial SOC of the galvanostatic discharge protocols is set at 0.8, where the true value is 1. The initial SOC of the CCCV protocol is set at 0.2, where the true value is 0. The initial SOC of dynamic current protocols is set at
0.5, where the true value is 1 for the LFPO cell and 0.7 for the NCM cells. Figs.~\ref{fig:VtclosedACC}-\ref{fig:VtclosedRC} plots $V_t$ under dynamic current protocols. After a short time oscillation, $V_t$ quickly corrects to the true value. The same phenomenon is observed under galvanostatic protocols in Figs.~\ref{fig:Vt_closedLFPO}-\ref{fig:Vt_closedNCM811}. As the current rate varies from 1C to 4C, the ambient temperature varies from 273 K to 313 K, and the correction scheme can always perform well. In addition, we note that the increment of the $V_T$ error at the end of discharge (as shown in Figs.~\ref{fig:Vt_LFPO}-\ref{fig:Vt_NCM811}) is also eliminated. Figs.~\ref{fig:xsCC_close_LFPO}-\ref{fig:xsCC_close_NCM811} plot the trajectories of $\theta_s$. The wrong initialized $\theta_s$ at the interface between the negative electrode and separator is quickly corrected to the true value, verifying the effectiveness of the proposed scheme.}

\section{Conclusions}\label{sec:con}
\textcolor{blue}{This paper proposes a simplified electro-chemical model along with a specific simulation framework that enables the in situ monitoring and online control of commonly used NCM and LFPO batteries. A bottom-up approach is designed to construct the model, which not only makes the model adaptive to variant working environments and materials but also reserves potential for future upgrades. Comprehensive numerical experiments validate the effectiveness and superiority of this work, which provides opportunities for degradation analysis and meticulous management of batteries in practice. However, there still remain three limitations of this work. First, the proposed model is derived from the P2D model; thus, it is suitable only for those batteries that can be described by the P2D model. Since LIB technology is under rapid development, the proposed model
might not be suitable for future LIBs with advanced material technologies and needs further upgrades. Second, a lumped thermal model is used to predict the cell temperature. However, sometimes the lack of information on the temperature distribution can lead to the unawareness of severe problems such as thermal runaway. Since the temperature distribution is closely related to the reaction rate distribution, we can use the reaction information as the signal to thermal runaway as a matter of expediency in this work. Regardless, it would be better to incorporate a thermal model that can estimate the spatial temperature with low complexity. Third, we should know the values of all the parameters involved in the model to implement this work in real-world applications. However, some of the parameters cannot be directly measured. Thus, a specific parameter identification method should be developed to ensure the practicability of this work. In future work, we plan to focus on addressing the above three challenges.}

\appendix

\printcredits

\bibliographystyle{model1-num-names}

\bibliography{cas-refs}

\end{document}